\documentclass[prb,aps,tightenlines,twocolumn,floatfix,superscriptaddress,showpacs,preprintnumbers,citeautoscript,longbibliography,10pt]{revtex4-1}
\usepackage{amsmath}
\usepackage{amssymb}
\usepackage{mathrsfs}
\usepackage{bm}
\usepackage{graphicx}
\usepackage{xfrac}
\usepackage{url}
\usepackage{tikz}
\usepackage{multirow}
\usepackage{mathtools}
\usepackage{braket}
\usepackage{courier}
\usepackage[version=4]{mhchem}
\usepackage[T1]{fontenc}
\usepackage{scalerel}
\usepackage{soul}
\usepackage{algorithm}
\usepackage{algpseudocode}
\usepackage{tabularx}
\usepackage{xspace}
\usepackage{calrsfs}

\AtBeginDocument{%
    \newwrite\bibnotes
    \def\bibnotesext{Notes.bib}
    \immediate\openout\bibnotes=\jobname\bibnotesext
    \immediate\write\bibnotes{@CONTROL{REVTEX41Control}}
    \immediate\write\bibnotes{@CONTROL{%
    apsrev41Control,author="08",editor="1",pages="1",title="0",year="1"}}
     \if@filesw
     \immediate\write\@auxout{\string\citation{apsrev41Control}}%
    \fi
}%

\algnewcommand{\Or}{\textbf{ or }}
\algnewcommand{\And}{\textbf{ and }}
\algnewcommand{\Break}{\textbf{break}}
\algnewcommand{\Continue}{\textbf{continue}}
\algrenewcommand\algorithmicforall{\textbf{foreach}}
\def\dd{\mathrm{d}}
\newcommand{\RN}[1]{\textup{\uppercase\expandafter{\romannumeral#1}}}
\newcommand{\BRN}[1]{\textbf{\textup{\uppercase\expandafter{\romannumeral#1}}}}
\definecolor{dartmouthgreen}{rgb}{0.05, 0.5, 0.06}
\newcommand{\tphin}[1]{\widetilde{\phi}_{#1}}
\newcommand{\tphi}[1]{\widetilde{\phi}_{#1}(\bm r)}
\newcommand{\tphib}[1]{\widetilde{\phi}_{#1}(\overline{\bm r})}
\newcommand{\trho}[1]{\widetilde{\rho}_{#1}(\bm r)}
\newcommand{\trhon}[1]{\widetilde{\rho}_{#1}}

\newcommand{\trhob}[1]{\widetilde{\rho}_{#1}(\overline{\bm r})}

\newcommand{\tv}[1]{\widetilde{v}_{#1}(\bm r)}

\newcommand{\tvb}[1]{\widetilde{v}_{#1}(\overline{\bm r})}
\newcommand{\tC}[1]{\widetilde{\bm C}_{#1}}
\newcommand{\gC}[1]{\bm C_{#1}}
\newcommand{\xmega}[1]{\Omega_{#1}}

\newcommand{\rbar}{\overline{\bm r}}
\newcommand{\sbar}{\overline{\bm s}}

\newcommand{\rpr}{R_{\rm pair}}
\newcommand{\rpe}{R_{\rm PE}}
\newcommand{\rme}{R_{\rm ME}}
\newcommand{\npe}{N_{\rm PE}}
\newcommand{\nme}{N_{\rm ME}}
\newcommand{\rpes}{R_{\rm PE}^{\rm s}}
\newcommand{\rmes}{R_{\rm ME}^{\rm s}}
\newcommand{\rpep}{R_{\rm PE}^{\rm ns}}
\newcommand{\rmep}{R_{\rm ME}^{\rm ns}}
\newcommand{\npes}{N_{\rm PE}^{\rm s}}
\newcommand{\nmes}{N_{\rm ME}^{\rm s}}
\newcommand{\npep}{N_{\rm PE}^{\rm ns}}
\newcommand{\nmep}{N_{\rm ME}^{\rm ns}}
\newcommand{\exx}{E_{\rm xx}}

\newcommand{\sxxele}{\sigma_{\rm xx}}
\newcommand{\sxx}{\bm \sigma_{\rm xx}}
\newcommand{\dxx}[1]{\widetilde{D}_{\rm xx}^{#1}(\bm r)}
\newcommand{\dxxb}[1]{\widetilde{D}_{\rm xx}^{#1}(\overline{\bm r})}

\newcommand{\lu}{\mathcal{L}}

\newcommand{\pe}[1]{\Theta(#1, \rpe)}
\newcommand{\me}[1]{\Theta(#1, \rme)}

\newcommand{\pep}[1]{\Theta(#1, \rpep)}

\newcommand{\mpi}[0]{\texttt{MPI}}
\newcommand{\omp}[0]{\texttt{OpenMP}}

\DeclareMathOperator{\sgn}{sgn}
\newcommand{\exxm}[0]{\texttt{exx}\xspace}
\newcommand{\ie}[0]{\textit{i.e.}, }
\newcommand{\eg}[0]{\textit{e.g.}, }
\newcommand{\via}[0]{\textit{via} }
\newcommand{\cf}[0]{\textit{cf}.\ }
\newcommand{\qe}[0]{\texttt{Quantum ESPRESSO} }
\newcommand{\pI}[0]{\textsc{paper-i}\xspace}

\newcommand{\DefineAuthor}[2]{%
  \expandafter\newcommand\csname #1note\endcsname[1]{%
    \textbf{\textcolor{#2}{\textbf{#1:} ##1}}}%
  \expandafter\newcommand\csname #1\endcsname[1]{
    \textbf{\textcolor{#2}{##1}}}
  \expandafter\newcommand\csname #1cancel\endcsname[1]{%
    \textbf{\textcolor{#2}{\sout{##1}}}}%
  \expandafter\newcommand\csname #1change\endcsname[2]{%
    \textbf{\textcolor{#2}{\sout{##1} ##2}}}%
  \newenvironment{#1text}{\color{#2}}{\color{black}}
}

\definecolor{dartmouthgreen}{rgb}{0.05, 0.5, 0.06}
\DefineAuthor{red}{red}
\DefineAuthor{blue}{blue}
\DefineAuthor{RAD}{red}
\DefineAuthor{HK}{blue}
\DefineAuthor{GEN}{dartmouthgreen}

\usetikzlibrary{calc}
\usetikzlibrary{positioning}
\usetikzlibrary{shapes,arrows}
\usetikzlibrary{arrows.meta}
\tikzstyle{if} = [diamond, draw, fill=blue!20, text width=4.5em, text badly centered, node distance=3cm, inner sep=0pt]
\tikzstyle{prg} = [rectangle, draw, fill=blue!40, text width=5em, text centered, rounded corners, minimum height=4em]
\tikzstyle{prgN} = [rectangle, draw, fill=brown!50, text width=5em, text centered, rounded corners, minimum height=4em]
\tikzstyle{sbrt} = [rectangle, draw, fill=blue!0, text width=15em, text centered, rounded corners, minimum height=5em]
\tikzstyle{sbrtR} = [rectangle, draw, red, fill=blue!0, text width=15em, text centered, rounded corners, minimum height=5em]
\tikzstyle{sbrtN} = [rectangle, draw, fill=brown!30, text width=6.2em, text centered, rounded corners, minimum height=4em]
\tikzstyle{mod} = [rectangle, draw, fill=yellow!20, text width=5em, text centered, rounded corners, minimum height=4em]
\tikzstyle{inp} = [circle, draw, fill=purple!40, text width=3.5em, text centered, rounded corners, minimum height=3em]
\tikzstyle{inpN} = [circle, draw, fill=red!40, text width=3.5em, text centered, rounded corners, minimum height=3em]
\tikzstyle{outp} = [circle, draw, fill=yellow!40, text width=3.5em, text centered, rounded corners, minimum height=3em]
\tikzstyle{line} = [draw, -latex']
\tikzstyle{set} = [draw, ellipse,fill=red!20, node distance=3cm, minimum height=2em]
\tikzstyle{var} = [draw, circle,fill=green!0, node distance=3cm,text width=5em,text centered, minimum height=2em] 
\tikzstyle{vary} = [draw, circle,fill=yellow!30, node distance=3cm,text width=5em,text centered, minimum height=2em] 
\tikzstyle{varg} = [draw, circle,fill=violet!30, node distance=3cm,text width=5em,text centered, minimum height=2em] 
\tikzstyle{varb} = [draw, circle,fill=brown!30, node distance=3cm,text width=5em,text centered, minimum height=2em] 
\tikzstyle{varN} = [draw, circle,fill=green!40, node distance=3cm,text width=3.5em,text centered, minimum height=2em] 
\tikzstyle{dummy} = [] 
\tikzstyle{comm} = [rectangle, draw, fill=cyan!20, text width=5em, text centered, rounded corners, minimum height=4em]

\begin{document} 
\title{Enabling Large-Scale Condensed-Phase Hybrid Density Functional Theory Based \\ \textit{Ab Initio} Molecular Dynamics II: Extensions to the Isobaric-Isoenthalpic and Isobaric-Isothermal Ensembles
}
\author{Hsin-Yu Ko}
\affiliation{Department of Chemistry and Chemical Biology, Cornell University, Ithaca, NY 14853, USA}
\author{Biswajit Santra}
\affiliation{Department of Physics, Temple University, Philadelphia, PA 19122, USA}
\author{Robert A. DiStasio Jr.}
\email{distasio@cornell.edu} 
\affiliation{Department of Chemistry and Chemical Biology, Cornell University, Ithaca, NY 14853, USA}
\date{\today}


\begin{abstract}
In the previous paper of this series [Ko,~H-Y. \textit{et al.} J. Chem. Theory Comput. 2020, \textbf{16}, 3757--3785], we presented a theoretical and algorithmic framework based on a localized representation of the occupied space that exploits the inherent sparsity in the real-space evaluation of the exact exchange (EXX) interaction in finite-gap systems.
This was accompanied by a detailed description of \exxm, a massively parallel hybrid \mpi{}/\omp{} implementation of this approach in \texttt{Quantum ESPRESSO} that enables linear-scaling hybrid DFT based \textit{ab initio} molecular dynamics (AIMD) in the microcanonical/canonical ($NVE$/$NVT$) ensembles of condensed-phase systems containing $500\mathrm{-}1000$~atoms (in fixed orthorhombic cells) with a wall time cost comparable to semi-local DFT.
In this work, we extend the current capabilities of \exxm to enable hybrid DFT based AIMD simulations of large-scale condensed-phase systems with general and fluctuating cells in the isobaric-isoenthalpic/isobaric-isothermal ($NpH$/$NpT$) ensembles.
Theoretical extensions to this approach include an analytical derivation of the EXX contribution to the stress tensor for systems in general simulation cells with a computational complexity that scales linearly with system size.
The corresponding algorithmic extensions to \exxm include optimized routines that: (\textit{i}) handle both static and fluctuating simulation cells with non-orthogonal lattice symmetries, (\textit{ii}) solve Poisson's equation in general/non-orthogonal cells \via an automated selection of the auxiliary grid directions in the Natan-Kronik representation of the discrete Laplacian operator, and (\textit{iii}) evaluate the EXX contribution to the stress tensor.
Using this approach, we perform a case study on a variety of condensed-phase systems (including liquid water, a benzene molecular crystal polymorph, and semi-conducting crystalline silicon) and demonstrate that the EXX contributions to the energy and stress tensor simultaneously converge with an appropriate choice of \exxm parameters. 
This is followed by a critical assessment of the computational performance of the extended \exxm module across several different high-performance computing (HPC) architectures \via case studies on: (\textit{i}) the computational complexity due to lattice symmetry during $NpT$ simulations of three different ice polymorphs (\ie ice I$h$, II, III), and (\textit{ii}) the strong/weak parallel scaling during large-scale $NpT$ simulations of liquid water.
We demonstrate that the robust and highly scalable implementation of this approach in the extended \exxm module is capable of evaluating the EXX contribution to the stress tensor with negligible cost ($< 1\%$) as well as all other EXX-related quantities needed during $NpT$ simulations of liquid water (with a very tight $150$~Ry planewave cutoff) in $\approx 5.2$~s (\ce{(H2O)128}) and $\approx 6.8$~s (\ce{(H2O)256}) per AIMD step.
As such, the extended \exxm module presented in this work brings us another step closer to routinely performing hybrid DFT based AIMD simulations of sufficient duration for large-scale condensed-phase systems across a wide range of thermodynamic conditions.
\end{abstract}

\maketitle

\section{Introduction \label{sec:intro}}

Molecular dynamics (MD) is a deterministic numerical simulation method for efficiently sampling high-dimensional potential energy surfaces (PES) in systems of importance throughout biology, chemistry, physics, and materials science~\cite{frenkel_understanding_2001,allen_computer_1989}.
Following the fundamental postulates of statistical mechanics, the trajectory of an MD simulation can be used to determine the thermodynamic properties of a system, as well as connect such macroscopic quantities to microscopic behavior. 
As such, MD simulations are commonly used to furnish detailed microscopic-level insight into a wide range of phenomena, including (but not limited to) the assembly and structure of large-scale nanostructures and materials~\cite{finocchi_microscopic_1992,johnson_probing_2008,trabuco_flexible_2008,zhao_mature_2013,martelli_local-order_2018}, chemical reactions and kinetics~\cite{bergsma_molecular_1987,van_gunsteren_computer_1990,craig_chemical_2005,van_voorhis_diabatic_2010,santra_root-growth_2018}, as well as complex biological processes~\cite{cheatham_observation_1996,sugita_replica-exchange_1999,karplus_molecular_2002,martelli_structural_2018}.
In practice, MD simulations of finite-sized systems are performed in the statistical mechanical ensemble corresponding to the thermodynamic conditions used to prepare and characterize the system of interest.
In the microcanonical ($NVE$) ensemble, for example, the particle number ($N$), volume ($V$), and total internal energy ($E$) of the system are kept constant, which corresponds to an isolated system under adiabatic conditions.
In canonical ($NVT$) MD simulations, the energy associated with endothermic and exothermic processes is exchanged with a thermostat at a fixed temperature ($T$), which allows one to account for thermal effects at constant $V$ (or constant system density, $N/V$).
To account for an externally applied pressure ($p$), a barostat can be introduced to facilitate sampling in the isobaric-isoenthalpic ($NpH$, when decoupled from a thermostat) and isobaric-isothermal ($NpT$, when coupled to a thermostat) ensembles, thereby enabling direct comparison to a larger swath of experiments (as most are performed at constant $p$ instead of constant $V$).
Another ensemble worth mention includes the grand canonical ($\mu VT$) ensemble, which fixes the chemical potential ($\mu$) and enables MD simulations of open systems in contact with thermal and particle reservoirs.
Since MD is an importance sampling technique, it can also be used to efficiently generate high-quality data (\eg positions, ionic and cell forces, etc.) that can be used to learn complex high-dimensional PES \via machine learning (ML) based approaches~\cite{han2017deep,zhang2018deepmd,zhang_end--end_2018,ko_isotope_2019}.

Assuming that the system is ergodic, the accuracy of a given MD simulation in predicting equilibrium properties is primarily governed by the quality of the ionic forces and stress tensor (or cell forces) used when propagating the corresponding equations of motion.
As such, a physically sound approach for obtaining these forces is given by first-principles based electronic structure theories, which are the foundation for \textit{ab initio} MD (AIMD) simulations~\cite{car_unified_1985,marx_ab_2009}.
With the AIMD technique, the nuclear PES is generated on-the-fly from the electronic ground state and does not require any empirical input, thereby allowing for a quantum mechanical treatment of structural, electronic/dielectric, and dynamical properties, as well as any potential chemical reactions that may occur~\cite{car_unified_1985}.
Due to its favorable balance between accuracy and computational cost, Kohn-Sham (KS) density functional theory (DFT)~\cite{hohenberg_inhomogeneous_1964,kohn_self-consistent_1965} is the predominant electronic structure theory in AIMD, especially when performing large-scale simulations of complex condensed-phase materials.
Within the KS-DFT framework, the total ground-state energy ($E$, which is not to be confused with the total internal energy of the system mentioned above) is comprised of the following terms: the KS (or mean-field) electronic kinetic energy ($E_{\rm kin}$), the external potential energy ($E_{\rm ext}$, which includes contributions from nucleus-electron and nucleus-nucleus interactions, as well as any other interactions with external fields), the Hartree potential energy ($E_{\rm H}$, the classical description of the electron-electron interactions), and the so-called exchange-correlation (xc) energy ($E_{\rm xc}$, which accounts for all remaining many-body electron correlation effects).
While DFT provides an exact solution for the ground-state density (and properties) in principle, the exact functional form for $E_{\rm xc}$ remains unknown to date and must be approximated in practice~\cite{parr_density-functional_1989,fiolhais_primer_2003,becke_perspective:_2014,mardirossian_thirty_2017,medvedev_density_2017,kepp_comment_2017,hammes-schiffer_conundrum_2017,medvedev_response_2017,lehtola_recent_2018}.

When treating condensed-phase systems (such as solids and liquids), the most commonly used approaches for computing $E_{\rm xc}$ are generalized gradient approximation (GGA) functionals such as those put forth by Perdew, Burke, and Ernzerhof (PBE)~\cite{perdew_generalized_1996} as well as Becke, Lee, Yang, and Parr (BLYP)~\cite{becke_density-functional_1988,lee_development_1988}, which express $E_{\rm xc}$ as a functional of the electron density, $\rho(\bm r)$, and its gradient, $\nabla \rho(\bm r)$.
Although such approaches are computationally efficient, the accuracy of a GGA functional is primarily limited by: (\textit{i}) its inability to fully describe non-local correlation effects such as dispersion (or van der Waals, vdW) interactions~\cite{klimes_perspective:_2012,grimme_dispersion-corrected_2016,hermann_first-principles_2017,berland_van_2015}, and (\textit{ii}) its propensity to suffer from self-interaction error (SIE), in which each electron spuriously interacts with itself~\cite{perdew_self-interaction_1981,cohen_insights_2008}.
Without a complete and physically sound description of dispersion/vdW interactions, GGA-DFT faces difficulties when determining the structure of liquid water~\cite{distasio_jr._individual_2014}, investigating drug-DNA binding~\cite{distasio_jr._collective_2012}, predicting the structures and relative stabilities of molecular crystal polymorphs~\cite{hoja_reliable_2018}, as well as quantifying the cohesion in asteroids~\cite{scheeres_scaling_2010,rozitis_cohesive_2014}.
In addition, the presence of SIE at the GGA-DFT level leads to $\rho(\bm r)$ that are typically too delocalized, which results in a number of shortcomings including (but not limited to) excessive proton delocalization in liquid water~\cite{zhang_first_2011,zhang_structural_2011,gaiduk_first-principles_2018}, inadequate descriptions of transition states and charge transfer complexes~\cite{grafenstein_impact_2003,lundberg_quantifying_2005,leblanc_pervasive_2018}, as well as overestimation of lattice parameters~\cite{marsman_hybrid_2008}.
To account for dispersion/vdW forces in GGA-DFT, a number of different approaches have been suggested in the literature~\cite{klimes_perspective:_2012,grimme_dispersion-corrected_2016,hermann_first-principles_2017,berland_van_2015}, which range from effective pairwise models~\cite{becke_exchange-hole_2007,tkatchenko_accurate_2009,grimme_consistent_2010,ferri_electronic_2015,caldeweyher_extension_2017} to more sophisticated many-body approaches~\cite{tkatchenko_accurate_2012,distasio_jr._collective_2012,distasio_jr._many-body_2014,ambrosetti_long-range_2014,blood-forsythe_analytical_2016} and fully non-local xc functionals~\cite{dion_van_2004,vydrov_nonlocal_2009,lee_higher-accuracy_2010}.
To mitigate the SIE, hybrid-GGA functionals~\cite{becke_densityfunctional_1993} admix a fraction of exact exchange (EXX) into $E_{\rm xc}$ as follows:
\begin{align}
    E_{\rm xc}^{\rm hybrid} =  a_{\rm x} \exx + \left( 1 - a_{\rm x} \right) E_{\rm x}^{\rm GGA} + E_{\rm c}^{\rm GGA} ,
    \label{eq:hyb}
\end{align}
where $0 < a_{\rm x} < 1$ is a constant, $\exx$ is the EXX energy, and $E_{\rm x}^{\rm GGA}$ and $E_{\rm c}^{\rm GGA}$ are the GGA exchange and correlation contributions to $E_{\rm xc}$, respectively.
When compared to evaluating $E_{\rm xc}$ at the GGA level, the computational complexity introduced by the EXX contribution in Eq.~\eqref{eq:hyb} is significantly higher.
As such, the efficient evaluation of $\exx$ is the key limitation to performing hybrid DFT based AIMD simulations of large-scale condensed-phase systems, and has triggered much attention in the community.~\cite{heyd_hybrid_2003,guidon_robust_2009,duchemin_scalable_2010,bylaska_parallel_2011,barnes_improved_2017,varini_enhancement_2013,guidon_auxiliary_2010,hu_interpolative_2017,dong_interpolative_2018,lin_adaptively_2016,hu_projected_2017,marzari_maximally_1997,wu_order-n_2009,marzari_maximally_2012,distasio_jr._individual_2014,gygi_compact_2009,gygi_efficient_2013,damle_compressed_2015,damle_computing_2017,damle_scdm-k:_2017,mountjoy_exact_2017,izmaylov_efficient_2006,guidon_ab_2008,guidon_robust_2009,guidon_auxiliary_2010,carnimeo_fast_2018,chawla_exact_1998,sorouri_accurate_2006,boffi_efficient_2016,mandal_enhanced_2018,mandal_speeding-up_2019,mandal_efficient_2020,mandal_achieving_2021}
For a more detailed summary of these approaches, we recommend the reader to the first paper in this series~\cite{paper1}, which will be referred to as \pI throughout this work.

As discussed in \pI~\cite{paper1}, a linear-scaling yet numerically accurate evaluation of $\exx$ can be accomplished for large-scale finite-gap condensed-phase systems by employing a localized representation of the occupied orbitals~\cite{wu_order-n_2009,distasio_jr._individual_2014} (\eg maximally localized Wannier functions (MLWFs))~\cite{marzari_maximally_1997,marzari_maximally_2012}.
In that work, we provided an in-depth discussion of the theoretical background, accuracy, and performance of a massively parallel implementation (the \exxm module) of this MLWF-based EXX approach in the pseudopotential- and planewave-based open-source \texttt{Quantum ESPRESSO (QE)} package~\cite{giannozzi_quantum_2009,giannozzi_advanced_2017}, and again refer the reader back to this work for additional details.
As briefly summarized below, this algorithm achieves $\mathcal{O}(N)$ scaling by using localized orbitals to exploit the natural sparsity in the EXX interaction in real space, \ie this quantum mechanical interaction is short-ranged and only occurs in regions of orbital overlap. 
Letting $\{\tphi{i}\}$ be the set of MLWFs obtained \via an orthogonal (unitary) transformation of the occupied KS eigenstates, $\{\phi_{i}(\bm r)\}$, \ie $\widetilde{\phi}_i (\bm r) = \sum_j U_{ij} \phi_j (\bm r)$, we first note that $\exx$ is invariant to such transformations and can be written as follows in the MLWF representation:
\begin{align}
  \exx &= - \sum_{ij}^{N_o} \int \dd \bm{r} \int \dd \bm{r}' \, \frac{\trho{ij} \trhon{ij}(\bm{r}')}{|\bm{r}-\bm{r}'|} ,
  \label{eq:exx_2int}
\end{align}
or equivalently,
\begin{align}
  \exx &= - \sum_{ij}^{N_o} \int \dd \bm{r} \, \trho{ij} \tv{ij} . 
  \label{eq:exx}
\end{align}
In these expressions (shown here without loss of generality for a closed-shell system with $N_o$ occupied orbitals), $\trho{ij}$ is the so-called MLWF-product density,
\begin{align}
  \trho{ij} \equiv \tphi{i} \tphi{j} ,
  \label{eq:rho_from_phi_mlwf}
\end{align}
and $\tv{ij}$ is the corresponding MLWF-product potential,
\begin{align}
  \tv{ij} \equiv \int \dd \bm{r}' \, \frac{\trhon{ij}(\bm{r}')}{|\bm{r}-\bm{r}'|} ,
  \label{eq:vxxGen_mlwf}
\end{align}
\ie the electrostatic potential felt by a test charge at $\bm{r}$ originating from the charge distribution $\trhon{ij}(\bm{r}')$.
Since the focus of this work is large-scale condensed-phase systems with finite gaps, the first Brillouin zone can be sampled at the $\Gamma$ point only; as such, we have the flexibility to work with real-valued KS orbitals (and MLWFs), and so $\trho{ij} = \trho{ji}$ and $\tv{ij} = \tv{ji}$.
In this work, we again follow \pI~\cite{paper1} by dressing all MLWF-specific quantities with tildes, and leaving quantities that are invariant to the MLWF representation unmodified (\eg $\exx$ in Eqs.~\eqref{eq:exx_2int} and \eqref{eq:exx}).
Since the MLWFs in finite-gap systems are exponentially localized in real space~\cite{kohn_analytic_1959,des_cloizeaux_analytical_1964,nenciu_existence_1983,marzari_maximally_1997,niu_theory_1991,panati_bloch_2013} and have a significantly smaller support than the entire simulation cell, $\Omega$, the use of MLWFs (or any other localized representation which spans the occupied space) allows us to exploit two levels of sparsity when computing $\exx$ (as well as all other EXX-related quantities, \textit{vide infra}). 
Considering the expression for $\exx$ in the MLWF representation, one can immediately see that a numerically accurate evaluation of Eq.~\eqref{eq:exx} \textit{only} requires contributions from overlapping pairs of MLWFs (\ie when $\trho{ij} \ne 0$).
Hence, the first level of computational savings in our approach originates from the fact that a given MLWF is exponentially localized and will only appreciably overlap with a finite number of neighboring MLWFs.
As such, the number of EXX pair interactions \textit{per orbital} becomes independent of system size (assuming a fixed system density), and the quadratic sum over MLWFs in Eq.~\eqref{eq:exx} can be replaced with a linear sum over overlapping pairs of MLWFs without loss of accuracy.
To harness the second level of computational savings, we define the MLWF-orbital domain corresponding to $\tphi{i}$ as $\Omega_i \equiv \{ \bm r \in \Omega \,\bm{\mid}\, | \tphi{i} | > \epsilon  \}$.
Hence, $\Omega_i$ encompasses the support of $\tphi{i}$ and thereby delineates the region of space where this MLWF is non-negligible.
In the above expression, we follow the approach outlined by Gygi and co-workers,~\cite{gygi_compact_2009,gygi_efficient_2013,dawson_performance_2015} and neglect the regions of space where $| \tphi{i} |$ is less than a small threshold $\epsilon$.
In analogy, we also define the MLWF-product domain corresponding to a pair of overlapping MLWFs, $\tphi{i}$ and $\tphi{j}$, as $\Omega_{ij} \equiv \Omega_i \cap \Omega_j$, which encompasses the support of $\trho{ij}$ (see Fig.~1 of \pI~\cite{paper1} for a schematic illustration of these domains).
Considering again the energy expression in Eq.~\eqref{eq:exx}, one can also see that a numerically accurate evaluation of the contribution to $\exx$ from each overlapping MLWF pair \textit{only} requires spatial integration over $\Omega_{ij}$ (given that $\epsilon$ is sufficiently small).
As such, the costly integration over $\Omega$ (\ie the entire simulation cell) in Eq.~\eqref{eq:exx} can be replaced with spatial integrals over system-size-independent $\Omega_{ij}$ domains.
By accounting for both of these sparsity levels, Eq.~\eqref{eq:exx} for $\exx$ can now be rewritten as the following working expression:
\begin{align}
  \exx = - \sum_{\braket{ij}} \int_{\Omega_{ij}} \dd \bm{r} \, \trho{ij} \tv{ij} ,
  \label{eq:exxGen_mlwf}
\end{align}
in which $\braket{ij}$ indicates that the summation over $i$ and $j$ only includes overlapping MLWF pairs and each integral is performed on the corresponding $\Omega_{ij}$ domain.
With a judicious choice of cutoff parameters (see \pI~\cite{paper1}), the \exxm module in \texttt{QE} is able to compute $\exx$ in a numerically accurate fashion at a computational cost that scales linearly with system size.

From Eq.~\eqref{eq:exxGen_mlwf}, it is clear that an accurate and efficient real-space evaluation of $\tv{ij}$ is of central importance to developing a numerically accurate and linear-scaling algorithm for computing $\exx$ (as well as all other EXX-related quantities, \textit{vide infra}) in large-scale condensed-phase systems.
In the \exxm algorithm,~\cite{paper1} this is accomplished by an efficient conjugate-gradient (CG) solution to Poisson's equation (PE) for $\tv{ij}$ in the near field,
\begin{equation}
  \nabla^2\tv{ij}=-4\pi\trho{ij} \qquad \bm r \in \Omega_{ij} ,
  \label{eq:pe}
\end{equation}
subject to boundary conditions provided by a sufficiently converged multipole expansion (ME) of $\tv{ij}$ in the far field,
\begin{equation}
  \tv{ij} = 4\pi \sum_{lm} \frac{Q_{lm}}{(2l+1)} \frac{Y_{lm}(\theta,\varphi)}{r^{l+1}} \qquad \bm r \notin \Omega_{ij} .
  \label{eq:me}
\end{equation}
In this expression, $\bm r = (r,\theta,\varphi)$ is given in spherical polar coordinates, $Y_{lm} (\theta,\varphi)$ are the spherical harmonics, and
\begin{equation}
  Q_{lm} = \int_{\Omega_{ij}} \dd\bm{r} \, Y_{lm}^{*}(\theta,\varphi) \, r^l \trho{ij}
  \label{eq:mepole}
\end{equation}
are the multipole moments of $\trho{ij}$. 
In addition to providing the boundary conditions required during the CG solution of the PE, the ME in Eq.~\eqref{eq:me} is also used when computing the EXX contribution to the wavefunction forces, which formally requires $\tv{ij}$ on $\Omega_{i}$ and $\Omega_{j}$ (see Sec.~\ref{sec:exx_wff}).

In \pI~\cite{paper1}, we presented a linear-scaling and numerically accurate algorithm for computing the EXX contribution to the energies and wavefunction forces in fixed orthorhombic cells, thereby enabling large-scale hybrid DFT based AIMD simulations in the $NVE$ and $NVT$ ensembles for a wide array of condensed-phase systems.
With access to high-performance computing (HPC) resources, the hybrid message-passing interface (\mpi{}) and open multi-processing (\omp{}) based implementation of \exxm in \texttt{QE} enables us to compute the EXX contributions to the energy and wavefunction forces for \ce{(H2O)256}, a condensed-phase system containing $> 750$~atoms, in approximately $2.4$~s on the IBM Blue Gene/Q architecture.
As such, the current \exxm module (and earlier pilot versions) has already enabled computational investigations into a number of important condensed-phase systems, including the electronic structure of semi-conducting solids~\cite{wu_hybrid_2009,chen_electronic_2011}, the structure and local order of ambient liquid water~\cite{distasio_jr._individual_2014,santra_local_2015}, the structural and dynamical properties of aqueous ionic solutions~\cite{bankura_systematic_2015,chen_hydroxide_2018}, the thermal properties of the pyridine-I molecular crystal~\cite{ko_thermal_2018}, as well as the subtle isotope effects on the structure of liquid water~\cite{ko_isotope_2019}.

In this work, we extend the capabilities of the \exxm module by deriving and implementing: (\textit{i}) the EXX contribution to the stress tensor (or cell forces), which is required for performing constant-pressure simulations, and (\textit{ii}) a stable and efficient representation for the Laplacian during the solution of the PE in general/non-orthogonal simulation cells.
In doing so, we enable accurate and efficient hybrid DFT based AIMD simulations of large-scale condensed-phase systems (with arbitrary symmetries) in the $NpH$ and $NpT$ (as well as $NVE$ and $NVT$) ensembles using the \exxm module.
Since most experiments are performed at constant $p$ and $T$ (instead of constant $V$ and $T$), this development will enable more sophisticated computational investigations into large-scale condensed-phase systems across a wider range of thermodynamic conditions.
The remainder of the manuscript is organized as follows.
In Sec.~\ref{sec:Theory}, we derive the EXX contribution to the stress tensor within the framework of our MLWF-based EXX approach, which is required for propagating the Car-Parrinello (CP) equations of motion under constant-pressure conditions~\cite{parrinello_crystal_1980,car_unified_1985,marx_ab_2009}.
In Secs.~\ref{sec:Implementation} and \ref{sec:Performance}, we provide a detailed discussion of the algorithmic extensions implemented in the \exxm module as well as its accuracy and performance when simulating condensed-phase systems in the $NpT$ ensemble.
The paper is then ended with some brief conclusions in Sec.~\ref{sec:conclusion}.

\section{Theory \label{sec:Theory}}

In this section, we expand the theoretical framework underlying our linear-scaling hybrid DFT approach~\cite{paper1} to enable constant-pressure simulations of condensed-phase systems with general/non-orthogonal cells.
We focus the discussion around the CP equations of motion (in conjunction with the Parrinello-Rahman barostat~\cite{parrinello_crystal_1980}) used to propagate the electronic, ionic, and cell degrees of freedom during constant-pressure simulations in the $NpH$ ensemble; with the introduction of an appropriately chosen thermostat (for the ionic degrees of freedom), this approach can easily be extended to sample the $NpT$ ensemble.
Although the scope of this discussion is limited to the CPMD variant of AIMD, which provides a computationally efficient scheme for propagating localized orbitals~\cite{sharma_ab_2003,iftimie_--fly_2004,thomas_field_2004}, a novel and cost-effective extension to enable Born-Oppenheimer MD (BOMD) using this approach will be addressed in a forthcoming paper.
When used in conjunction with second-order damped dynamics (SODD)~\cite{tassone_acceleration_1994} (or other global optimization techniques such as CG) on the ionic and cell degrees of freedom, variable-cell (VC) optimizations (in the absence of thermal and nuclear quantum fluctuations) are also possible with the approach presented herein.

\subsection{Index Conventions \label{sec:notation}}

Following \pI, we will utilize the following conventions for the indices encountered in this work:
\begin{itemize}
    \item $i$, $j$, $k$: indices for the $N_o$ occupied orbitals (or MLWFs)
    \item $a$, $b$, $c$: indices corresponding to the Cartesian directions ${\bm x}$, ${\bm y}$, and ${\bm z}$
    \item $\alpha$, $\beta$, $\gamma$: indices corresponding to the cell (lattice) vectors $\bm L_{1}$, $\bm L_{2}$, and $\bm L_{3}$
    \item $I$, $J$, $K$: indices for the $N_A$ ions
    \item $p$, $q$: indices for the points on the real-space grid (with $p$ not to be confused with the pressure)
    \item $l,m$: indices for spherical harmonics
\end{itemize}

\subsection{EXX-Based CPMD in the $NpH$ Ensemble \label{sec:hybrid_aimd}}

\subsubsection{Equations of Motion \label{sec:eom}}

In constant-pressure CPMD simulations, fictitious dynamics are introduced on the $N_o$ occupied KS orbitals $\left\{ \phi_i \left( \bm r \right) \right\}$ and simulation cell tensor $\bm h$ \via artificial/fictitious masses $\mu$ (not to be confused with the chemical potential) and $W$, respectively.
In this work, $\bm h$ is a $3 \times 3$ matrix defined as $\bm h \equiv ( \bm L_1 \,\, \bm L_2 \,\, \bm L_3 )$ or $\bm h_{a\alpha} \equiv (\bm L_{\alpha})_a$, where $\bm L_{1}$, $\bm L_{2}$, and $\bm L_{3}$ are the corresponding cell (lattice) vectors.
The simulation cell volume will be denoted by $V = \text{det}\,(\bm h)$.
Constant-pressure ($NpH$) CPMD simulations with the Parrinello-Rahman barostat~\cite{parrinello_crystal_1980} are governed by the following equations of motion for the electronic, ionic, and cell degrees of freedom:~\cite{marx_ab_2009} 
\begin{align}
  \mu \ddot{\phi}_i (\bm r) &= -\left( \frac{\delta E}{\delta \phi^*_i (\bm r)} \right) + \sum_{j} \Lambda_{ij} \phi_j (\bm r) \label{eq:cpE} \\
  M_I \ddot{\bm S}_I        &= - \bm h^{-1} \left( \nabla_{\bm R_I} E \right) - M_I \mathcal{G}^{-1} \dot{\mathcal{G}} \dot{\bm S}_{I} \label{eq:cpI} \\
  W \ddot{\bm h}            &= \left( \bm \Pi - p \bm{1} \right) \left( \bm h^{T} \right)^{-1}V , \label{eq:cpC}
\end{align}
in which Newton's dot notation was used to indicate time derivatives, $E$ is the total ground-state DFT energy (including the nuclear-nuclear repulsion), $-(\delta E/\delta \phi^*_i (\bm r))$ is the force acting on the $i$-th occupied KS wavefunction, $\Lambda_{ij}$ is a Lagrange multiplier enforcing orthonormality in $\left\{ \phi_i (\bm r) \right\}$, $-\nabla_{\bm R_I} E$ is the force acting on the $I$-th ion (located at $\bm R_I$ with mass $M_I$), $\mathcal{G} = \bm h^{T}\bm h$ is the so-called metric tensor, $\bm \Pi$ is the total internal stress tensor, $p$ is the applied (external) pressure, and $\bm 1$ is the identity matrix.
For the fluctuating simulation cells encountered in constant-pressure CPMD, it is more convenient to work in crystal (fractional) coordinates $\bm S_I$ for the ions, which are independent of the dynamical variables associated with the cell degrees of freedom, and are related to the Cartesian coordinates \via $\bm R_I = \bm h \bm S_I$ or $\bm S_I = \bm h^{-1} \bm R_I$.

The components of the $3 \times 3$ total internal stress tensor, $\bm \Pi$, can be further decomposed into kinetic (kin) and potential (pot) contributions as follows~\cite{marx_ab_2009}:
\begin{align}
  \Pi_{ab} &= \Pi_{ab}^{\rm kin} + \Pi_{ab}^{\rm pot} .
  \label{eq:tot_stress}
\end{align}
In this expression, the kinetic contribution ($\Pi_{ab}^{\rm kin}$) originates from the ionic kinetic energy \via
\begin{align}
  \Pi^{\rm kin}_{ab}  &= \frac{1}{V} \sum_I M_{I}\sum_{\alpha\beta} h_{a\alpha} \dot{S}_{I\alpha} \dot{S}_{I\beta} h_{b\beta} ,
  \label{eq:kin_stress}
\end{align}
while the potential contribution ($\Pi_{ab}^{\rm pot}$) arises from cell derivatives, $\sigma^{a\alpha}$, of the ionic potential energy (\ie the DFT energy):
\begin{align}
  \Pi^{\rm pot}_{ab} &= -\frac{1}{V}\sum_{\alpha}\left(\frac{\partial E}{\partial h_{a\alpha}}\right) h_{b\alpha} \equiv -\frac{1}{V}\sum_{\alpha}\sigma^{a\alpha} h_{b\alpha} .
  \label{eq:cel_for_stress}
\end{align}

\subsubsection{EXX Contribution to the Wavefunction Forces \label{sec:exx_wff}}

Since the explicit functional dependence of $\exx$ on the total one-electron density, $\rho(\bm r) \equiv 2 \sum_i |\phi_{i}(\bm r)|^2 = 2 \sum_i |\tphi{i}|^2$, is unknown, one would need to use special methods like the optimized effective potential (OEP) technique~\cite{kummel_orbital-dependent_2008} to derive the EXX contribution to the wavefunction forces within a strict KS-DFT scheme.
In this work, we instead adopt a generalized KS-DFT scheme by allowing for an orbital-dependent xc potential, as this approach (which is currently standard practice in the field) yields the same ground-state energies as the OEP formalism at a fraction of the computational cost.
Given the working expression for $\exx$ in Eq.~\eqref{eq:exxGen_mlwf}, the EXX contribution to the wavefunction forces (which is needed to propagate the electronic degrees of freedom in Eq.~\eqref{eq:cpE}) can be derived in a straightforward manner (see \pI~\cite{paper1} for more details).
In the MLWF representation, the wavefunction force acting on the $i$-th MLWF, $\dxx{i} = -(\delta \exx / \delta \widetilde{\phi}_{i}^*(\bm r))$, takes on the following form:
\begin{align}
  \dxx{i} = \sum_{j} \tv{ij} \tphi{j} \equiv \sum_{j} \dxx{ij} ,
  \label{eq:Dxx}
\end{align}
in which the sum only includes $\tphi{j}$ that overlap with $\tphi{i}$.
Here, we again follow \pI by dressing all MLWF-specific quantities with tildes, and leaving quantities that are invariant to the MLWF representation unmodified.
Since $\tphi{j}$ is exponentially localized, a numerically accurate evaluation of $\dxx{ij}$ in Eq.~\eqref{eq:Dxx} only requires the action of $\tv{ij}$ on $\tphi{j}$ over the system-size-independent $\xmega{j}$ domain, \ie where $\tphi{j}$ is non-negligible. 
Taken together with the fact that the number of overlapping MLWF pairs is also system-size-independent (for a given $i$), the entire set of $\{\dxx{i}\}$ can therefore be evaluated in linear time.
From Eq.~\eqref{eq:Dxx}, it is again clear that an accurate and efficient real-space evaluation of $\tv{ij}$---on compact and system-size independent domains---is the cornerstone of our linear-scaling MLWF-based EXX approach.

\subsubsection{EXX Contribution to the Stress Tensor \label{sec:Pi_xx}}

The remaining quantity needed to propagate the equations of motion during constant-pressure CPMD simulations at the hybrid DFT level is the EXX contribution to the stress tensor in Eq.~\eqref{eq:cpC}. 
As seen in Eqs.~\eqref{eq:tot_stress}--\eqref{eq:cel_for_stress}, the EXX contribution is only present in the potential part of $\bm{\Pi}$, and arises from $\sigma^{a\alpha} = (\partial E / \partial h_{a\alpha})$, the derivative of the DFT energy with respect to the cell tensor (\ie the so-called cell derivatives).
As such, the EXX contribution to $\bm \Pi^{\rm pot}$ in Eq.~\eqref{eq:cel_for_stress} requires evaluation of $\sxxele^{a\alpha} = (\partial \exx /\partial h_{a\alpha})$, which takes on the following form (\cf Eq.~\eqref{eq:exx_2int}):
\begin{equation}
  \sxxele^{a\alpha} = -\sum_{ij}\frac{\partial}{\partial h_{a\alpha}} \int \dd \bm r \int \dd \bm r' \, \frac{\trho{ij}\widetilde{\rho}_{ij}(\bm r')}{\left| \bm r - \bm r' \right|} .
  \label{eq:exx_deriv}
\end{equation}
To compute these cell derivatives, it is again more convenient to work in crystal coordinates, as was done above in the $NpH$ equations of motion for the ionic degrees of freedom (see Eq.~\eqref{eq:cpI}).
For the electrons, the transformation between crystal coordinates $\bm s$ and Cartesian coordinates $\bm r$ is completely analogous, and is given by $\bm r = \bm h \bm s$ or $\bm s = \bm h^{-1} \bm r$.
Since the Jacobian for this transformation is given by $\text{det}\,(\dd \bm r / \dd \bm s) = \text{det}\,(\bm h) = V$, the relationship between an MLWF in Cartesian and crystal coordinates is $\tphi{i} = \widetilde{\phi}_i (\bm s) / \sqrt{V}$, from which it follows that:
\begin{equation}
  \trho{ij} = \frac{1}{V} \, \widetilde{\rho}_{ij} (\bm s) .
  \label{eq:rho_s2r}
\end{equation}
Using this expression and the fact that $\dd \bm r = V \dd \bm s$, we can transform Eq.~\eqref{eq:exx_deriv} into crystal coordinates, namely,
\begin{equation}
  \sxxele^{a\alpha} = -\sum_{ij}\frac{\partial}{\partial h_{a\alpha}} \int \dd \bm s \int \dd \bm s' \, \frac{\widetilde{\rho}_{ij}(\bm s)\widetilde{\rho}_{ij}(\bm s')}{\left| {\bm h} \left( {\bm s} - {\bm s'} \right) \right|} ,
  \label{eq:exx_drv_Sspace_key}
\end{equation}
in which all factors of $V$ (arising from the transformations of the MLWF-product densities and differentials) have canceled.
Since crystal coordinates are independent of the dynamical variables associated with the cell, the only remaining dependence on $\bm h$ is in the denominator of the integrand in Eq.~\eqref{eq:exx_drv_Sspace_key}.
Letting $\Delta \bm s = \bm s -\bm s'$, we can now perform the relevant derivative as follows:
\begin{align}
  \frac{\partial \left| {\bm h} \Delta \bm s \right|^{-1}}{\partial h_{a\alpha}} &= \frac{\partial (\Delta \bm s^{T} \bm h^{T}\bm h \Delta \bm s)^{-1/2}}{\partial h_{a\alpha}} \nonumber \\
  &= -\frac{\sum_{\beta} \left( \Delta s_{\alpha} h_{a \beta} \Delta s_{\beta} \right)}{\left| {\bm h} \Delta \bm s \right|^3} .
  \label{eq:exx_drv_Sspace}
\end{align}
Plugging this expression into Eq.~\eqref{eq:exx_drv_Sspace_key} and transforming back to Cartesian coordinates yields:
\begin{align}
  \sxxele^{a\alpha} &= \sum_{ij} \sum_{b} \left(h^{-1}\right)_{\alpha b} \nonumber \\
  &\times \int \dd \bm r \int \dd \bm r' \, \widetilde{\rho}_{ij}(\bm r) \widetilde{\rho}_{ij}(\bm r') \frac{\Delta r_{b} \Delta r_{a}}{\left| \Delta\bm r \right|^{3}} ,
  \label{eq:exx_drv_Sspace_tmp}
\end{align}
in which $\Delta \bm r = \bm r - \bm r'$.
Further reduction of this expression is possible by splitting the integrand into two terms \via $\Delta r_{b} = r_{b} - r'_{b}$, using the fact that $\Delta \bm r' = \bm r' - \bm r = - \Delta \bm r$, and then noticing that these terms are equivalent after swapping the $\bm r$ and $\bm r'$ dummy variables.
After doing so, we can now write Eq.~\eqref{eq:exx_drv_Sspace_tmp} in the following intermediate form:
\begin{align}
  \sxxele^{a\alpha} &= 2\sum_{ij} \sum_{b} \left( h^{-1} \right)_{\alpha b} \nonumber \\
  & \times \int \dd \bm r \int \dd \bm r' \, \widetilde{\rho}_{ij}(\bm r) \widetilde{\rho}_{ij}(\bm r') \frac{r_{b} \Delta r_a}{\left| \Delta \bm r \right|^3} .
  \label{eq:exx_drv_Sspace_tmp2}
\end{align}
This expression can be further simplified by separating the integrand as follows:
\begin{align}
  \sxxele^{a\alpha} &= 2\sum_{ij} \sum_{b} \left( h^{-1} \right)_{\alpha b} \nonumber \\
  &\times \int \dd \bm r \, r_{b} \, \widetilde{\rho}_{ij}(\bm r) \left[ \int \dd \bm r' \, \widetilde{\rho}_{ij}(\bm r') \frac{\Delta r_a}{\left| \Delta \bm r \right|^{3}} \right] ,
\end{align}
and then realizing that the term inside the square brackets (to within a sign) is the derivative of $\tv{ij}$ with respect to the $a$-th Cartesian component, \ie 
\begin{align}
  \frac{\partial \, \tv{ij}}{\partial \, r_a} &= \frac{\partial}{\partial \, r_a} \int \dd \bm r' \, \frac{\widetilde{\rho}_{ij}(\bm r')}{\left| \bm r - \bm r' \right|} \nonumber \\
  &= - \int \dd \bm r' \, \widetilde{\rho}_{ij}(\bm r') \frac{\Delta r_{a}}{\left| \Delta \bm r \right|^{3}} ,
  \label{eq:exx_dvdr}
\end{align}
where we have used Eq.~\eqref{eq:vxxGen_mlwf}.
As such, we now arrive at the final expression for the EXX cell derivatives needed during hybrid DFT based CPMD simulations in the $NpH$ (or $NpT$) ensemble:
\begin{align}
  \sxxele^{a\alpha} &= -2\sum_{ij} \sum_{b} \left( h^{-1} \right)_{\alpha b} \nonumber \\ 
  &\times \int \dd \bm r \, r_{b} \, \widetilde{\rho}_{ij}(\bm r) \left( \frac{\partial \, \tv{ij}}{\partial \, r_a} \right) .
  \label{eq:exx_drv_sep_trick}
\end{align}
In analogy to the working expression for $\exx$ in Eq.~\eqref{eq:exxGen_mlwf}, a linear-scaling and numerically accurate evaluation of Eq.~\eqref{eq:exx_drv_sep_trick} is also possible by: (\textit{i}) replacing the quadratic sum over MLWFs with a linear sum over overlapping MLWF pairs ($\sum_{ij} \rightarrow \sum_{\braket{ij}}$), and (\textit{ii}) performing the spatial integrals over system-size-independent $\Omega_{ij}$ domains instead of $\Omega$ (\ie the entire simulation cell).
Doing so leads us to the following working expression for $\sxxele^{a\alpha}$ in our MLWF-based EXX approach (\cf Eq.~\eqref{eq:exxGen_mlwf}):
\begin{align}
  \sxxele^{a\alpha} &= -2\sum_{\braket{ij}} \sum_{b} \left( h^{-1} \right)_{\alpha b} \nonumber \\ 
  &\times \int_{\Omega_{ij}} \dd \bm r \, r_{b} \, \widetilde{\rho}_{ij}(\bm r) \left( \frac{\partial \, \tv{ij}}{\partial \, r_a} \right) .
  \label{eq:sxx_working}
\end{align}
From Eq.~\eqref{eq:sxx_working}, it is clear that once $\tv{ij}$ is evaluated (which is also required for computing $\exx$ and $\{\dxx{i}\}$), its gradient provides the remaining ingredients needed to compute the EXX contribution to the stress tensor \via Eq.~\eqref{eq:cel_for_stress}, \ie
\begin{align}
    \left(\Pi^{\rm pot}_{\rm xx}\right)_{ab} &= -\frac{1}{V}\sum_{\alpha}\sxxele^{a\alpha} h_{b\alpha} .
    \label{eq:pixx_working}
\end{align}

\begin{center}
  \textit{Isotropic Constraints on $\bm \Pi^{\rm pot}_{\rm xx}$}
\end{center}

In general, all components of $\bm \Pi$ (\ie the full stress tensor) are utilized when propagating the equations of motion for the cell degrees of freedom during constant-pressure CPMD simulations (see Eq.~\eqref{eq:cpC}).
In certain cases, however, constraints can be applied to $\bm \Pi$ which allow one to maintain desired lattice symmetries, avoid shear stress in fluids, and/or suppress phase transitions during constant-pressure CPMD simulations.
For instance, it is common practice to enforce isotropic constraints on $\bm \Pi$ during $NpH$ (or $NpT$) simulations of solids or liquids in simple cubic cells.
In order to do so, the off-diagonal components of $\bm \Pi$ are set to zero, and the diagonal components are replaced by the \textit{internal} pressure of the system, \ie
\begin{align}
  \Pi_{ab} &= p^{\rm int} \delta_{ab}  \qquad \text{(isotropic)} 
\end{align}
in which $p^{\rm int}$ is equivalent to the isotropic average of $\bm \Pi$,
\begin{align}
  p^{\rm int} \equiv \frac{1}{3} \sum_{a} \Pi_{aa} = \frac{1}{3} {\rm Tr} \, \bm \Pi .
\end{align}
From the EXX point of view, this is tantamount to replacing $\bm \Pi^{\rm pot}_{\rm xx}$ in Eq.~\eqref{eq:pixx_working} with
\begin{align}
  \left(\Pi^{\rm pot}_{\rm xx}\right)_{ab} &= p_{\rm xx}^{\rm int}\delta_{ab}  \qquad \text{(isotropic)}
\end{align}
where
\begin{align}
  p_{\rm xx}^{\rm int} &= \frac{1}{3} {\rm Tr} \, \bm \Pi_{\rm xx}^{\rm pot}
  \label{eq:pxx_working}
\end{align}
is the EXX contribution to the internal pressure.
By applying an equal (isotropic) cell force along each lattice vector, the simulation cell is not subjected to shear stress and remains simple cubic throughout the MD trajectory.

When performing such isotropic $NpH$ (or $NpT$) simulations of solids or liquids in simple cubic cells (with side length $L$ and $h_{a\alpha} = L \delta_{a\alpha}$), $p_{\rm xx}^{\rm int}$ does not even require evaluating all of the diagonal components of $\bm \Pi_{\rm xx}^{\rm pot}$, and can be simplified as follows (\cf Eqs.~\eqref{eq:pixx_working} and \eqref{eq:pxx_working}):
\begin{align}
    p_{\rm xx}^{\rm int} &= -\frac{1}{3V} \sum_{a}\sum_{\alpha} \sxxele^{a\alpha} h_{a\alpha} \nonumber \\
    &= -\frac{1}{3V} \sum_{a}\sum_{\alpha} \sxxele^{a\alpha} L \delta_{a\alpha} 
    = -\frac{L}{3V} \sum_{a} \sxxele^{aa} \nonumber \\
    &= -\frac{L}{3V} {\rm Tr} \, \sxx \qquad \text{(simple cubic)} .
    \label{eq:liq_press_simp}
\end{align}
Since $\partial h_{a\alpha}/\partial L = \partial \left( L \delta_{a\alpha} \right) /\partial L = \delta_{a\alpha}$ for a simple cubic cell, the trace over cell derivatives in Eq.~\eqref{eq:liq_press_simp} is equivalent to the derivative of $\exx$ with respect to $L$, \ie
\begin{align}
   \left( \frac{\partial \exx}{\partial L} \right) &= \sum_{a} \sum_{\alpha} \left( \frac{\partial \exx}{\partial h_{a\alpha}} \right) \left( \frac{\partial h_{a\alpha}}{\partial L} \right) \nonumber \\
   &= \sum_{a} \sum_{\alpha} \left( \frac{\partial \exx}{\partial h_{a\alpha}} \right) \delta_{a\alpha} = \sum_{a} \left( \frac{\partial \exx}{\partial h_{aa}} \right) \nonumber \\
   &= {\rm Tr} \, \sxx \qquad \text{(simple cubic)} ,
   \label{eq:dexxdl_to_tr_sxx}
\end{align}
which allows us to write $p_{\rm xx}^{\rm int}$ in the following alternative form:
\begin{align}
    p_{\rm xx}^{\rm int} &= -\frac{L}{3V} \left( \frac{\partial \exx}{\partial L} \right) \qquad \text{(simple cubic)} .
    \label{eq:liq_press_simp_L}
\end{align}
Since $\bm r = L \bm s$ in a simple cubic cell, the evaluation of $(\partial \exx/\partial L)$ can be further simplified as follows (\cf Eqs.~\eqref{eq:exx_deriv}--\eqref{eq:exx_drv_Sspace_key}):
\begin{align}
  \left(\frac{\partial \exx}{\partial L}\right) &= - \sum_{ij} \frac{\partial}{\partial L} \int \dd \bm r \int \dd \bm r' \, \frac{\widetilde{\rho}_{ij}(\bm r)\widetilde{\rho}_{ij}(\bm r')}{\left| {\bm r} - {\bm r'} \right|} \nonumber \\
  &= - \sum_{ij}\frac{\partial}{\partial L} \int \dd \bm s \int \dd \bm s' \, \frac{\widetilde{\rho}_{ij}(\bm s)\widetilde{\rho}_{ij}(\bm s')}{L \left| {\bm s} - {\bm s'} \right|} \nonumber \\
  &=  \frac{1}{L} \sum_{ij} \int \dd \bm s \int \dd \bm s' \, \frac{\widetilde{\rho}_{ij}(\bm s)\widetilde{\rho}_{ij}(\bm s')}{L \left| {\bm s} - {\bm s'} \right|} \nonumber \\
  &= \frac{1}{L} \sum_{ij} \int \dd \bm r \int \dd \bm r' \, \frac{\widetilde{\rho}_{ij}(\bm r)\widetilde{\rho}_{ij}(\bm r')}{\left| {\bm r} - {\bm r'} \right|} \nonumber \\
  &= -\frac{\exx}{L} \qquad \text{(simple cubic)} .
  \label{eq:dexxdL}
\end{align}
By combining Eqs.~\eqref{eq:liq_press_simp_L} and \eqref{eq:dexxdL}, we arrive at the following expression for $p_{\rm xx}^{\rm int}$ in a simple cubic cell:
\begin{equation}
  p_{\rm xx}^{\rm int} = -\frac{L}{3V} \left(\frac{\partial \exx}{\partial L}\right)  = \frac{\exx}{3V} \qquad \text{(simple cubic)} .
  \label{eq:liq_press}
\end{equation}
As such, the EXX contribution to $p^{\rm int}$ (as well as $\bm \Pi$) is trivial, and only requires evaluation of $\exx$ when performing isotropic $NpH$ (or $NpT$) simulations of solids or liquids in simple cubic cells at the hybrid DFT level of theory.

\section{Implementation and Algorithmic Details \label{sec:Implementation}}

In \pI~\cite{paper1}, we presented a massively parallel implementation of our linear-scaling MLWF-based EXX algorithm (\ie the \exxm module), which enabled hybrid DFT based AIMD simulations of large-scale condensed-phase systems with fixed orthorhombic unit cells in the $NVE$ and $NVT$ ensembles.
In this section, we describe an algorithmic extension to the \exxm module that enables such hybrid DFT simulations in the $NpH$ and $NpT$ (as well as the $NVE$ and $NVT$) ensembles for systems with general/non-orthogonal cells.
To do so, we first briefly review the \exxm module (Sec.~\ref{Impl:exx_module}) and then describe our extensions to \exxm, which includes algorithms that: (\textit{i}) handle fluctuating simulation cells with non-orthorhombic lattice symmetries (Secs.~\ref{subsec:impl_spdomain_npt} and \ref{subsec:impl_PS}), and (\textit{ii}) compute the previously derived (Sec.~\ref{sec:Pi_xx}) analytical evaluation of the EXX contribution to the stress tensor (Sec.~\ref{subsec:impl_cell_forces}).

\subsection{Review of the \exxm Module \label{Impl:exx_module}}

\begin{figure}[t!]
  \centering
  \includegraphics[width=\linewidth]{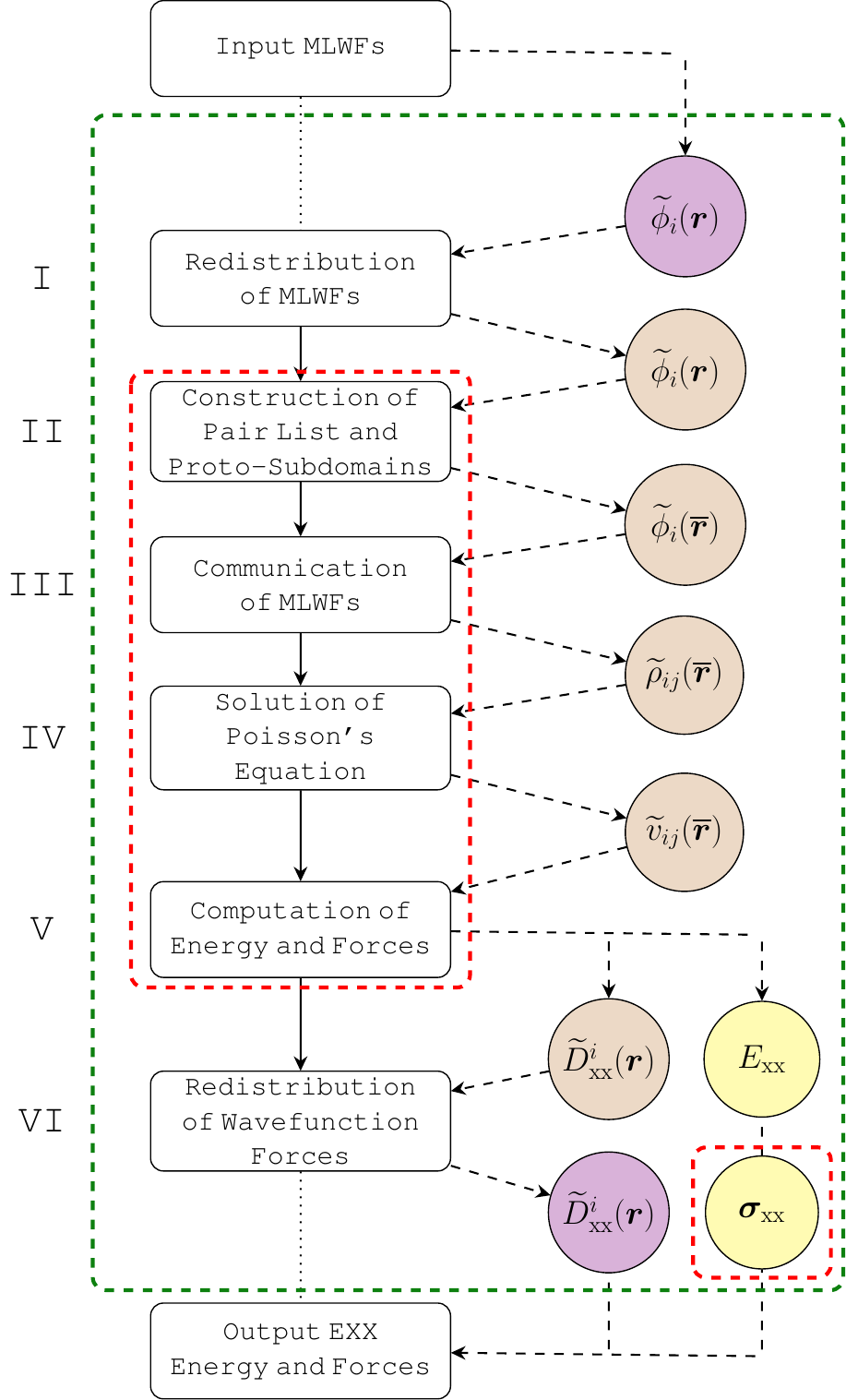}
   \caption{
   Flowchart of the \exxm module (dashed green box) in \texttt{QE} with extensions (dashed red boxes) for performing constant-volume ($NVE$/$NVT$) and constant-pressure ($NpH$/$NpT$) MLWF-based hybrid DFT simulations in general/non-orthogonal cells.
   As summarized in the main text, the input required by \exxm includes the set of MLWFs at the current CPMD step ($\{ \tphi{i} \}$), while the output produced by the extended \exxm module includes: the EXX contributions to the energy ($\exx$), wavefunction forces ($\{ \dxx{i}$ \}), and cell derivatives ($\sxx$).
   Purple (brown) circles denote that a given quantity is represented according to the the default \texttt{GRID} (customized~\cite{paper1} \texttt{ORBITAL}) data distribution scheme in \texttt{QE}.
   Pale yellow circles denote data that are globally broadcast \via \mpi{} during the execution of the \exxm module.
   For a detailed description of each step, see Secs.~III C 1--III C 6 in \pI~\cite{paper1} as well as Secs.~\ref{Impl:exx_module}--\ref{subsec:impl_cell_forces} in the current manuscript.
   }
   \label{fig:flowchart}
\end{figure}

In this section, we briefly review the implementation of our linear-scaling MLWF-based EXX algorithm in \exxm, a standalone module which has been integrated (\via a portable input/output interface) with the MLWF-enabled semi-local DFT routines in the \texttt{CP} module of \texttt{QE}~\cite{giannozzi_advanced_2017}.
To enable hybrid DFT simulations of large-scale condensed-phase systems using this approach, the \exxm module employs a dual-level \mpi{}/\omp{} parallelization scheme, which is able to exploit both internode and intranode HPC resources.
As depicted in the flowchart in Fig.~\ref{fig:flowchart}, the main input required for the \exxm module includes the current set of MLWFs at each CPMD step, $\{ \tphi{i} \}$, while the output produced by \exxm includes the corresponding EXX contributions to the energy ($\exx$), wavefunction forces ($\{\dxx{i}\}$), and cell derivatives/stress tensor ($\sxx$, see Sec.~\ref{subsec:impl_cell_forces}).
Given the capability to generate ``on-the-fly'' MLWFs during CPMD simulations, it should be reasonably straightforward to integrate the \exxm module into other (periodic) DFT codebases.
Since the only input requirement of \exxm is an orthonormal set of sufficiently localized orbitals, the use of alternative localization schemes (\eg recursive subspace bisection (RSB)~\cite{gygi_compact_2009,gygi_efficient_2013}, selected columns of the density matrix (SCDM)~\cite{damle_compressed_2015,damle_computing_2017,damle_scdm-k:_2017}, and condensed-phase Pipek-Mezey (PM)~\cite{jonsson_theory_2017}) are also possible with slight modifications to the code.

\textbf{Step I (Redistribution of MLWFs).}
As mentioned above, the input to the \exxm module is $\{ \tphi{i} \}$, the current set of MLWFs at a given CPMD step.
In \texttt{QE}, the so-called \texttt{GRID} scheme is employed when distributing the data corresponding to real-space quantities such as $\{ \tphi{i} \}$; in this scheme, each of the $N_{\rm proc}$ \mpi{} processes holds the data corresponding to \textit{all} $N_{o}$ MLWFs on \textit{a subset of} the real-space grid (see Sec.~III A, Sec.~III B, and Fig.~3 in \pI~\cite{paper1}).
To efficiently utilize massively parallel HPC resources, Step I of the \exxm module redistributes the $\{ \tphi{i} \}$ data from the \texttt{GRID} scheme to an alternative \texttt{ORBITAL} data distribution scheme, in which each \mpi{} process now holds the data corresponding to \textit{a subset of} MLWFs across the \textit{entire} real-space grid (see Sec.~III B, Sec.~III C 1, and Fig.~3 in \pI~\cite{paper1}).
In the \exxm module, the assignment of MLWFs across the pool of available \mpi{} processes is governed by $\zeta \equiv N_{\rm proc} / N_{o}$, \ie the ratio of \mpi{} processes to MLWFs; when $\zeta = 1$ (which is a common mode for running \exxm), each \mpi{} process, $P_{i}$, is assigned a single MLWF, $\tphin{i}$.
When $N_{\rm proc} < N_{o}$ (\ie $\zeta < 1$, less computational resources), multiple MLWFs are assigned to each $P_{i}$; although the \exxm module allows for any (rational) value of $\zeta < 1$, a balanced distribution of MLWFs across \mpi{} processes is only (currently) possible when $N_{\rm proc}$ is an exact divisor of $N_{o}$.
When $N_{\rm proc} > N_{o}$ (\ie $\zeta > 1$, HPC resources), each MLWF is assigned to multiple \mpi{} processes; in this case, the current \exxm module only allows for integer values for $\zeta > 1$.
Unless otherwise specified, we will assume that $\zeta = 1$ throughout the remainder of this work.

\textbf{Step II (Construction of Pair List and Proto-Subdomains).}
With the $\{\tphi{i}\}$ now distributed according to the \texttt{ORBITAL} scheme, the \exxm module enters Step II, and generates a unique list of overlapping $\braket{ij}$ MLWF pairs to avoid redundant computation (see Sec.~III C 2, Algorithm~1, Fig.~4, and Fig.~5 in \pI~\cite{paper1}).
Each overlapping $\braket{ij}$ pair is determined based on the criteria that $| \tC{i}-\tC{j} | \le \rpr$, \ie the distance between two MLWF centers ($\tC{i} = \braket{\tphin{i}|\bm{r}|\tphin{i}}$ and $\tC{j} = \braket{\tphin{j}|\bm{r}|\tphin{j}}$) must be less than or equal to a user-defined radial distance cutoff ($\rpr$).
The \exxm module then constructs the so-called unique MLWF-pair list, $\lu$, which determines how the computational workload will be distributed across the pool of available \mpi{} processes, and therefore defines the computation and communication protocol in our algorithm.
In constructing $\lu$, the \exxm module removes $\braket{ij}$ and $\braket{ji}$ pair redundancy (which minimizes the overall computational workload), and then attempts to balance the workload among \mpi{} processes while keeping the number of interprocess communication events minimal.
During Step II, the \exxm module also generates two concentric spherical proto-subdomains, $\pe{\gC{0}}$ and $\me{\gC{0}}$, which will be used later when computing each $\tv{ij}$ \via the solution to Poisson's equation in the near field (PE) and a multipolar expansion in the far field (ME).
Centered around $\gC{0}$ (the grid-resolved center of $\Omega$), the sizes of these spherical proto-subdomains are determined by user-defined radii, \ie $\rpe \in \{ \rpes , \rpep \}$ and $\rme \in \{\rmes, \rmep\}$ for $\braket{ii}$ (self-, s) and $\braket{ij}$ (non-self, ns) pairs; judicious choices for these parameters determine the accuracy and performance of the \exxm module (see Sec.~III C 2 and Fig~5, as well as Sec.~IV, Fig.~6, and Fig.~7 in \pI~\cite{paper1}).
For each point in these proto-subdomains, we store the \textit{local} (relative) Cartesian coordinates ($\rbar = \bm r - \gC{0}$) as well as the \textit{global} grid point indices ($\bm g^0$) along the three lattice directions ($\bm L_1$, $\bm L_2$, $\bm L_3$).
Based on these stored quantities, the $\pe{\gC{0}}$ and $\me{\gC{0}}$ proto-subdomains will be used (during future steps) to generate the pair-specific $\pe{\gC{ij}}$ and $\me{\gC{ij}}$ subdomains \via a rigid translation from $\gC{0}$ to $\gC{ij}$, the grid-resolved midpoint of $\tC{i}$ and $\tC{j}$ (see Sec.~III C 2, Algorithm~2, and Fig.~5 in \pI~\cite{paper1}).

\textbf{Step III (Communication of MLWFs).}
For each overlapping $\braket{ij}$ pair in $\lu$ (computed above in Step II), the \mpi{} process $P_{j}$ (which holds $\tphi{j}$ according to the \texttt{ORBITAL} scheme) first off-loads $\tphi{j}$ onto the $\me{\gC{ij}}$ subdomain, and then sends this orbital to $P_{i}$.
With $\tphib{j}$ on $\me{\gC{ij}}$ and $\tphi{i}$ stored locally according to the \texttt{ORBITAL} scheme, $P_{i}$ now computes $\trhob{ij}$ on the smaller $\pe{\gC{ij}}$ subdomain (a formal subset of $\me{\gC{ij}}$) by multiplying these two MLWFs (see Sec.~III C 3, Fig.~4, and Fig.~5 in \pI~\cite{paper1}).

\textbf{Step IV (Solution of Poisson's Equation).}
In Step IV, each \mpi{} process $P_{i}$ will first compute the far-field MLWF-product potential ($\tvb{ij}$) on $\me{\gC{ij}}\setminus\pe{\gC{ij}}$ \via a ME (see Eqs.~\eqref{eq:me}--\eqref{eq:mepole}).
Each $P_{i}$ then computes the near-field $\tvb{ij}$ by solving the PE on $\pe{\gC{ij}}$ (Eq.~\eqref{eq:pe}, with boundary conditions provided by the far-field $\tvb{ij}$) using a finite-difference representation of the Laplacian operator~\cite{fornberg_generation_1988} in conjunction with an iterative conjugate-gradient (CG) solver that has been efficiently parallelized over $N_{\rm thread}$ \omp{} threads (see Sec.~III C 4 and Fig.~10 in \pI~\cite{paper1}).

\textbf{Step V (Computation of Energy and Forces).}
With $\tvb{ij}$ on $\me{\gC{ij}}$ for each $\braket{ij}$ pair (constructed using the combined near- and far-field solutions computed in Step IV), $P_{i}$ then computes the $\braket{ij}$ contribution to the EXX energy ($\exx$) and wavefunction forces ($\dxxb{ij}$ and $\dxxb{ji}$).
Following Eq.~\eqref{eq:exxGen_mlwf}, $\exx$ is evaluated on $\pe{\gC{ij}}$ (a fixed-size spherical representation for $\Omega_{ij}$), and is accumulated \via a straightforward \texttt{MPI\_SUM} over the partial $\braket{ij}$ contributions computed on each \mpi{} process.
With $\tvb{ij}$ in hand, $P_{i}$ is also well-positioned to compute both $\dxxb{ij} = \tvb{ij} \tphib{j}$ and $\dxxb{ji} = \tvb{ij} \tphib{i}$, which are required for $\dxx{i}$ and $\dxx{j}$, the total wavefunction forces acting on $\tphi{i}$ and $\tphi{j}$ (see Eq.~\eqref{eq:Dxx}). 
Both of these contributions are evaluated on $\me{\gC{ij}}$, a fixed-size spherical domain that should be chosen to be large enough (\via the user-defined $\rme$ parameter) to cover the relevant sectors of both $\dxx{ij}$ and $\dxx{ji}$.
Since the far-field $\tv{ij}$ is dipolar at lowest order (due to the vanishing monopole associated with $\trho{ij}$), this quantity decays as $1/r^2$ for $i \neq j$; as such, a judicious choice for $\rme$ ensures rapid convergence in the $\braket{ij}$ (and $\braket{ji}$) contributions to the wavefunction forces (see Sec.~II C and Fig.~7 in \pI~\cite{paper1}).
After computing both $\dxxb{ij}$ and $\dxxb{ji}$, $\dxxb{ij}$ is locally accumulated on $P_{i}$ to form $\dxx{i}$, while $\dxxb{ji}$ is shipped back (\via \mpi{}) to $P_{j}$, where it is accumulated to form $\dxx{j}$ (see Sec.~III C 5 and Fig.~4 in \pI~\cite{paper1}).

\textbf{Step VI (Redistribution of Wavefunction Forces).}
At this stage, all EXX-related quantities have been evaluated; $\exx$ has been accumulated and broadcast to all \mpi{} processes, while $\{\dxx{i}\}$ is now stored in the \texttt{ORBITAL} data distribution scheme.
For compliance with the \texttt{CP} module in \texttt{QE}, $\{\dxx{i}\}$ is redistributed from the \texttt{ORBITAL} to the \texttt{GRID} scheme in this last step (see Sec.~III A, Sec.~III B, Sec.~III C 6, and Fig.~3 in \pI~\cite{paper1}).

In order to extend our MLWF-based approach to enable constant-volume ($NVE$/$NVT$) and constant-pressure ($NpH$/$NpT$) hybrid DFT simulations of condensed-phase systems described by general/non-orthogonal cells, we have made a series of modifications to the \exxm module.
Each of these modifications are described in detail below, and are delineated by the red dashed boxes in the \exxm flowchart provided in Fig.~\ref{fig:flowchart}.
In Sec.~\ref{subsec:impl_spdomain_npt}, we describe our modifications to Step II and Step III, which deal with proto-subdomain construction for (potentially fluctuating) simulation cells with general lattice symmetries.
Our extensions to Step IV, which enable an efficient CG solution of the PE on non-orthogonal real-space domains, are detailed in Sec.~\ref{subsec:impl_PS}. 
In Sec.~\ref{subsec:impl_cell_forces}, we present the needed extensions to Step V during constant-pressure CPMD simulations, \ie analytical evaluation of the EXX contribution to the stress tensor \via the cell derivatives ($\sxx$), as derived above in Sec.~\ref{sec:Pi_xx}.

\subsection{Extension of the \exxm Module: Subdomains in Constant-Pressure CPMD \label{subsec:impl_spdomain_npt}}

In this section, we describe our modifications to Step II and Step III of the \exxm module regarding the construction and selection of proto-subdomains during constant-volume and constant-pressure CPMD simulations of condensed-phases systems with general/non-orthogonal cells.

\subsubsection{Proto-Subdomain Construction for General/Non-Orthogonal Simulation Cells \label{sec:subdomain_gen}}

Treatment of general/non-orthogonal cells is a fairly straightforward extension to the orthorhombic case discussed previously (see Sec.~III C 2 and Algorithm 2 of \pI~\cite{paper1}), and requires the following two distinctions.
For one, the lattice vectors ($\bm L_{1}, \bm L_{2}, \bm L_{3}$) no longer coincide with the Cartesian directions (which are labelled using Roman indices $a,b,c$), and therefore require a distinct index convention (\ie Greek indices $\alpha,\beta,\gamma$) as defined in Sec.~\ref{sec:notation}.
In addition, the transformation between Cartesian and crystal coordinates requires the full cell tensor, \ie $\bm{r} = \bm{h} \bm{s}$ (as opposed to the simpler $r_{a} = |\bm L_{a}| s_{a}$ in the orthorhombic case).
For a general/non-orthogonal cell with $N_{\rm{grid},\alpha}$ equispaced grid points along each of the $\bm L_\alpha$ lattice vectors (with grid spacing $\delta\xi_{\alpha}=|\bm L_\alpha|/N_{\rm{grid},\alpha}$), the global grid index along $\bm L_\alpha$ is given by $g_\alpha = N_{\rm{grid},\alpha} \bm s_{\alpha}$.

Given user-defined values for $\rpe$ and $\rme$ (for both self (s) and non-self (ns) cases as discussed in Sec.~IV of \pI~\cite{paper1}), the \exxm module now employs a general/non-orthogonal variant of Algorithm~2 in \pI~\cite{paper1} during the construction of the $\pe{\gC{0}}$ and $\me{\gC{0}}$ proto-subdomains, each of which contains $\npe \in \{ \npes , \npep \}$ and $\nme \in \{ \nmes , \nmep \}$ grid points, respectively.
When compared to the original algorithm for orthorhombic simulation cells, the only difference lies in the use of the full cell tensor (instead of the lattice dimensions) during the computation of the $\bm g^0_{\rm PE}$ and $\bm g^0_{\rm ME}$ global grid indices.
In practice, this leads to a revised assignment of $\bm g^0_{\rm PE}[q']$ and $\bm g^0_{\rm ME}[q'']$ with $\texttt{NINT} \left[ N_{\rm{grid},\alpha} \left( \bm h^{-1} \bm r \right)_{\alpha} \right]$ (for $\alpha = 1,2,3$) instead of the original form, \ie $\texttt{NINT} \left[ N_{\rm{grid},a} r_{a}/|\bm L_a| \right]$ (for $a = 1,2,3$).
As such, the resultant proto-subdomains reflect the symmetry of the underlying (general/non-orthogonal) simulation cell.
\noindent Following the same conventions defined in \pI~\cite{paper1}, the $\pe{\gC{0}}$ and $\me{\gC{0}}$ proto-subdomains are again stored by the modified \exxm module as a set of local (relative) Cartesian coordinates in a $3\times\nme$ double-precision array,
\begin{align}
\hspace{-0.075in}\rbar[q] &= \left.
  \begin{cases}
    \rbar_{\rm PE}[q] & q = 1,\ldots,N_{\rm PE} \\
    \rbar_{\rm ME}[q-N_{\rm PE}] & q = N_{\rm PE}+1,\ldots,N_{\rm ME}
  \end{cases}
  \right\} ,
  \label{eq:rbar}
\end{align}
and a set of global grid indices in a $3\times\nme$ integer array,
\begin{align}
\hspace{-0.09in}\bm g^0[q] &= \hspace{-0.01in}\left.
  \begin{cases}
    \bm g^0_{\rm PE}[q] & q = 1,\ldots,N_{\rm PE} \\
    \bm g^0_{\rm ME}[q-N_{\rm PE}] & q = N_{\rm PE}+1,\ldots,N_{\rm ME}
  \end{cases}
  \right\} .
  \label{eq:g0}
\end{align}

With such compact representations of the $\pe{\gC{0}}$ and $\me{\gC{0}}$ proto-subdomains, the \exxm module is now positioned to construct the $\pe{\gC{ij}}$ subdomain (for computing the $\braket{ij}$ contributions to $\exx$ and $\sxx$) as well as the $\me{\gC{ij}}$ subdomain (for computing $\dxxb{ij}$ and $\dxxb{ji}$).
As discussed in Sec.~III C 3 of \pI~\cite{paper1}, these subdomains can be conveniently obtained \via a rigid translation of the $\pe{\gC{0}}$ and $\me{\gC{0}}$ proto-subdomains from $\gC{0}$ to $\gC{ij}$ (which is a crucial operation when communicating the MLWFs among \mpi{} processes in Step III of the \exxm module).
For general/non-orthogonal cells, the component of the required grid translation vector ($\bm \tau^{ij}$) along a lattice vector $\bm L_{\alpha}$ is evaluated \via an intermediate mapping to crystal coordinates ($\bm s = \bm h^{-1} \left( \gC{ij} - \gC{0} \right)$) and given by:
\begin{align}
  \tau_{\alpha}^{ij} &= \texttt{NINT} \left[ N_{\rm{grid},\alpha} \left[ \bm h^{-1} \left( \gC{ij} - \gC{0} \right) \right]_{\alpha} \right] \nonumber \\
                     &= \texttt{NINT} \left[ \frac{ |\bm L_{\alpha}| \left[ \bm h^{-1} \left( \gC{ij} - \gC{0}\right) \right]_{\alpha}}{\delta\xi_{\alpha}} \right] ,
\end{align}
where we have used the fact that $\delta\xi_{\alpha}=|\bm L_\alpha|/N_{\rm{grid},\alpha}$.
Application of $\bm \tau^{ij}$ to a given proto-subdomain leaves the radius ($\rpe$ or $\rme$) and local Cartesian coordinates ($\rbar$) unchanged, and simply offsets the global grid indices as follows:
\begin{align}
  g^{ij}_{\alpha}[q] = \texttt{MOD} \left[ g^{0}_{\alpha}[q] + \tau^{ij}_{\alpha}, \, N_{\rm{grid},\alpha} \right] ,
  \label{eq:l2g}
\end{align}
thereby resulting in a subdomain ($\pe{\gC{ij}}$ or $\me{\gC{ij}}$) that is centered at $\gC{ij}$ and has the symmetry of the underlying simulation cell.

\subsubsection{Proto-Subdomain Selection During Constant-Pressure CPMD Simulations \label{sec:subdmain_npt}}

During an MLWF-based CPMD simulation of an insulating system, the band gap is not expected to have substantial variations; as such, individual MLWF spreads will fluctuate, but the size/extent of the support associated with these exponentially decaying functions will remain essentially constant throughout the trajectory.
Here, we note that this assumption may break down (to varying extents) for small-gap and/or substantially inhomogeneous systems~\cite{dawson_performance_2015} (\eg solvated semiconducting nanoparticles, water-semiconductor interfaces, surface adsorption of gas-phase molecules, etc) as well as systems undergoing bond breaking and formation; in such cases, the use of MLWF-specific subdomains will be necessary to ensure a sufficiently converged evaluation of all EXX-related quantities, and will therefore be addressed in future versions of \exxm.
For fixed-cell simulations (\eg $NVE$/$NVT$), the size of the $\pe{\gC{ij}}$ and $\me{\gC{ij}}$ subdomains (\ie the translated $\pe{\gC{0}}$ and $\me{\gC{0}}$ proto-subdomains) are kept fixed throughout CPMD simulations by the \exxm module; for most systems (not including the pathological examples listed above), this choice results in a high-fidelity evaluation of $\exx$ and $\{\dxx{i}\}$ (as well as $\sxx$, \textit{vide infra}).
As such, all proto-subdomain related quantities in \exxm, which include the radii ($\rpe \in \{\rpes, \rpep\}$ and $\rme \in \{\rmes, \rmep\}$), the number of local grid points ($\npe \in \{\npes, \npep\}$ and $\nme \in \{\nmes, \nmep\}$), the local (relative) Cartesian coordinates ($\{ \rbar \}$), and the global grid indices ($\{ \bm g^0 \}$), are pre-computed prior to the first MD step and fixed throughout the simulation.

When a fluctuating cell is employed (\eg during $NpH$/$NpT$ simulations), the size and shape of $\Omega$ can vary significantly, while $\{ \Omega_{i} \}$ (again for non-pathological systems) is expected to retain a similar size/extent (but potentially a different shape) throughout the MD trajectory.
As such, we are now faced with the question of how to define the $\pe{\gC{ij}}$ and $\me{\gC{ij}}$ subdomains during constant-pressure simulations with \exxm.
In this work, we consider two common subdomain choices for CPMD simulations with fluctuating cells.
As a first option, the subdomains could be chosen such that the radii (\ie $\rpe$ and $\rme$) are fixed throughout the simulation; this leads to fixed quasi-spherical subdomain shapes with varying numbers of points (\ie $\npe$ and $\nme$) as the cell fluctuates.
Algorithmically speaking, the use of fixed $\rpe$ and $\rme$ has the disadvantages of: (\textit{i}) requiring the computation of $\{ \rbar \}$ and $\{ \bm g^0 \}$ during each CPMD step, (\textit{ii}) introducing an imbalance in the computational workload and associated memory requirements between CPMD steps, and (\textit{iii}) complicating the extrapolation schemes used for the $\tvb{ij}$ initial guess during the iterative solution of the PE.

To combat these algorithmic issues, we have opted to employ an alternative option in \exxm---choosing subdomains with a \textit{fixed number of grid points} throughout constant-pressure CPMD simulations, with $\npe$ and $\nme$ values determined by the \textit{initially chosen} proto-subdomains.
More specifically, we retain the following (initial) proto-subdomain related quantities throughout a given $NpH$/$NpT$ simulation: the number of grid points ($\npe$ and $\nme$), the global grid indices ($\{\bm g^0\}$), and the relative \textit{scaled} (not Cartesian) coordinates ($\{ \sbar \} = \{ \bm h_{0}^{-1} \rbar \}$, where $\bm h_{0}$ is the initial cell tensor).
In other words, the subdomains employed in our approach do not have fixed radii, and are therefore no longer (necessarily) quasi-spherical in shape; instead, these subdomains deform with the underlying fluctuating cell.
In doing so, this scheme directly addresses all of the algorithmic disadvantages that accompany the use of subdomains with fixed radii.
For one, there is no need for the additional computational overhead associated with computing $\{\rbar\}$ and $\{\bm g^0\}$ by screening $\Omega$ at each CPMD step; in this case, $\{\rbar\}$ is straightforwardly obtained \via $\{\rbar\} = \{\bm h \sbar\}$ (where $\bm h$ is the current cell tensor) and $\{\bm g^0\}$ is simply stored in memory.
In addition, the complications associated with workload/memory imbalances as well as extrapolation schemes (for the PE guess) are largely eliminated with the use of a fixed number of grid points in each subdomain.

In the presence of severely anisotropic cell fluctuations (\eg as one might encounter during a phase transition with large uniaxial strain), this approach should be further modified to ensure that the substantially deformed subdomains still provide adequate support for evaluating $\exx$, $\{\dxx{i}\}$, and $\sxx$.
This can be accomplished with the re-assembly (from scratch) of appropriately sized quasi-spherical proto-subdomains based on a pre-defined strain criteria or a given stride (\eg every $1000$ CPMD steps) throughout the simulation. 
Doing so would ensure a sufficiently converged evaluation of all EXX-related quantities and still retain all of the algorithmic advantages mentioned above.

It is also worth noting that both of these subdomain choices (\ie fixed radii or fixed number of points) are subject to Pulay-like errors~\cite{marx_ab_2009} during constant-pressure CPMD simulations.
Such errors originate from the use of discrete Laplacian representations---the accuracy of which is governed by the grid point spacing ($\{ \delta\xi_{\alpha} \}$) in $\Omega$---during the solution of the PE.
In the \exxm module, the accumulation of such errors is largely mitigated by the default use of a sufficiently accurate finite-difference representation of the Laplacian operator (\ie with an associated error of $\mathcal{O}(\delta\xi_{\alpha}^{6})$, \textit{vide infra}), which can be reduced even further (at linear computational cost) by simply employing a higher-order stencil (see Sec.~\ref{subsec:impl_PS}).

\subsection{Extension of the \exxm Module: Solving Poisson's Equation in an Arbitrary Simulation Cell \label{subsec:impl_PS}}

In this section, we describe the extensions introduced in \exxm to enable the solution of the PE for each overlapping $\braket{ij}$ MLWF-pair (\ie $\nabla^2 \tvb{ij} = -4\pi \trhob{ij}$, see Eq.~\eqref{eq:pe}) in condensed-phase systems described by general/non-orthogonal simulation cells.
Throughout this discussion, we will therefore consider the most general case in which the lattice vectors ($\{ \bm L_1, \bm L_2, \bm L_3\}$) are non-orthogonal and therefore not necessarily aligned with the standard unit Cartesian directions ($\{ \widehat{\bm e}_{x}, \widehat{\bm e}_{y}, \widehat{\bm e}_{z}\}$), as one would encounter with orthorhombic (\eg simple cubic) cells.

While a ME about $\gC{ij}$ (which is used to obtain the boundary conditions for the PE as well as the far-field solution for $\tvb{ij}$) can be straightforwardly computed using Eqs.~\eqref{eq:me}--\eqref{eq:mepole}, the near-field solution for $\tvb{ij}$ requires a discrete representation for the Laplacian operator when computing numerical second derivatives during the solution of the PE.
Since the subdomains employed in the \exxm module are coincident with the underlying real-space grid, it is most computationally efficient to employ a discrete representation for the Laplacian that is aligned with $\bm L_1$, $\bm L_2$, and $\bm L_3$.
To proceed, we employ the unit lattice vectors as the basis for this tilted (non-Cartesian) space, \ie $\widehat{\bm{L}}_{\alpha} \equiv \bm L_{\alpha} / \left| \bm L_{\alpha} \right|$ for $\alpha \in \{1,2,3\}$, such that a given position vector $\bm{v} \in \mathbb{R}^3$ can be written using either Cartesian ($\bm{r} = \{ r_x,r_y,r_z \}$) or tilted/non-Cartesian ($\bm{\xi} = \{ \xi_1,\xi_2,\xi_3 \}$) coordinates.
Direct solution of the PE on these subdomains (\ie without the need for interpolation of $\trhob{ij}$ and $\tvb{ij}$ to and from an auxiliary Cartesian grid) will therefore require a coordinate transformation that connects the Laplacian operator in these two representations \via the corresponding Jacobian matrix ($\bm J = \partial \bm \xi / \partial \bm r$).

Since the tilted/non-Cartesian coordinates (which use the \textit{unit} lattice vectors as a basis) are related to crystal coordinates (which use the lattice vectors as a basis) for any arbitrary position vector $\bm{v}$, namely,
\begin{align}
\bm{v} = \sum_{\alpha}s_{\alpha} \bm{L}_{\alpha} = \sum_{\alpha}s_{\alpha} |\bm{L}_{\alpha}| \widehat{\bm{L}}_{\alpha} = \sum_{\alpha} \xi_{\alpha} \widehat{\bm{L}}_{\alpha} , 
   \label{eq:s_to_xi}
\end{align}
one sees that $\xi_{\alpha} = |\bm L_{\alpha}| s_{\alpha} = |\bm L_{\alpha}|\sum_{a} (h^{-1})_{\alpha a} r_{a}$.
Using this relationship, one can derive an explicit expression for $\bm{J}$ as follows:
\begin{align}
  J_{a\alpha} &= \left( \frac{\partial \xi_{\alpha}}{\partial r_{a}} \right) = \frac{\partial}{\partial r_{a}} \left[ |\bm L_{\alpha}| \sum_{b} \left( h^{-1}\right)_{\alpha b} r_{b}\right] \nonumber \\
  &= |\bm L_{\alpha}| \sum_{b} \left( h^{-1}\right)_{\alpha b} \delta_{b a} = | \bm L_{\alpha}| \left( h^{-1}\right)_{\alpha a} .
  \label{eq:jacb}
\end{align}
With the Jacobian in Eq.~\eqref{eq:jacb}, the Cartesian gradient operator, $\bm{\nabla}_{\bm{r}} \equiv \left( \partial/\partial r_x, \partial/\partial r_y, \partial/\partial r_z \right)$, can be written in terms of the directional derivatives along the (unit) lattice vectors, $\bm \nabla_{\bm \xi} \equiv \left( \partial/\partial \xi_1, \partial/\partial \xi_2, \partial/\partial \xi_3 \right)$, \via $\bm{\nabla}_{\bm{r}} = \bm{J} \bm{\nabla}_{\bm{\xi}}$.
These expressions can in turn be used to derive the desired form for the Laplacian operator, \ie
\begin{align}
  \nabla^2_{\bm r} &= \bm \nabla_{\bm r} \cdot \bm \nabla_{\bm r} = \sum_{a} \left( \nabla_{\bm r} \right)_{a} \left( \nabla_{\bm r} \right)_{a} \nonumber \\
  &= \sum_{a} \Bigg[ \sum_{\alpha} J_{a \alpha} \left( \nabla_{\bm \xi} \right)_{\alpha} \Bigg] \Bigg[ \sum_{\beta} J_{a \beta} \left( \nabla_{\bm \xi} \right)_{\beta} \Bigg] \nonumber \\
  &= \sum_{\alpha\beta} F_{\alpha\beta} \left( \nabla_{\bm \xi} \right)_{\alpha} \left( \nabla_{\bm \xi} \right)_{\beta} ,
  \label{eq:lapG_}
\end{align}
in which $F_{\alpha\beta} \equiv \sum_{a} J_{a\alpha} J_{a\beta}$ is an element of the symmetric $\bm{F} = \bm{J}^{T}\bm{J}$ matrix.

Using the Clairaut-Schwarz theorem, the Laplacian in Eq.~\eqref{eq:lapG_} can be further split into a sum over pure ($\partial^2/\partial \xi_\alpha^2$) and mixed ($\partial^2/\partial \xi_\alpha \partial \xi_\beta$) second partial derivatives as follows:
\begin{align}
  \nabla^2_{\bm r} &= \sum_{\alpha} \left[ F_{\alpha\alpha} \frac{\partial^2}{\partial \xi_{\alpha}^2} + 2 \sum_{\beta > \alpha} F_{\alpha \beta} \frac{\partial^2}{\partial \xi_{\alpha}\partial \xi_{\beta}} \right] .
  \label{eq:lapG}
\end{align}
The pure derivatives in Eq.~\eqref{eq:lapG} can be straightforwardly represented by standard central-difference formulae along each of the lattice vectors; at a given grid point, $\bm{\xi}_{0}$, these pure derivatives are evaluated using the following working expression (shown here for a generic function, $f(\bm{\xi})$, along $\bm{L}_{\alpha}$): 
\begin{align}
  \left.\frac{\partial^2 f(\bm \xi)}{\partial \xi_{\alpha}^2} \right|_{\bm \xi = \bm \xi_{0}} = \sum_{q=-n}^{n} w_{q}\frac{  f(\bm \xi_{0} + q \, \delta \xi_{\alpha} \widehat{\bm L}_{\alpha})}{ \delta \xi_{\alpha}^2} .
  \label{eq:d2_1}
\end{align}
In this expression, the sum is over the $n$ neighboring grid points (along $\bm{L}_{\alpha}$) located on each side of $\bm{\xi}_0$, and $w_{q}=w_{-q}$ is the central-difference coefficient~\cite{fornberg_generation_1988} for the $q$-th neighboring grid point.
As such, the finite-difference representation of a pure second derivative results in a ($2n+1$)-point stencil along the given grid direction with an associated discretization error of $\mathcal{O} \left( \delta \xi_{\alpha}^{2n} \right)$.
The default option in \exxm is $n=3$ with a discretization error of $\mathcal{O}\left( \delta\xi_{\alpha}^6 \right)$, as this choice furnishes well-converged values for all EXX-related quantities~\cite{wu_order-n_2009,distasio_jr._individual_2014,paper1}. 
In this case, the corresponding central-difference coefficients~\cite{fornberg_generation_1988} are given by: $w_{0} = -49/18$, $w_{1} = +3/2 = w_{-1}$, $w_{2} = -3/20 = w_{-2}$, and $w_{3} = +1/90 = w_{-3}$.

While the pure derivatives in Eq.~\eqref{eq:lapG} can be accurately and efficiently evaluated using standard central-difference techniques, there is considerable flexibility when evaluating the mixed derivatives in this expression. 
Here, we remind the reader that direct calculation of each mixed derivative $\partial^2 /\partial \xi_\alpha \partial \xi_\beta$ in Eq.~\eqref{eq:lapG} would require consecutive finite-difference evaluations of the $\partial /\partial \xi_\alpha$ and $\partial /\partial \xi_\beta$ first derivatives.
However, the number of stencil points in such an approach would scale quadratically with $n$,~\cite{brandt_multigrid_1999,natan_real-space_2008} and would therefore result in a substantially more expensive EXX algorithm for non-orthogonal simulation cells.

\subsubsection{The Natan-Kronik (NK) Representation of $\nabla^2$: Elimination of Mixed Derivatives via Auxiliary Grid Directions \label{subsub:nk_lapl}}

To alleviate this quadratic complexity, we follow the approach proposed by Natan, Kronik, and coworkers~\cite{natan_real-space_2008}, which has roots in earlier work by Brandt and Diskin~\cite{brandt_multigrid_1999} (in the 2D theory of sonic flow), and will be referred to as NK throughout the remainder of the manuscript.
Before describing the NK approach for treating 3D general/non-orthogonal cells (as well as our algorithmic implementation for dealing with fluctuating cells during constant-pressure simulations in \exxm), we first review the core idea behind the NK approach, \ie the use of grid-resolved auxiliary direction(s) to eliminate the computationally expensive mixed derivative(s) in Eq.~\eqref{eq:lapG}.

\begin{figure}[t!]
  \includegraphics[width=0.8\linewidth]{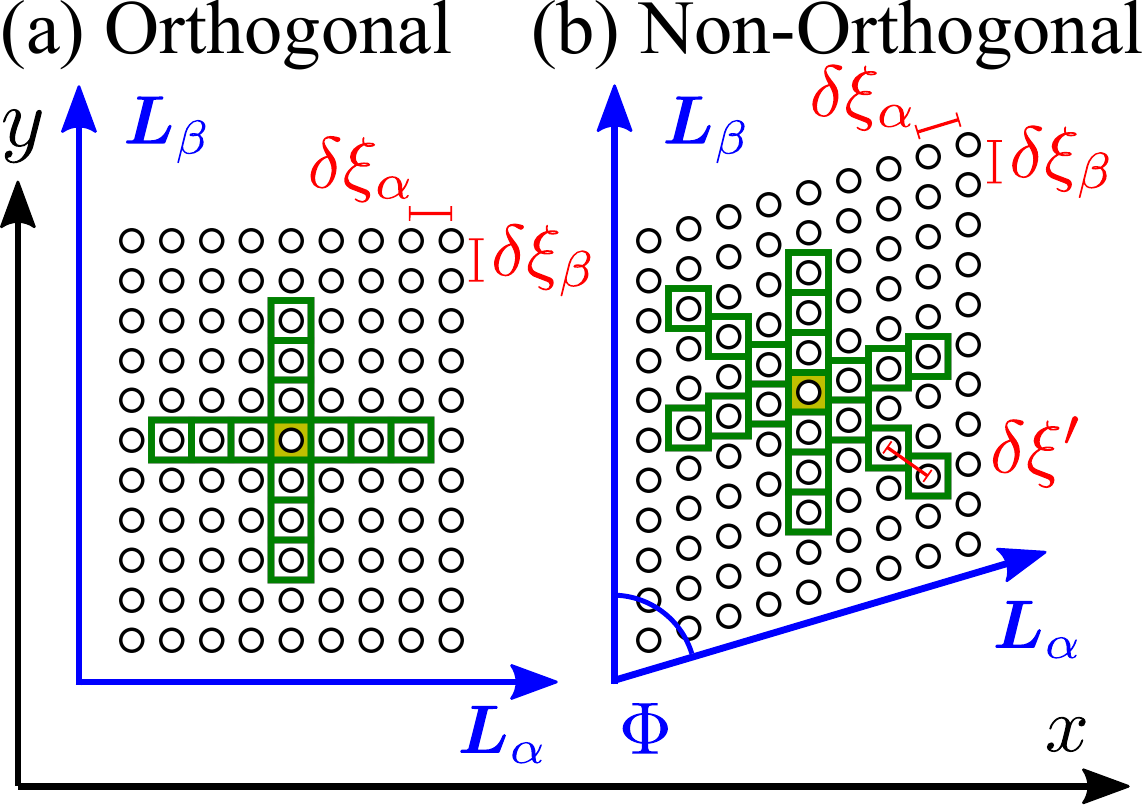}
  \caption{
  Graphical depiction of the grid-resolved directions used in the NK representation of the 2D Laplacian ($\nabla^2_{\bm r}$) in (a) an orthogonal ($\bm L_{\alpha} \perp \bm L_{\beta}$) and (b) a non-orthogonal ($\bm L_{\alpha} \not\perp \bm L_{\beta}$) cell.
  Each of these discretized Laplacians is centered at a given grid point ($\bm{\xi}_0$, highlighted in yellow), and represented by a finite-difference stencil which covers the neighboring $\pm n$ grid points (shown here for $n=3$) in each required derivative direction (see Eq.~\eqref{eq:d2_1}). 
  In the 2D orthogonal case, the NK Laplacian takes on the standard form for $\nabla^2_{\bm r}$, and includes pure derivatives $\partial^2/\partial \xi_{\alpha}^2$ and $\partial^2/\partial \xi_{\beta}^2$ along lattice vectors $\bm L_{\alpha}$ and $\bm L_{\beta}$ (with corresponding grid spacings $\delta\xi_{\alpha}$ and $\delta\xi_{\beta}$).
  In the 2D non-orthogonal case, the NK Laplacian (in addition to $\partial^2/\partial \xi_{\alpha}^2$ and $\partial^2/\partial \xi_{\beta}^2$) replaces the mixed derivative $\partial^2/\partial \xi_{\alpha}\partial \xi_{\beta}$ with a pure derivative $\partial^2/\partial \xi'^2$ along a grid-resolved \textit{auxiliary} direction $\widehat{\bm{L}}'$ (with grid spacing $\delta\xi'$).
  To maximize the accuracy of this finite-difference Laplacian, the nearest-neighbor grid-resolved direction (\ie with the smallest possible $\delta\xi'$) is chosen as $\widehat{\bm{L}}'$: when the angle $\Phi$ between $\bm{L}_{\alpha}$ and $\bm{L}_{\beta}$ is obtuse (acute), $\widehat{\bm{L}}'$ is chosen to be the grid-resolved bisector of $\Phi$ (the supplementary angle to $\Phi$, as shown above).
  }
  \label{fig:grid}
\end{figure}
To do so, we will first consider the simplest non-orthogonal case, a 2D simulation cell with lattice vectors $\bm L_{\alpha} \not\perp \bm L_{\beta}$.
In this case, $F_{\alpha\beta}$ is the only non-zero off-diagonal term in $\bm F$ (see Eq.~\eqref{eq:lapG_}), and hence $\partial^2/\partial \xi_\alpha \partial \xi_\beta$ is the only mixed partial derivative in Eq.~\eqref{eq:lapG}.
In what follows, we will assume that $\delta\xi_{\alpha} \approx \delta\xi_{\beta}$ (\ie the grid spacings in the $\bm L_{\alpha}$ and $\bm L_{\beta}$ directions are approximately equivalent), which is typically enforced by the planewave (kinetic energy) cutoff and the FFT algorithm.
Under this assumption, the NK approach (for a 2D non-orthogonal cell) involves choosing a \textit{single} unit auxiliary direction ($\widehat{\bm{L}}'$) that meets the following criteria: (\textit{i}) $\widehat{\bm{L}}'$ is non-axial, \ie distinct from $\widehat{\bm{L}}_{\alpha}$ and $\widehat{\bm{L}}_{\beta}$, (\textit{ii}) $\widehat{\bm{L}}'$ lies in the plane defined by $\widehat{\bm{L}}_{\alpha}$ and $\widehat{\bm{L}}_{\beta}$ (or equivalently, $\bm{L}_{\alpha}$ and $\bm{L}_{\beta}$), (\textit{iii}) $\widehat{\bm{L}}'$ is coincident with the underlying real-space grid (\ie $\widehat{\bm{L}}'$ is grid-resolved), and (\textit{iv}) $\widehat{\bm{L}}'$ corresponds to the nearest-neighbor grid direction (\ie $\widehat{\bm{L}}'$ has the smallest possible grid spacing, $\delta\xi'$).
To ensure that all four of these criteria are satisfied, $\widehat{\bm{L}}'$ can be written in the following compact form:
\begin{align}
    \widehat{\bm L}' &= \frac{\widehat{\bm L}_{\alpha}+\kappa \widehat{\bm L}_{\beta}}{\left|\widehat{\bm L}_{\alpha}+\kappa \widehat{\bm L}_{\beta}\right|} \equiv \frac{\widehat{\bm L}_{\alpha}+\kappa \widehat{\bm L}_{\beta}}{d} ,
    \label{eq:nk2d_aux}
\end{align}
in which $\kappa$ is defined as
\begin{align}
    \kappa \!\equiv\! - \! \left( \frac{\delta\xi_{\beta}}{\delta\xi_{\alpha}} \right) \sgn \! \left[ \widehat{\bm L}_{\alpha} \cdot \widehat{\bm L}_{\beta} \right]
    \!=\! - \! \left(\frac{\delta\xi_{\beta}}{\delta\xi_{\alpha}}\right) \sgn \! \left[ \cos \Phi \right] .
    \label{eq:kappa_def}
\end{align}
When the angle $\Phi$ between $\widehat{\bm L}_{\alpha}$ and $\widehat{\bm L}_{\beta}$ is obtuse (acute), this convention for $\kappa$ makes $\widehat{\bm L}'$ the grid-resolved bisector of $\Phi$ (the supplementary angle to $\Phi$), as depicted in Fig.~\ref{fig:grid}.
This choice for $\widehat{\bm{L}}'$ also has the smallest possible $\delta\xi'$, which allows us to retain the highest degree of accuracy (at a given discretization order) in the finite-difference representation of $\nabla^2_{\bm r}$ (see Eq.~\eqref{eq:d2_1}).

With these expressions in hand, the first and second partial derivatives with respect to $\xi'$ (the coordinate associated with $\widehat{\bm{L}}'$) take on the following form:
\begin{align}
  \frac{\partial}{\partial \xi'} &= \frac{1}{d} \left( \frac{\partial}{\partial \xi_{\alpha}} +\kappa \frac{\partial}{\partial \xi_{\beta}} \right) ,
  \label{eq:nk2d_aux_d}
\end{align}
and
\begin{align}
    \frac{\partial^2 }{\partial {\xi'}^2} &= \frac{1}{d^2} \left( \frac{\partial^2}{\partial \xi_{\alpha}^2} + \kappa^2 \frac{\partial^2}{\partial \xi_{\beta}^2} + 2 \kappa\frac{\partial^2}{\partial \xi_{\alpha}\xi_{\beta}} \right) .
  \label{eq:nk2d_aux_d2}
\end{align}
Eq.~\eqref{eq:nk2d_aux_d2} can then be rearranged to express the mixed partial derivative,
\begin{align}
  \frac{\partial^2 }{\partial \xi_{\alpha}\partial \xi_{\beta}} = \frac{1}{2\kappa} \left( -\frac{\partial^2 }{\partial {\xi_{\alpha}^2}} - \kappa^2 \frac{\partial^2}{\partial {\xi_{\beta}^2}} + d^2 \frac{\partial^2 }{\partial {\xi'}^2 } \right) ,
  \label{eq:mix_to_sec}
\end{align}
as a linear combination of pure derivatives along the $\widehat{\bm{L}}_{\alpha}$ and $\widehat{\bm{L}}_{\beta}$ unit lattice vectors, as well as the $\widehat{\bm{L}}'$ unit auxiliary vector.~\cite{natan_real-space_2008}
After plugging Eq.~\eqref{eq:mix_to_sec} into Eq.~\eqref{eq:lapG}, one arrives at the NK Laplacian with a total of $N_{\rm pure} = 3$ pure derivatives, \ie
\begin{align}
  \nabla^2_{\bm r} &= \left( F_{\alpha\alpha} - \frac{F_{\alpha\beta}}{\kappa} \right)\frac{\partial^2}{\partial \xi_{\alpha}^2} + \left( F_{\beta\beta} - \kappa F_{\alpha\beta} \right)\frac{\partial^2}{\partial \xi_{\beta}^2} \nonumber \\
  &+\frac{F_{\alpha\beta}d^2}{\kappa}\frac{\partial^2}{\partial {\xi'}^2} ,
  \label{eq:lapG_nk2d}
\end{align}
each of which can now be accurately and efficiently evaluated using standard central-difference techniques (see Eq.~\eqref{eq:d2_1}), and the computationally expensive ($\mathcal{O}(n^{2})$) direct evaluation of the mixed derivative is completely avoided.

For the general 3D case, up to three lattice vectors can be mutually non-orthogonal.
For each pair of non-orthogonal lattice vectors, the corresponding off-diagonal element in $\bm{F}$ will be non-zero, thereby necessitating the corresponding mixed derivative in Eq.~\eqref{eq:lapG}.
In this work, we follow the original NK prescription~\cite{natan_real-space_2008,NKnote} in which the pure derivatives in Eq.~\eqref{eq:lapG} are always evaluated along the unit lattice directions, $\{ \widehat{\bm{L}}_{1}, \widehat{\bm{L}}_{2}, \widehat{\bm{L}}_{3} \}$.
Evaluation of the mixed derivative(s) in Eq.~\eqref{eq:lapG} will then require one ($N_{\rm aux} = 1$) to three ($N_{\rm aux} = 3$) additional grid-resolved unit auxiliary directions, $\{ \widehat{\bm{L}}'_{p} \}$, each of which can be written as a linear combination of the unit lattice vectors:
\begin{equation}
  \widehat{\bm{L}}'_{p} = \sum_{\alpha} a_{p\alpha} \widehat{\bm{L}}_{\alpha} \qquad p = 1, 2, \ldots , N_{\rm aux} .
  \label{eq:nk3d_mu}
\end{equation}
In this expression, the expansion coefficients, $\{ a_{p\alpha} \}$, are then chosen to satisfy all of the requirements of the 3D NK approach (see Sec.~\ref{subsub:nk_lapl_our_algorithm}). 

Following the procedure described above, we take the first and second partial derivatives of Eq.~\eqref{eq:nk3d_mu} with respect to $\xi'_{p}$ (the coordinate associated with $\widehat{\bm{L}}'_{p}$), which yields:
\begin{align}
  \frac{\partial}{\partial \xi'_{p}} &= \sum_{\alpha} a_{p\alpha} \frac{\partial}{\partial \xi_{\alpha}} ,
  \label{eq:nk3d_aux_d}
\end{align}
and
\begin{align}
    \frac{\partial^2 }{\partial {{\xi'_{p}}^2} } &=
    \sum_{\alpha}\left( a_{p\alpha}^2 \frac{\partial^2}{\partial \xi_{\alpha}^2} + 2 \sum_{\beta > \alpha} a_{p\alpha} a_{p\beta} \frac{\partial^2}{\partial \xi_{\alpha} \partial \xi_{\beta}}\right) .
  \label{eq:nk3d_aux_d2}
\end{align}
Unlike Eq.~\eqref{eq:nk2d_aux_d2}, the pure derivative along a given auxiliary direction in Eq.~\eqref{eq:nk3d_aux_d2} generally contains contributions from more than one mixed derivative (since $a_{p\alpha}a_{p\beta}$ is generally non-vanishing). 
To address this issue, the NK approach seeks to find a linear combination, $\sum_{p} b_{p} (\partial^2 /\partial {{\xi'_{p}}^2})$, that has the same \textit{mixed} derivative contribution as that in Eq.~\eqref{eq:lapG}, \ie
\begin{align}
  \sum_{p} b_{p} \, \sum_{\mathclap{\alpha , \, \beta > \alpha}} a_{p\alpha} a_{p\beta} \frac{\partial^2}{\partial \xi_{\alpha} \partial \xi_{\beta}} = \sum_{\mathclap{\alpha , \, \beta > \alpha}} F_{\alpha \beta} \frac{\partial^2}{\partial \xi_{\alpha}\partial \xi_{\beta}} .
  \label{eq:nk3d_replace_mixed_d}
\end{align}
This expression can be written in matrix form as:
\begin{align}
    \!\!
    \left[
    \begin{array}{ccc}
       a_{11}a_{12} & a_{21}a_{22} & a_{31}a_{32} \\
       a_{11}a_{13} & a_{21}a_{23} & a_{31}a_{33} \\
       a_{12}a_{13} & a_{22}a_{23} & a_{32}a_{33} \\
    \end{array}
    \right]
    \!\!
    \left[
    \begin{array}{c}
       b_{1}  \\
       b_{2}  \\
       b_{3}  \\
    \end{array}
    \right]
    \! \equiv
    \bm{M} \bm{b}
     = \!\!
    \left[
    \begin{array}{c}
         F_{12}  \\
         F_{13}  \\
         F_{23}  \\
    \end{array}
    \right] ,
  \label{eq:nk3d_Mb_f}
\end{align}
the solution of which ($\bm{b}$) can be used to eliminate the mixed derivatives in Eq.~\eqref{eq:lapG}, and derive the following working expression for the 3D NK Laplacian,
\begin{align}
  \nabla^2_{\bm r} &= \sum_{\alpha} \left(F_{\alpha\alpha} - \sum_{p} b_{p} a_{p\alpha}^2 \right) \!\! \frac{\partial^2}{\partial \xi_\alpha^2} + \sum_{p} b_p \frac{\partial^2}{\partial {\xi'_{p}}^2} ,
  \label{eq:lapG_mod_nk3d}
\end{align}
following the analogous procedure used above to derive Eqs.~\eqref{eq:mix_to_sec} and \eqref{eq:lapG_nk2d} for the 2D non-orthogonal case.
In this expression, each pure derivative can again be accurately and efficiently evaluated using standard central-difference techniques (see Eq.~\eqref{eq:d2_1}), thereby avoiding the computationally expensive ($\mathcal{O}(n^{2})$) direct evaluation of the mixed derivatives.
Using the approach outlined here,~\cite{NKnote} the number of auxiliary directions ($N_{\rm aux}$) is typically equal to the number ($N_{\rm off}$) of non-zero off-diagonal elements in $\bm F$, thereby leading to a total of $N_{\rm pure} = N_{\rm off} + 3$ pure derivatives in Eq.~\eqref{eq:lapG_mod_nk3d} (and a corresponding central-difference stencil which contains $N_{\rm stcl} = 2 n N_{\rm pure}+1$ points \via Eq.~\eqref{eq:d2_1}).
In the orthorhombic case, $N_{\rm off} = 0$ and the NK Laplacian in Eq.~\eqref{eq:lapG_mod_nk3d} reduces to the standard Laplacian with $N_{\rm pure} = 3$ pure derivatives along the lattice directions.

\subsubsection{Algorithmic Implementation of the NK Scheme for Fluctuating Simulation Cells \label{subsub:nk_lapl_our_algorithm}}

During constant-pressure ($NpH$/$NpT$) AIMD simulations, the size and shape of the cell will constantly change due to instantaneous fluctuations and/or on-going phase transitions throughout the trajectory.
During such fluctuations, the number of auxiliary directions required to evaluate Eq.~\eqref{eq:lapG_mod_nk3d} could range from zero (\eg orthorhombic) to three (\eg triclinic).
As such, we have implemented an automated algorithm in \exxm (executed at the beginning of each MD step) that chooses a set of auxiliary directions which meets all of the requirements of the NK approach (see Algorithm~\ref{alg:nk3d_aux_sel}) and holds for the 2D and 3D non-orthogonal cases described above (\ie Eqs.~\eqref{eq:lapG_nk2d} and \eqref{eq:lapG_mod_nk3d}).
%
In particular, this algorithm identifies a set of grid-resolved unit auxiliary directions, $\{ \widehat{\bm{L}}'_{p} \}$, that satisfy the following criteria: (\textit{i}) each $\widehat{\bm{L}}'_{p}$ is non-axial, \ie distinct from the $\widehat{\bm{L}}_1$, $\widehat{\bm{L}}_2$, and $\widehat{\bm{L}}_3$ unit lattice directions; (\textit{ii}) each $\widehat{\bm{L}}'_{p}$ has the minimum possible grid spacing $\delta\xi'_{p}$; and (\textit{iii}) the $\bm{M}$ matrix constructed using $\{ \widehat{\bm{L}}'_{p} \}$ is non-singular (see Eq.~\eqref{eq:nk3d_Mb_f}).

Input into Algorithm~\ref{alg:nk3d_aux_sel} is $\widehat{\bm{v}}$, a list containing $N_{\rm q} \gg 3$ \textit{candidate} non-axial auxiliary directions, $\widehat{\bm{\ell}}'_{q}= \sum_{\alpha} \widetilde{a}_{q\alpha} \widehat{\bm{L}}_{\alpha}$, each of which has been sorted (in ascending order) by grid spacing ($\delta\widetilde{\xi}{}'_{q}$); by providing this list as input, criterion (\textit{i}) is automatically satisfied.
To generate $\widehat{\bm{v}}$, we start from a reference grid point ($\bm{\xi}_{0}$) and sweep through surrounding shells of grid points ($\mathbb{G_S} = \{ \bm \xi \mid \texttt{NINT}[ \, \max_{\alpha} | (\xi_{\alpha} - (\xi_{0})_{\alpha}) / \delta\xi_{\alpha} | \, ] = \mathbb{S} \}$ for $\mathbb{S} = 1, 2, \ldots$) to locate $\{ \widehat{\bm{\ell}}'_{q} \}$.~\cite{UpperGnote}
Defining $\delta \xi_{>}$ as the largest spacing seen in the first shell (\ie $\delta \xi_{>} \equiv \max_{\bm \xi \in \mathbb{G}_{1}} | \bm{\xi} - \bm{\xi}_{0} |$), the search stops at the $\mathbb{S}$-th shell if $| \bm{\xi} - \bm{\xi}_{0}| \, > \delta \xi_{>} \, \forall \, \bm{\xi} \in \mathbb{G_S}$; doing so efficiently ensures that we do not miss any of the first $N_{\rm q}$ candidate auxiliary directions.
In practice, \exxm uses a default value of $N_{\rm q} = 15$, which is larger than the $10$ non-axial grid points in $\mathbb{G}_{1}$~\cite{UpperGnote} and should suffice for almost all cases; if necessary, $N_{\rm q}$ can be increased for simulation cells with very small ($\approx 0^\circ$) or very large ($\approx 180^\circ$) angles between lattice directions.
\begin{algorithm}[H]
  \begin{algorithmic}
    \State \textit{Input}: $\widehat{\bm{v}}[q] = \sum_{\alpha} \widetilde{a}_{q\alpha} \widehat{\bm{L}}_\alpha$ \, ($q = 1, 2, \ldots, N_{\rm q}$)
    \State $p \gets 1$; $\bm{a} \gets [\bm 0]_{3\times3}$; $\bm{M} \gets [\bm 0]_{3\times3}$
    \For{($q = 1, N_{\rm q}$)}
      \State $\{a_{p1},a_{p2},a_{p3}\} \gets \{\widetilde{a}_{q1},\widetilde{a}_{q2},\widetilde{a}_{q3}\}$  \Comment{propose candidate}
      \State $\{M_{1p},M_{2p},M_{3p}\} \gets \{a_{p1}a_{p2}, a_{p1}a_{p3}, a_{p2}a_{p3}\}$
      \If {($p == 1$)}
        \State \texttt{is\_accepted} $\gets$ \texttt{TRUE}
      \ElsIf {($p == 2$)}
        \State \texttt{is\_accepted} $\gets$ ($\bm{M}_{:,1} \nparallel \bm{M}_{:,2}$)
      \ElsIf {($p == 3$)}
        \State \texttt{is\_accepted} $\gets$ ($\det \bm{M} \neq 0$)
      \EndIf
      \If {(\texttt{is\_accepted})}
        \State $p \gets p+1$
        \If {($p > 3$)}
          \State \Break
        \EndIf
      \EndIf
    \EndFor
    \caption{Choice of NK Auxiliary Directions}
    \label{alg:nk3d_aux_sel}
  \end{algorithmic}
\end{algorithm}

In a loop over $\widehat{\bm{\ell}}'_{q}$ in $\widehat{\bm{v}}$, Algorithm~\ref{alg:nk3d_aux_sel} now seeks to find the set of auxiliary directions, $\{ \widehat{\bm{L}}'_{p} \}$, which satisfy the remaining criteria (\textit{ii}) and (\textit{iii}).
Since $\widehat{\bm{\ell}}'_{1}$ (the first element in $\widehat{\bm{v}}$) has the smallest grid spacing, this candidate auxiliary direction is automatically assigned to be $\widehat{\bm{L}}'_{1}$; algorithmically speaking, this corresponds to setting $a_{1\alpha} = \widetilde{a}_{1\alpha}$ for $\alpha =1,2,3$ and populating $\bm{M}_{:,1}$ (\ie the first column of $\bm M$).
After successfully identifying $\widehat{\bm{L}}'_{1}$, the loop then continues to the next element of $\widehat{\bm{v}}$ in the search for $\widehat{\bm{L}}'_{2}$.
In a loop over $q$ (which runs from $2,3,\ldots$), $\widehat{\bm{\ell}}'_{q}$ becomes the proposed candidate for $\widehat{\bm{L}}'_{2}$, \ie $a_{2\alpha}$ is \textit{temporarily} assigned to $\widetilde{a}_{q\alpha}$ and $\bm{M}_{:,2}$ is populated accordingly. 
If $\bm{M}_{:,2}$ is non-parallel to $\bm{M}_{:,1}$ (determined \via the Cauchy-Schwarz inequality), then $\widehat{\bm{\ell}}'_{q}$ is assigned to be $\widehat{\bm{L}}'_{2}$; if not, the loop continues to the next element in $\widehat{\bm{v}}$.
After successfully identifying $\widehat{\bm{L}}'_{2}$, the loop then continues to the next element of $\widehat{\bm{v}}$ in the search for $\widehat{\bm{L}}'_{3}$, \ie $a_{3\alpha}$ is \textit{temporarily} assigned to $\widetilde{a}_{q\alpha}$ and $\bm{M}_{:,3}$ is again populated accordingly. 
If $\bm{M}$ is non-singular (\ie $\det \bm{M} \neq 0$), then $\widehat{\bm{\ell}}'_{q}$ is assigned to be $\widehat{\bm{L}}'_{3}$ and Algorithm~\ref{alg:nk3d_aux_sel} terminates; if not, the loop continues to the next element in $\widehat{\bm{v}}$.

Upon successful execution, the output of Algorithm~\ref{alg:nk3d_aux_sel} is $\{ \widehat{\bm{L}}'_{p} \}$, the final set of auxiliary directions (which satisfies all of the criteria given above), and $\bm{M}$, which can be trivially inverted to obtain $\bm{b}$ \via Eq.~\eqref{eq:nk3d_Mb_f}.
With $\{ \widehat{\bm{L}}'_{p} \}$ and $\bm{b}$ in hand, the NK Laplacian in Eq.~\eqref{eq:lapG_mod_nk3d} can now be evaluated, allowing for a computationally efficient treatment of non-orthogonal cells during constant-pressure simulations in \exxm (see Sec.~\ref{performance:latt_sym_ices} for a detailed computational timings profile of CPMD simulations of ice I$h$, II, and III at the hybrid DFT level using this approach).

\subsection{Extension of the \exxm Module: Computation of the EXX Contribution to the Stress Tensor \label{subsec:impl_cell_forces}}

Using Algorithm~\ref{alg:nk3d_aux_sel} in conjunction with the NK representation for the Laplacian (see Sec.~\ref{subsec:impl_PS}), the \exxm module is now equipped to solve the PE for systems with fluctuating and non-orthogonal simulation cells.
For each overlapping $\braket{ij}$ pair, the \exxm module leverages this new capability to compute the corresponding MLWF-product potential ($\tvb{ij}$) during Step IV (see Fig.~\ref{fig:flowchart}).
This quantity is the cornerstone of our MLWF-based EXX approach, and is required for evaluating all of the EXX-related contributions ($\exx$, $\{\dxx{i}\}$, $\sxx$) to the CPMD equations of motion in Eqs.~\eqref{eq:cpE}--\eqref{eq:cpC}.
Since the evaluation of $\exx$ and $\{\dxx{i}\}$ have been discussed extensively in \pI,~\cite{paper1} we focus the following discussion on the extensions to \exxm needed for computing $\sxx$ \via Eq.~\eqref{eq:sxx_working} during Step V (see Fig.~\ref{fig:flowchart}).
In this working expression, one can immediately see that a numerically accurate evaluation of the $\braket{ij}$ contribution to $\sxx$ only requires integration over $\Omega_{ij}$ (in analogy to the evaluation of $\exx$ \via Eq.~\eqref{eq:exxGen_mlwf}).
In fact, once the gradient of $\tvb{ij}$ is evaluated (\textit{vide infra}), the computation of $\sxx$ follows a similar procedure to that used for $\exx$ (see Sec.~III C 5 of \pI~\cite{paper1}): (\textit{i}) for each overlapping $\braket{ij}$ pair, integration over the $\pe{\gC{ij}}$ subdomain on a given \mpi{} process is efficiently parallelized over $N_{\rm thread}$ \omp{} threads; (\textit{ii}) partial summations over the $\braket{ij}$ pairs assigned to each \mpi{} process are then accumulated \via \texttt{MPI\_REDUCE} (using the \texttt{MPI\_SUM} operation) to form $\sxx$ with minimal associated communication (\ie $3\times3$ double-precision numbers per \mpi{} process).

Since the integral needed to evaluate each $\braket{ij}$ contribution to $\sxx$ is restricted to the $\pe{\gC{ij}}$ subdomain, each component of the Cartesian gradient of $\tv{ij}$ in Eq.~\eqref{eq:sxx_working} (\ie $\partial \, \tv{ij}/\partial \, r_{a}$) only needs to be evaluated on $\pe{\gC{ij}}$ as well.
With the Jacobian derived in Eq.~\eqref{eq:jacb}, the Cartesian gradient operator, $\bm{\nabla}_{\bm{r}}$, can be written in terms of the (pure) directional derivatives along the unit lattice vectors, $\bm \nabla_{\bm \xi}$, \via $\bm{\nabla}_{\bm{r}} = \bm{J} \bm{\nabla}_{\bm{\xi}}$; as such, there is no need to introduce auxiliary lattice directions as done above when using the NK representation of $\nabla^2_{\bm{r}}$.
In analogy to Eq.~\eqref{eq:d2_1}, the derivatives in Eq.~\eqref{eq:sxx_working} can be accurately and efficiently evaluated using standard central-difference formulae along each of the lattice vectors (shown here for a generic function, $f(\bm{\xi})$, along $\bm{L}_{\alpha}$): 
\begin{align}
  \left.\frac{\partial f(\bm \xi)}{\partial \xi_{\alpha}} \right|_{\bm \xi = \bm \xi_{0}} = \sum_{q=-n}^{n} w_{q}\frac{  f(\bm \xi_{0} + q \, \delta \xi_{\alpha} \widehat{\bm L}_{\alpha})}{ \delta \xi_{\alpha}} .
  \label{eq:d1_1}
\end{align}
In this expression, the sum is over the $n$ neighboring grid points located on each side of $\bm{\xi}_0$ (along $\bm{L}_{\alpha}$), and the corresponding anti-symmetric ($2n+1$)-point stencil uses the following central-difference coefficients (with $w_{q}=-w_{-q}$):~\cite{fornberg_generation_1988} $w_{0} = 0$, $w_{1} = +3/4$, $w_{2} = -3/20$, and $w_{3} = +1/60$.
The default option in \exxm is $n=3$ with a discretization error of $\mathcal{O}\left( \delta\xi_{\alpha}^{2 n} \right) = \mathcal{O}\left( \delta\xi_{\alpha}^6 \right)$, as this choice furnishes well-converged values for $\exx$ and $\{ \dxx{i} \}$~\cite{wu_order-n_2009,distasio_jr._individual_2014,paper1} as well as $\sxx$.

Here, we stress to the reader that Eq.~\eqref{eq:sxx_working} provides an analytical expression for $\sxx=\partial \exx / \partial \bm{h}$ (\ie the cell derivatives of $\exx$), and the finite-difference evaluation of $\partial \, \tv{ij}/\partial \, r_{a}$ (\via Eq.~\eqref{eq:d1_1}) is needed since $\tv{ij}$ is not analytical and only known on the real-space grid.
As such, the approach for computing $\sxx$ in \exxm is simultaneously more accurate and more computationally efficient than numerical differentiation of $\exx$ with respect to $\bm{h}$ (which would require perturbing each element of $\bm{h}$ by $\pm \delta$ and then re-computing $\exx$ for each of these cell displacements).
Unlike the numerical differentiation of $\exx$ with respect to $\bm{h}$, which requires $2\mathrm{-}12 \times$ the cost of evaluating $\exx$ (depending on the number of non-zero elements in $\bm{h}$), the computational complexity of evaluating Eq.~\eqref{eq:sxx_working} in \exxm is comparable to a \textit{single} application of the Laplacian during the CG solution of the PE.
As such, computation of the EXX contribution to the stress tensor (\via $\sxx$) only requires a small fraction of the cost associated with computing $\exx$; for all of the simulations performed in this work, the cost associated with $\sxx$ was $<1\%$ of the wall time spent in the \exxm module.

\section{Accuracy and Performance \label{sec:Performance}}

In this section, we critically assess the accuracy and computational performance of \exxm, which uses a dual-level \mpi{}/\omp{} parallelization scheme to exploit both internode and intranode HPC resources during hybrid DFT simulations of large-scale condensed-phase systems.
We will focus on the extensions to \exxm introduced in this work (see Sec.~\ref{sec:Implementation}) that enable constant-pressure ($NpH/NpT$) simulations at the hybrid DFT level for general/non-orthogonal cells using the \texttt{CP} module of \texttt{QE}~\cite{giannozzi_advanced_2017}.
We begin by exploring the accuracy of the extended \exxm module when computing $\exx$ and $\sxx$ for a variety of condensed-phase systems, including ambient liquid water, a benzene molecular crystal polymorph, and semi-conducting crystalline silicon, in Sec.~\ref{performance:accuracy}.
We then study the effects of lattice symmetry on computational complexity in the \exxm module \via a detailed case study of three different ice polymorphs (I$h$, II, and III) in Sec.~\ref{performance:latt_sym_ices}.
In particular, we perform and analyze a series of short $NpT$ CPMD simulations on these ice phases (in conjunction with specific angular constraints on each cell tensor) to investigate how the number of non-orthogonal cell directions affects the performance of \exxm. 
In Sec.~\ref{performance:scalability_liq}, we investigate the computational performance and parallel scaling of \exxm during constant-pressure simulations of large-scale condensed-phase systems \via a strong- and weak-scaling analysis of liquid water (\ie ranging from \ce{(H2O)64} to \ce{(H2O)256}) in the $NpT$ ensemble (in analogy to that performed in \pI~\cite{paper1} in the $NVT$ ensemble).
In all cases, the performance of \exxm will be examined across a wide array of HPC architectures, including \textit{Mira} IBM Blue Gene/Q, \textit{Cori} Haswell, and \textit{Cori} KNL.

\subsection{Accuracy of the EXX Contributions to the Energy and Cell Forces \label{performance:accuracy}}

In \pI~\cite{paper1}, we used a snapshot of ambient liquid water (\ie \ce{(H2O)64} at the equilibrium density, $85$~Ry planewave cutoff) to determine the default \exxm parameters used in \texttt{QE}.
Here, we remind the reader that there are five key parameters used when performing a hybrid DFT calculation with \exxm (see Sec.~\ref{Impl:exx_module}):
(\textit{i}) $\rpr$, a radial cutoff used to determine whether or not two MLWFs, $\tphin{i}$ and $\tphin{j}$, are an overlapping $\braket{ij}$ pair based on their center-to-center distance (\ie $| \tC{i}-\tC{j} | \le \rpr$);
(\textit{ii})--(\textit{iii}) $\rpes$ and $\rpep$, the radii of the fixed-size spherical domains over which Poisson's equation (Eq.~\eqref{eq:pe}) is solved for the near-field potential ($\tv{}$) for self (s, $\braket{ii}$) and non-self (ns, $\braket{ij}$) overlapping pairs;
and
(\textit{iv})--(\textit{v}) $\rmes$ and $\rmep$, the outer radii of the concentric spherical shells (with inner radii $\rpes$ and $\rpep$) over which the multipole expansion (Eqs.~\eqref{eq:me}--\eqref{eq:mepole}) is performed for the far-field potential ($\tv{}$) for the self and non-self overlapping pairs.
In doing so, we demonstrated that these parameters govern both the accuracy and performance of \exxm, and judicious choices for each ensured rapid convergence of $\exx$ and $\{\dxx{i}\}$ for \ce{(H2O)64} (see Figs.~6 and 7 as well as Secs.~IV A 1 and IV A 2 in \pI~\cite{paper1}).

\begin{figure}[t!]
  \includegraphics[width=\linewidth]{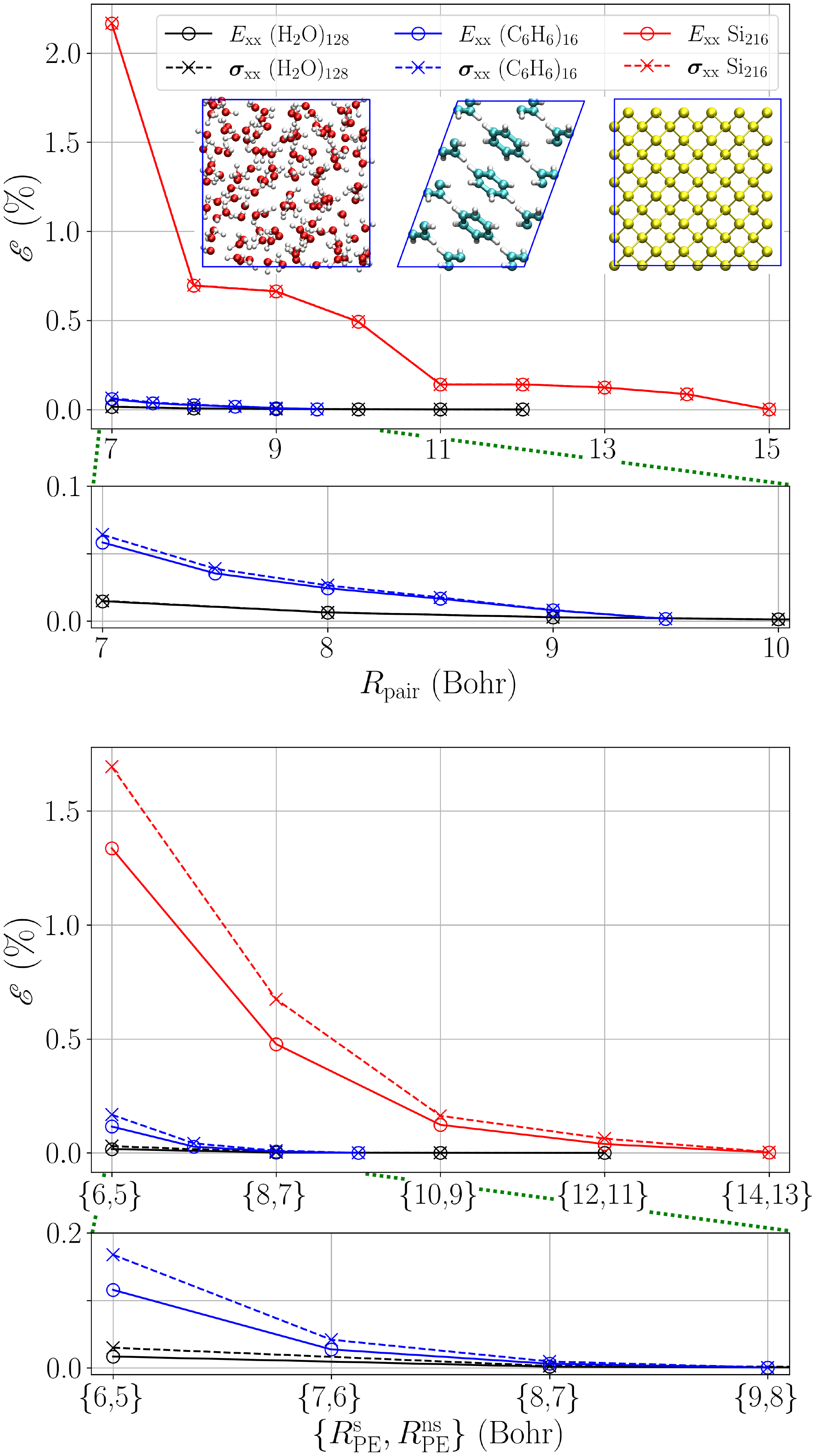}
  \caption{
  Convergence of $\exx$ (open circles, $\circ$) and $\sxx$ (crosses, $\times$) in \ce{(H2O)128} (black), \ce{(C6H6)16} (blue), and \ce{Si216} (red) as a function of $\rpr$ (\textit{top panel}) and $\{ \rpes, \rpep \}$ (\textit{bottom panel}). As described in the main text, all other \exxm parameters were set to the maximum allowed values during each convergence study.
  }
  \label{fig:transferability}
\end{figure}

To test the new capabilities of the extended \exxm module (\ie $\exx$ and $\sxx$ for general/non-orthogonal cells) as well as the transferability of the default \exxm parameters, we now investigate the convergence of $\exx$ and $\sxx$ on three different condensed-phase systems: ambient liquid water, a benzene molecular crystal polymorph, and semi-conducting crystalline silicon.
For consistency with \pI~\cite{paper1}, we included a snapshot of ambient liquid water at the equilibrium density; however, we have doubled the system size to \ce{(H2O)128} ($L = 15.645$~\AA{}) to investigate any finite-size effects on the \exxm parameters from using \ce{(H2O)64} in \pI~\cite{paper1} and increased the planewave cutoff from $85$~Ry to $150$~Ry (a typical setting employed during constant-pressure $NpH$/$NpT$ simulations of aqueous systems). 
To go beyond liquid water, we also carried out a case study on the monoclinic benzene-II molecular crystal polymorph~\cite{fourme_redetermination_1971}---a non-orthogonal and anisotropic system with a similar band gap.
To do so, we considered a \ce{(C6H6)16} system (with $a = 10.834$~\AA{}, $b = 10.752$~\AA{}, $c = 15.064$~\AA{}, $\alpha=\gamma=90^\circ$, and $\beta =110^\circ$) constructed from a $2 \times 2 \times 2$ supercell of the experimentally assigned unit cell (which contains two benzene molecules).~\cite{fourme_redetermination_1971} 
During all calculations on \ce{(C6H6)16}, we used a $110$~Ry planewave cutoff.
As an even more stringent test on the \exxm module, we also considered semi-conducting crystalline silicon---a system with a significantly smaller band gap and therefore substantially more diffuse (less localized) MLWFs.
In this case, we constructed a cubic \ce{Si216} snapshot (with $L = 16.29$~\AA{}) as a $3 \times 3 \times 3$ supercell of the classic eight-atom diamond structure in a cubic unit cell.
During all calculations on \ce{Si216}, we used a $35$~Ry planewave cutoff.
A graphical depiction of each of these three systems can be found in the inset to Fig.~\ref{fig:transferability}.

Using the procedure outlined in Sec.~IV A  of \pI~\cite{paper1} to determine the default \exxm parameter values, we first performed a series of reference single-point energy calculations on each of these systems at the PBE0~\cite{perdew_rationale_1996,adamo_toward_1999} level.
This was accomplished by self-consistently solving for the electronic ground state with \textit{all} EXX parameters in the \exxm module set to their largest possible values: $\rpr$, $\rmes$, and $\rmep$ are set to the radius of the largest sphere that can be contained within each simulation cell; $\rpes = \rpep = \rmes - n \max_{\alpha} \{ \delta\xi_{\alpha} \}$ (with $n = 3$), which provides us with a thin shell (halo region) on the real-space grid needed for the PE boundary conditions.
The $\exx$ and $\sxx$ values obtained from these calculations are then used as reference values (\ie $\exx^{\rm ref}$ and $\sxx^{\rm ref}$) to gauge the accuracy of the \exxm module when computing these quantities using different parameter values.
In analogy to the previously used error metric for $\dxx{i}$ (see Eq.~(36), Fig.~7, and Sec.~IV A 2 in \pI~\cite{paper1}), we define the relative errors in $\exx$ and $\sxx$ as follows:
\begin{align}
    \mathcal{E}(\exx) &= \frac{ \left| \left| \exx - \exx^{\rm ref} \right| \right|_1}{ \left| \left| \exx^{\rm ref} \right| \right|_1} \label{eq:error_exx}\\
    \mathcal{E}(\sxx) &= \frac{\left| \left| \sxx - \sxx^{\rm ref} \right| \right|_1 }{\left| \left| \sxx^{\rm ref} \right| \right|_1}, \label{eq:error_sxx}
\end{align}
in which $|| \boldsymbol{\cdot} ||_1$ denotes the $1$-norm of the inserted quantity.
Based on these relative error definitions, we first investigated the accuracy of the default EXX parameters (which were determined using \ce{(H2O)64} at the equilibrium density, $85$~Ry planewave cutoff) when computing $\exx$ and $\sxx$ for the three systems described above (see Table~\ref{tab:transferability}).
When assessing the accuracy of the default \exxm parameters for these systems, we again follow \pI~\cite{paper1} by computing $\exx$ and $\sxx$ using the converged MLWFs obtained during the corresponding reference calculations; a more detailed (and fully self-consistent) investigation of these parameters in anisotropic/heterogeneous systems will be addressed in a forthcoming paper in this series.
%
%
\begin{table}[t!]
  \caption{
  Relative $1$-norm errors in $\exx$ and $\sxx$ using the default \exxm parameters in \texttt{QE} (\ie $\rpr = 8.0$~Bohr, $\rpes = 6.0$~Bohr, $\rpep = 5.0$~Bohr, $\rmes = 10.0$~Bohr, and $\rmep = 7.0$~Bohr) for three different condensed-phase systems.
  All errors were computed using Eqs.~\eqref{eq:error_exx} and \eqref{eq:error_sxx} with respect to the reference values for these quantities obtained with all \exxm parameters set to their largest possible values.
  See text for more details.
  }
  \begin{tabular}{c|cccc}
  \hline
  \hline
                         & \ce{(H2O)64} & \ce{(H2O)128} & \ce{(C6H6)16} & \ce{Si216} \\
                         & (cubic)      & (cubic)       & (monoclinic)  & (cubic)    \\
  \hline
$\mathcal{E}(\exx)$ (in \%) & $0.02^{a}$ &$0.02$       &   $0.13$      &  $1.72$    \\
$\mathcal{E}(\sxx)$ (in \%) & --  &$0.03$       &   $0.19$      &  $2.29$    \\
  \hline
  \hline
  \multicolumn{5}{l}{\scriptsize{$^a$System used to determine default \exxm parameters in \pI.~\cite{paper1} }} \\ 
  \end{tabular}
  \label{tab:transferability}
\end{table}
%
%

As depicted in Table~\ref{tab:transferability}, the default \exxm parameters reproduce $\exx$ with very high fidelity for \ce{(H2O)128}---in this case, the accuracy of \exxm is equivalent to that found in \pI~\cite{paper1} for \ce{(H2O)64}, \ie $\mathcal{E}(\exx) \approx 0.02$\%.
As expected, the default parameters determined in \pI~\cite{paper1} seem to be well-converged for ambient liquid water with respect to both system size and basis set size.
When applied to the \ce{(C6H6)16} molecular crystal, we find that the default \exxm parameters are also quite transferable, yielding $\mathcal{E}(\exx) \approx 0.10$\%.
In this case, we attribute the slight decrease in accuracy to the increased variability in the MLWF spreads in the benzene molecular crystal, which contains \ce{C-H} $\sigma$-bonds (which have a similar spread to the MLWFs in liquid water) as well as a set of more diffuse \ce{C-C} $\tau$-bonds; as such, converging $\exx$ in this system will require (on average) a slightly larger support for $\trho{ij}$ during the solution of Poisson's equation (\textit{vide infra}).
In \ce{Si216}, the MLWFs are significantly more delocalized than those in both liquid water and the benzene molecular crystal due to the smaller band gap in this semi-conductor~\cite{kohn_analytic_1959}.
As such, the default \exxm parameters now yield a more sizable error of $\mathcal{E}(\exx) \approx 1.72$\%, as tight convergence of $\exx$ in this more challenging system will require an increase in $\rpes$ and $\rpep$ (to provide a larger support for the more diffuse $\trho{ij}$) as well as $\rpr$ (to account for the increased number of overlapping MLWF pairs).
%
%
\noindent Here, we also note that $\mathcal{E}(\sxx)$ is slightly (but consistently) larger than $\mathcal{E}(\exx)$ in all three of these cases; this systematic trend will be discussed below.

Since the accuracy required during an $\exx$ (or $\sxx$) calculation will depend on the system and/or application, we now perform a systematic study of how these quantities converge in \ce{(H2O)128}, \ce{(C6H6)16}, and \ce{Si216} as a function of the \exxm parameters.
Following the procedure outlined in \pI~\cite{paper1}, we again start with the converged MLWFs obtained during the reference calculations described above (in which all \exxm parameters were set to their largest possible values). 
We then track how $\exx$ and $\sxx$ converge with respect to: (\textit{i}) changes in $\rpr$ while keeping all other \exxm variables at their reference values (Fig.~\ref{fig:transferability}, top panel), and (\textit{ii}) simultaneous changes in $\rpes$ and $\rpep$ while again keeping all other \exxm variables at their reference values (Fig.~\ref{fig:transferability}, bottom panel).
As mentioned above, a more detailed (and fully self-consistent) investigation of these parameters (for a number of different anisotropic/heterogeneous systems) will be addressed in a forthcoming paper in this series.
As depicted in Fig.~\ref{fig:transferability}, the convergence behavior of $\mathcal{E}(\exx)$ in \ce{(H2O)128} is essentially identical to that in \ce{(H2O)64} (\cf Fig.~6 in \pI~\cite{paper1}).
For \ce{(C6H6)16}, we find that $\exx$ rapidly converges with both $\rpr$ and $\{ \rpes, \rpep \}$; in this case, increasing the radii used during the solution of Poisson's equation (to account for the more diffuse $\tau$ bonds on each benzene ring) is more important than increasing $\rpr$ when tight convergence (\ie $\mathcal{E}(\exx) < 0.10\%$) is desired.
For \ce{Si216}, the convergence of $\exx$ with respect to both $\rpr$ and $\{ \rpes, \rpep \}$ is slower due to the substantially more delocalized MLWFs in this semi-conducting system.
In this case, $\mathcal{E}(\exx)$ originates from the need for increased real-space domains during the solution of Poisson's equation (primary contribution) as well as the inclusion of more distant overlapping MLWF pairs (secondary but still sizable contribution).
Here, we find that systematically (and simultaneously) increasing both $\rpes$ and $\rpep$ leads to a smooth and exponential decay in $\mathcal{E}(\exx)$, reflecting the exponential decay rate of the MLWFs in this finite-gap system~\cite{kohn_analytic_1959}.
As $\rpr$ was increased, the observed decreases in $\mathcal{E}(\exx)$ are tiered (as opposed to the smoother decay seen in \ce{(H2O)128}), reflecting the crystalline structure in this atomic solid.
Even in this more challenging system, the MLWFs are still exponentially localized and therefore have a finite support in real space; as such, the \exxm module can still furnish $\exx$ to a pre-defined accuracy level---albeit with additional computational cost---by simply increasing the \exxm parameters beyond their default values. 
For example, ``chemical accuracy'' (\ie $1$~kcal/mol) in the PBE0 binding energy of \ce{Si216} can be achieved by increasing $\{\rpes, \rpep\}$ from $\{6.0, 5.0\}$~Bohr (default values) to $\{8.0, 7.0\}$~Bohr while leaving $\rpr$ at $8.0$~Bohr (default value).
When compared against the default setting in \exxm, the use of these more accurate parameters leads to an $\approx 50\%$ increase in the computational cost; however, this is still a significant speed-up and $\approx 60 \times$ less than the cost of the reference calculation.

Since the calculation of $\exx$ and $\sxx$ uses the same overlapping MLWF pairs (\cf Eqs.~\eqref{eq:exxGen_mlwf} and \eqref{eq:sxx_working}), the accuracy of these quantities will primarily be governed by $\{\rpes,\rpep\}$, \ie the coverage of $\trho{ij}$ during the solution of Poisson's equation.
As expected, we found that $\mathcal{E}(\exx)$ and $\mathcal{E}(\sxx)$ exhibited a similar convergence rate with respect to these \exxm parameters for all three systems considered herein, although $\mathcal{E}(\sxx)$ was consistently slightly larger than $\mathcal{E}(\exx)$ in all cases (see Table~\ref{tab:transferability} and Fig.~\ref{fig:transferability}, top panel).
While a small portion of this difference is due to inherent limitations when comparing relative errors in scalar and matrix quantities (\cf Eqs.~\eqref{eq:error_exx} and \eqref{eq:error_sxx}), the difference between $\mathcal{E}(\sxx)$ and $\mathcal{E}(\exx)$ is more pronounced for smaller $\{\rpes, \rpep\}$ values.
Hence, we attribute this difference to the larger intrinsic error when computing the integrand of $\sxx$ in Eq.~\eqref{eq:sxx_working}, which involves a displacement-weighted derivative of the MLWF-product potential, \ie $r_{b} \left( \frac{\partial \, \tv{ij}}{\partial \, r_a} \right)$, as opposed to the integrand of $\exx$ in Eq.~\eqref{eq:exxGen_mlwf}, which only involves $\tv{ij}$ itself.

Here, we note in passing that the need to scan for the set of optimal parameters in \exxm will be largely eliminated in a forthcoming paper in this series, where the entire \exxm module will be restructured based on variable-size supports for each MLWF. By intrinsically accounting for the size/shape/extent of each MLWF and treating each overlapping MLWF pair according to a user-defined level of accuracy, this restructured \exxm module will be able to treat challenging systems like crystalline \ce{Si} as well as complex multi-component/multi-phase systems without the need to sacrifice accuracy for computational performance (or vice versa).

\subsection{Computational Complexity due to Lattice Symmetry \label{performance:latt_sym_ices}}

For a more consistent comparison with the analysis of \exxm in \pI~\cite{paper1}, we now refocus our discussion on aqueous systems (\ie ice and liquid water) while assessing the computational performance of the extended \exxm module when treating general/non-orthogonal systems and using larger basis sets (\eg as needed during constant pressure simulations with fluctuating cells).
To explore the effects of lattice symmetry on computational complexity, we first carried out a detailed case study on the I$h$, II, and III polymorphs of ice.
More specifically, we performed and analyzed short (\ie $50$~steps) $NpT$ CPMD simulations on these ice phases (in conjunction with specific angular constraints on each lattice) to investigate how the number of non-orthogonal cell directions affects the performance of \exxm. 
As a first case, we considered the orthorhombic/tetragonal/cubic lattice systems, in which evaluation of the NK Laplacian in Eq.~\eqref{eq:lapG_mod_nk3d} is the simplest and requires $N_{\rm aux} = 0$ auxiliary grid directions (\ie $N_{\rm pure} = 3$ pure derivatives along the lattice directions).
In this case, we chose Ice III as the example system (which is tetragonal in the absence of thermal fluctuations) and applied a series of angular constraints ($\bm{L}_{1}\perp\bm{L}_{2}$, $\bm{L}_{1}\perp\bm{L}_{3}$, and $\bm{L}_{2}\perp\bm{L}_{3}$) to maintain orthogonality among all lattice vectors during the short $NpT$ simulation. 
As a second case, we considered the monoclinic/hexagonal/rhombohedral lattice systems, in which evaluation of the NK Laplacian requires $N_{\rm aux} = 1$ auxiliary grid direction (for a total of $N_{\rm pure} = 4$ pure derivatives). 
In this case, Ice I$h$ was chosen as the example system (which is hexagonal in the absence of thermal fluctuations), and $\bm{L}_{1}\perp\bm{L}_{2}$ and $\bm{L}_{2}\perp\bm{L}_{3}$ angular constraints were applied during the $NpT$ simulation to maintain $N_{\rm pure} = 4$.
As a third case, we considered the triclinic lattice system, in which evaluation of the NK Laplacian is the most complex and requires $N_{\rm aux} = 3$ auxiliary grid directions (for a total of $N_{\rm pure} = 6$ pure derivatives).
Here, we employed ice II as the example system; although this polymorph is rhombohedral in the absence of thermal fluctuations, we started the $NpT$ simulation with ice II in a triclinic cell.
We then allowed the $NpT$ simulation to proceed without any angular constraints to mimic the cell fluctuations of a triclinic system with $N_{\rm pure} = 6$ pure derivatives (rather than $N_{\rm pure} = 4$ for a perfect rhombohedral lattice).
By including these three cases (with $N_{\rm pure} = 3,4,6$), this study essentially covers all seven 3D lattice systems,~\cite{NKnote,NKnote2} and will now be used to evaluate the performance of \exxm.

Computational timings for each of these ice phases were generated using an in-house development version of \texttt{QE} (that is based on \texttt{v5.0.2})~\cite{note_our_code_repo} at the PBE0~\cite{perdew_rationale_1996,adamo_toward_1999} hybrid DFT level.
Each ice polymorph was modeled using a simulation cell containing \ce{(H2O)96} (each with $N_{o} = 4 \times N_{\rm water} = 384$~MLWFs) with initial snapshots taken from $NpT$ simulations of ice I$h$, II, and III at the corresponding experimental triple point (\ie $p = 2.1$~kBar and $T = 238$~K).
In ice I$h$ and III, proton disorder was introduced using an algorithm that enforces the Bernal--Fowler ice rules~\cite{bernal_theory_1933} as well as the additional constraint of vanishing polarization.~\cite{hayward_unit_1997,cota_computer_1977}
For the proton-ordered ice II phase, the supercell was made by directly replicating the unit cell containing \ce{(H2O)12} provided in Ref.~\onlinecite{santra_accuracy_2013}.
With the angular constraints described directly above, we performed a series of short CPMD simulations in the $NpT$ ensemble (at the same $p$ and $T$) for a duration of $50$~steps.
The pressure was controlled using a Parrinello--Rahman barostat~\cite{parrinello_crystal_1980} and the temperature was maintained by attaching massive Nos\'{e}--Hoover chain thermostats~\cite{martyna_nosehoover_1992,tobias_molecular_1993} (each with a chain length of $4$) to the ionic degrees of freedom.
All $NpT$ simulations were performed at the $\Gamma$-point only and employed a planewave kinetic energy cutoff of $150$~Ry; the corresponding CPMD equations of motion (Eqs.~\eqref{eq:cpE}--\eqref{eq:cpC}) were integrated using the standard Verlet algorithm and a time step of $2.0$~au ($\approx 0.05$~fs).
Planewave kinetic energies were modified following Bernasconi \textit{et al.}~\cite{bernasconi_first-principle-constant_1995} to maintain a constant planewave kinetic energy cutoff of $130$~Ry throughout each $NpT$ simulation.~\cite{footnote_Bernasconi}
To ensure an adiabatic separation between the electronic and nuclear degrees of freedom, the fictitious electronic mass was set to $\mu = 100$~au; in addition, the nuclear mass of deuterium was used for each hydrogen atom.
To improve the stability of the fictitious electron dynamics, mass preconditioning~\cite{tassone_acceleration_1994} was applied to all Fourier components of the electronic (pseudo-)wavefunctions with a kinetic energy $>25$~Ry.
The Hamann-Schl{\" u}ter-Chiang-Vanderbilt (HSCV) type norm-conserving pseudopotentials~\cite{hamann_norm-conserving_1979,vanderbilt_optimally_1985} provided by the \texttt{Qbox} package~\cite{gygi_architecture_2008} were used to treat the interactions between the valence electrons and the ions. 
All \exxm related parameters were set to the default values determined in \pI~\cite{paper1}, \ie $\rpr=8.0$~Bohr, $\rpes=6.0$~Bohr, $\rpep=5.0$~Bohr, $\rmes = 10.0$~Bohr, and $\rmep = 7.0$~Bohr.

All timings were obtained using $1536$ nodes (\ie $\zeta \equiv N_{\rm proc} / N_{o} = 4$) on the following HPC architectures: \textit{Mira} IBM Blue Gene/Q, \textit{Cori} Haswell, and \textit{Cori} KNL (see Table~\ref{tab:ice123}).
In all cases, the reported timings were obtained using one process per node for the internode \mpi{} parallelization (first parallelization level) and all available physical cores per node (\ie $16$ for \textit{Mira} IBM Blue Gene/Q, $32$ for \textit{Cori} Haswell, and $68$ for \textit{Cori} KNL) for the intranode \omp{} parallelization (second parallelization level).
Task-group parallelization (with $N_{\rm tg} = 4$) was also employed to improve the computational efficiency associated with the 3D \texttt{FFT} operations in the non-\exxm portions of \texttt{QE}.
Hyperthreading was fully activated on each physical core except for \textit{Cori} KNL, where hyperthreading was deactivated due to performance degradation in both the \exxm and non-\exxm modules in \texttt{QE}.
\begin{table}[ht!]
    \caption{
    Computational timings profile for $NpT$ CPMD simulations of ice I$h$, II, and III (each modeled by \ce{(H2O)96}) at the hybrid PBE0 level on \textit{Mira} IBM Blue Gene/Q, \textit{Cori} Haswell, and \textit{Cori} KNL using the extended \exxm module in \texttt{QE}.
    These $NpT$ simulations cover all seven 3D lattice systems,~\cite{NKnote,NKnote2} which have been grouped into three different categories according to the number of auxiliary grid directions ($N_{\rm aux}$) used in the NK Laplacian in Eq.~\eqref{eq:lapG_mod_nk3d}; the listed angular constraints were applied throughout each simulation to maintain the targeted $N_{\rm aux}$ value. 
    All timings (in s/step) were averaged over $50$ CPMD steps and correspond to the mean wall times associated with computing the EXX contribution to the stress tensor ($\braket{t_{\exxm}^{\rm stress}}$), solving the PE for all overlapping MLWF pairs ($\braket{t_{\exxm}^{\rm PE}}$), and running through the entire \exxm module ($\braket{t_{\exxm}}$); also shown are the $\braket{t_{\exxm}^{\rm PE}}/\braket{t_{\exxm}}$ ratios.
    Other relevant properties include: the total number of pure derivatives ($N_{\rm pure} = N_{\rm aux} + 3$), the number of stencil points in the finite-difference representation of the NK Laplacian ($N_{\rm stcl} = 2 n N_{\rm pure} + 1$, shown here for $n = 3$), the number of grid points in each $\pep{\gC{ij}}$ (Poisson) subdomain ($\npep$, shown here for non-self $\braket{ij}$ pairs only), the average number of CG iterations ($\braket{N_{\rm CG}}$) required to solve each PE, the condition number ($\mathcal{K}$) of the sparse PE operator (\ie $-\nabla^2$), and the average number of overlapping $\braket{ij}$ pairs assigned to each \mpi{} process ($\braket{N_{\rm pair}}$).
    All timings were obtained with $\zeta = 4$, $N_{\rm tg} = 4$, $1536$ nodes (using one \mpi{} process and all available physical cores per node).
    }
    \centering
    \begin{tabular}{ c || c  c  c }
    \hline\hline
    \multirow{2}{*}{\textbf{Lattice}}   & \textbf{Orthorhombic}     & \textbf{Monoclinic}    & \multirow{3}{*}{\textbf{Triclinic}} \\
    \multirow{2}{*}{\textbf{System(s)}} & \textbf{Tetragonal}       & \textbf{Hexagonal}     & \\     
                                        & \textbf{Cubic}            & \textbf{Rhombohedral}  & \\     
    \hline
    Example                             & Ice III  & Ice I$h$ & Ice II \\
    \begin{tabular}{c}Angular\\Constraints \end{tabular}
  & $\Bigg[$\begin{tabular}{c}$\bm{L}_{1}\perp\bm{L}_{2}$\\$\bm{L}_{1}\perp\bm{L}_{3}$\\$\bm{L}_{2}\perp\bm{L}_{3}$ \end{tabular}$\Bigg]$
  & $\Bigg[$\begin{tabular}{c}$\bm{L}_{1}\perp\bm{L}_{2}$\\$\bm{L}_{2}\perp\bm{L}_{3}$ \end{tabular}$\Bigg]$
  & $\Bigg[$\textit{None}$\Bigg]$\\
    $N_{\rm aux}$           & $0$               & $1$               & $3$            \\
    $N_{\rm pure}$             & $3$               & $4$               & $6$            \\
    $N_{\rm stcl}$          & $19$              & $25$              & $37$           \\
    ${\npep}^a$             & ${\sim}280{,}000$ & ${\sim}292{,}000$ & ${\sim}415{,}000$ \\ 
    $\braket{N_{\rm CG}}^a$ & $121$             & $89$              & $126$         \\
    $\mathcal{K}^a$         & ${\sim}3{,}100$   & ${\sim}2{,}400$   & ${\sim}3{,}200$       \\
    $\braket{N_{\rm pair}}$ & $7.0$             & $4.3$             & $6.5$         \\
    \hline
  &\multicolumn{3}{c}{\textit{Mira} IBM Blue Gene/Q (s/step)} \\
    \hline
    $\braket{t_{\exxm}^{\rm stress}}$                  & $0.02$ & $0.01$ & $\phantom{0}0.02$ \\
    $\braket{t_{\exxm}^{\rm PE}}$                      & $2.00$ & $1.30$ & $\phantom{0}3.40$ \\
    $\braket{t_{\exxm}}$                               & $5.92$ & $3.83$ & $10.25$           \\
    $\braket{t_{\exxm}^{\rm PE}} / \braket{t_{\exxm}}$ & $0.34$ & $0.34$ & $\phantom{0}0.33$ \\
    \hline
 &\multicolumn{3}{c}{\textit{Cori} Haswell (s/step)} \\
    \hline
    $\braket{t_{\exxm}^{\rm stress}}$                  & $0.01$ & $0.01$ & $\phantom{0}0.02$ \\
    $\braket{t_{\exxm}^{\rm PE}}$                      & $1.09$ & $0.58$ & $\phantom{0}1.48$ \\
    $\braket{t_{\exxm}}$                               & $3.24$ & $2.17$ & $\phantom{0}4.96$ \\
    $\braket{t_{\exxm}^{\rm PE}} / \braket{t_{\exxm}}$ & $0.34$ & $0.27$ & $\phantom{0}0.30$ \\
    \hline
 &\multicolumn{3}{c}{\textit{Cori} KNL$^{b}$ (s/step)} \\
    \hline
    $\braket{t_{\exxm}^{\rm stress}}$                  & $0.02$ & $0.01$ & $\phantom{0}0.03$ \\
    $\braket{t_{\exxm}^{\rm PE}}$                      & $1.59$ & $1.01$ & $\phantom{0}2.96$ \\
    $\braket{t_{\exxm}}$                               & $4.85$ & $3.89$ & $\phantom{0}7.88$ \\
    $\braket{t_{\exxm}^{\rm PE}} / \braket{t_{\exxm}}$ & $0.33$ & $0.26$ & $\phantom{0}0.38$ \\
    \hline\hline
    \multicolumn{4}{l}{\scriptsize{$^a$The architecture-dependence of the \texttt{FFT} algorithm leads to slight}} \\ 
    \multicolumn{4}{l}{\scriptsize{variations in $\npep$, $\braket{N_{\rm CG}}$, and $\mathcal{K}$; \textit{Cori} values are provided.}} \\
    \multicolumn{4}{l}{\scriptsize{$^b$Using \texttt{OMP\_PROC\_BIND = true} and \texttt{OMP\_PLACES = cores}.}}
    \end{tabular}
    \label{tab:ice123}
\end{table}

For each ice phase (and on each HPC architecture), we found that the wall time associated with computing the EXX contribution to the stress tensor ($\braket{t_{\exxm}^{\rm stress}}$) was $<0.5\%$ of the average wall time spent in the \exxm module ($\braket{t_{\exxm}}$).
This is not surprising as the evaluation of Eq.~\eqref{eq:sxx_working} is comparable to a single CG step during the solution of the PE.
As such, we will focus our discussion below on the more significant computational cost associated with solving the PE ($\braket{t_{\exxm}^{\rm PE}}$).
Since the real-space grids employed during these $NpT$ simulations were based on a planewave cutoff of $150$~Ry (which is needed for fluctuating-cell simulations), both $\braket{t_{\exxm}^{\rm PE}}$ and $\braket{t_{\exxm}}$ will be larger than that found during fixed-cell $NVT$ simulations in \exxm with a more conventional cutoff of $\approx 85$~Ry.
In all cases, $\braket{t_{\exxm}^{\rm PE}}$ comprises $\approx 30\%$ of $\braket{t_{\exxm}}$, and this finding is quite consistent with the detailed performance analysis of \exxm in \pI~\cite{paper1}, in which $\braket{t_{\exxm}}$ was (approximately) split evenly between computation, communication, and processor idling during large-scale $NVT$ simulations of liquid water (\ce{(H2O)64}--\ce{(H2O)256}) with $\zeta=4$.
On each HPC architecture, we find that $\braket{t_{\exxm}^{\rm PE}}$ and $\braket{t_{\exxm}}$ follow the same trend, in which ice I$h$ has the least computational cost, followed by ice III, and then ice II.

As discussed in Sec.~\ref{subsec:impl_PS}, the first factor that will affect the performance of \exxm during $NpT$ simulations is the number of grid points in the finite-difference (stencil) representation of $\nabla^2$ ($N_{\rm stcl} = 2 n N_{\rm pure}+1$), which directly depends on the total number of pure derivatives ($N_{\rm pure} = N_{\rm aux} + 3$) in the NK Laplacian (see Eq.~\eqref{eq:lapG_mod_nk3d}).
For typical condensed-phase systems such as liquid water, $n=3$ (with a discretization error of $\mathcal{O}\left( \delta\xi^6 \right)$) is sufficiently converged when computing all EXX-related quantities~\cite{wu_order-n_2009,distasio_jr._individual_2014,paper1}; with this choice for $n$, $N_{\rm stcl}=19,25,37$ for the (angularly constrained) $NpT$ simulations of ice III, I$h$, and II reported in Table~\ref{tab:ice123}.
The second factor that will affect performance is the number of grid points in the Poisson subdomain for each overlapping MLWF pair.
Since there are significantly more non-self than self pairs, the computational cost associated with solving the PE is dominated by the non-self pairs~\cite{paper1}; as such, we only report the number of grid points in each $\pep{\gC{ij}}$ subdomain.
While the number of points ($\npep$) in the PE subdomain is similar for ice III (${\sim}280{,}000$) and ice I$h$ (${\sim}292{,}000$), the noticeably larger $\npep$ in ice II (${\sim}415{,}000$) originates from the underlying real-space grid assignment by the \texttt{FFT} algorithm in \texttt{QE}.
Although the grid spacings along the lattice vectors are comparable among these three ice phases (due to the identical planewave cutoff), the lattice vectors in ice II (unlike III and I$h$) do \textit{not} correspond to the grid directions with minimal spacings; as such, the presence of the non-axial grid direction with minimal spacing (\ie the grid-resolved trigonal axis, which is one of the auxiliary grid directions in the NK Laplacian identified using Algorithm~\ref{alg:nk3d_aux_sel}) leads to a denser grid and hence the larger apparent $\npep$ in ice II.
Since $N_{\rm stcl} \times \npep$ is the total number of floating-point operations required for computing the action of the Laplacian over the $\pep{\gC{ij}}$ subdomain (\ie the left-hand side of Eq.~\eqref{eq:pe}, this quantity can be taken as a proxy for the computational cost \textit{per} CG iteration when solving the PE. 
However, this quantity is not necessarily a robust sole predictor of the computational timings in \exxm; in fact, this measure would predict that $NpT$ simulations of ice III would be similar (or slightly more efficient) than ice I$h$ and substantially more efficient than ice II, which is in contrast to the timings reported in Table~\ref{tab:ice123}.

To account for this discrepancy, two additional factors need to be taken into consideration, \ie the average number of CG iterations required to solve each PE ($\braket{N_{\rm CG}}$) and the average number of overlapping MLWF pairs assigned to each \mpi{} process ($\braket{N_{\rm pair}}$).
Since $\braket{N_{\rm CG}}$ is largely governed by the condition number ($\mathcal{K}$), which is the ratio between the largest and smallest eigenvalues of the sparse NK Laplacian ($-\nabla^2$), we also provide $\mathcal{K}$ values in Table~\ref{tab:ice123} corresponding to the first snapshot in each $NpT$ simulation.
Here, we find that the NK Laplacian is more well-conditioned for ice I$h$ ($\mathcal{K}{\sim}2{,}400$) than ice III (${\sim}3{,}100$) and ice II (${\sim}3{,}200$); as a result, the CG solution of the PE in ice I$h$ needed the least number of iterations ($\braket{N_{\rm CG}} = 89$), while ice III and ice II had larger but similar $\braket{N_{\rm CG}}$ values of $121$ and $126$, respectively.
Quite interestingly, the NK Laplacian in the non-orthogonal ice I$h$ and ice II cases seem to be relatively well-conditioned when compared to the orthogonal ice III case, despite the fact that $N_{\rm stcl}$ and $\npep$ are significantly larger for both ice I$h$ and ice II.
This finding highlights the strength of the NK approach (as well as our automated fluctuating-cell extension in Algorithm~\ref{alg:nk3d_aux_sel}) when treating systems with non-orthogonal simulation cells, as the selection of auxiliary directions is a non-trivial procedure that can lead to severe numerical instabilities if done incorrectly.
Taking ice II as an example, choosing the grid-resolved obtuse-angle bisector for each pair of lattice vectors as the three auxiliary directions (\ie a na\"ive 3D generalization of the non-orthogonal 2D NK procedure outlined in Eqs.~\eqref{eq:nk2d_aux}--\eqref{eq:kappa_def} and depicted in Fig.~\ref{fig:grid}) leads to a Laplacian that is no longer negative semi-definite; as a result, the CG solution to the PE requires an excessively large number of iterations if and when it converges.

Since $\braket{N_{\rm pair}}$ is roughly proportional to the total number of overlapping MLWF pairs in the system (which is determined by the $|\tC{i}-\tC{j}| < \rpr$ criterion), $\braket{N_{\rm pair}}$ for the lower-density ice I$h$ phase ($\braket{N_{\rm pair}} = 4.3$) is significantly less than that found in the higher-density ice III ($7.0$) and ice II ($6.5$) phases.
With this information in hand, it is now clear why \exxm-based $NpT$ simulations of ice I$h$ have the lowest $\braket{t_{\exxm}^{\rm PE}}$ among the ice phases.
Although ice I$h$ has intermediate values for $N_{\rm stcl}$ and $\npep$ (and hence an intermediary computational cost per CG iteration), this ice phase has the lowest $\braket{N_{\rm pair}}$ (due to its relatively lower density) and the lowest $\braket{N_{\rm CG}}$ (due to its relatively lower $\mathcal{K}$ value); as such, each CPMD step will requires CG solutions to the least number of PEs and the solution to each PE requires the least number of CG iterations.
To explain why $\braket{t_{\exxm}^{\rm PE}}$ for ice II is larger than ice III (in which both $\braket{N_{\rm CG}}$ and $N_{\rm pair}$ are similar), we again reiterate that ice II has the largest $N_{\rm stcl}$ and $\npep$ values, and therefore requires the largest number of floating-point operations per CG step.
Although a more detailed analysis of the communication and processor idling would be required to fully explain the total \exxm timings during these $NpT$ simulations, we can still justify the $\braket{t_{\exxm}}$ ordering among these ice phases by noting that: (\textit{i}) the $\braket{t_{\exxm}^{\rm PE}} / \braket{t_{\exxm}}$ ratio is $\approx 30\%$ for all three ice phases, and (\textit{ii}) the communication overhead is roughly proportional to $\nmep$ (which is proportional to $\npep$).
Since ice II has the largest values for $\braket{t_{\exxm}^{\rm PE}}$ and $\npep$, both computation and communication costs will be largest for this ice phase; with computation and communication comprising a majority of $\braket{t_{\exxm}}$, the increased wall times observed across all three HPC architectures are not only reasonable but expected for $NpT$ simulations of this higher-density (and non-orthogonal) ice polymorph.

From this discussion, it is clear that $N_{\rm aux}$ (or $N_{\rm pure} = N_{\rm aux} + 3$) governs $N_{\rm stcl}$, and hence modulates (in conjunction with $\npep$) the number of floating-point operations during each step in the iterative CG solution to the PE.
In the ice II case presented above, we intentionally performed the $NpT$ simulation \textit{without} angular constraints to showcase a triclinic lattice with $N_{\rm aux}=3$ (or $N_{\rm pure} = 6$), thereby allowing for non-constrained microscopic cell fluctuations in ice II.
In doing so, the computational cost of this simulation was $\approx 50\%$ higher than one in which ice II would be constrained to maintain rhombohedral symmetry with $N_{\rm aux}=1$ (or $N_{\rm pure} = 4$), \ie the naturally-occurring and macroscopically-observed lattice symmetry for this ice phase.
In addition to the application of angular constraints to change $N_{\rm stcl}$ (\via $N_{\rm stcl} = 2 n N_{\rm pure} + 1 = 2 n (N_{\rm aux} + 3) + 1 $), alternative cell choices may also be used to control the size/extent (and hence computational complexity) of the NK Laplacian.
For instance, a hexagonal (or rhombohedral) lattice with $N_{\rm pure} = 4$ can be transformed into an orthorhombic lattice with $N_{\rm pure} = 3$; for the well-known hexagonal case (with $\bm L_1 \not\perp \bm L_2$), one can construct an orthorhombic (super-)cell with lattice vectors $\{\bm L'_1, \bm L'_2, \bm L'_3\}$ such that $\bm L'_1 = \bm L_1$, $\bm L'_2 = 2 \bm L_2 + \bm L_1$, and $\bm L'_3 = \bm L_3$.
However, this reduction in $N_{\rm pure}$ (and hence $N_{\rm stcl}$) is accompanied by the increased complexity of dealing with a simulation cell containing twice as many atoms; while such an increase in system size may be cumbersome for AIMD simulations, the additional degrees of freedom can also prove useful when describing the proton disorder in a system like ice I$h$.

\subsection{Parallel Scaling and Performance \label{performance:scalability_liq}}

Having discussed the computational complexity associated with different lattice symmetries, we now move on to assess the performance and parallel scaling of the extended \exxm module when applied to large-scale $NpT$ simulations of liquid water.
In close analogy to the critical assessment of \exxm during $NVT$ simulations of liquid water in Sec.~IV B of \pI~\cite{paper1}, this section will focus on the internode (\mpi{}) parallelization level \via a strong-scaling analysis (in which the number of processing elements is varied for a fixed problem size) and a weak-scaling analysis (in which the problem size is varied for a fixed ratio of problem size to number of processing elements).
We will also briefly discuss the intranode (\omp{}) parallelization level (which is particularly relevant for $NpT$ simulations using relatively large planewave basis sets) as well as the general performance of \exxm across several different HPC architectures (\eg \textit{Mira} IBM Blue Gene/Q, \textit{Cori} Haswell, and \textit{Cori} KNL).

Unless otherwise specified, the computational timings for each of the following liquid water simulations were obtained using the same planewave/pseudopotential/CPMD settings and \exxm parameters as those employed above for the ice I$h$, II, and III phases in Sec.~\ref{performance:latt_sym_ices}.
In contrast with the ice simulations (in which the system size was fixed at \ce{(H2O)96}, $T=238$~K, $p=2.1$~kBar, and $\zeta=4$), we follow the same profiling procedure given in Sec.~IV B of \pI,~\cite{paper1} by performing a series of $12$ different EXX-based CPMD simulations of liquid water at $T=300$~K and $p=1.0$~bar, in which: (\textit{i}) the system size was varied to include $N_{\rm water}=64,128,256$ water molecules (each of which has $N_{o} = 4 \times N_{\rm water}$ MLWFs), and (\textit{ii}) the number of processing elements ($N_{\rm proc}$ \mpi{} processes) was varied by changing $\zeta = N_{\rm proc}/N_{o}$.
Initial snapshots for each liquid water system were prepared following the equilibration procedure detailed in \pI~\cite{paper1}; in the $NpT$ simulations performed in this work, all instantaneous cell fluctuations were constrained to maintain simple cubic symmetry (\ie $\bm L_{1} \perp \bm L_{2}$, $\bm L_{1} \perp \bm L_{3}$, $\bm L_{2} \perp \bm L_{3}$, and $|\bm L_{1}| = |\bm L_{2}| = |\bm L_{3}|$) with $N_{\rm aux}=0$ and $N_{\rm pure}=3$.
Strong- and weak-scaling test were performed on \textit{Mira} IBM Blue Gene/Q (using $\zeta \in \{\sfrac{1}{2},1,2,4\}$ and $N_{\rm water} \in \{64, 128, 256\}$), with an additional assessment of the extended \exxm module on \textit{Cori} Haswell and \textit{Cori} KNL (using $\zeta = 1$ and $N_{\rm water} = 128$).
In each case, we again use one process per node for the internode \mpi{} parallelization and all available physical cores per node for the intranode \omp{} parallelization (with the hyperthreading settings described in Sec.~\ref{performance:latt_sym_ices}); following the discussion in Sec.~IV B of \pI,~\cite{paper1} the highest possible task-group parallelization level was employed (\via $N_{\rm tg} \in \{1,2,4,8\}$) for the 3D \texttt{FFT} operations in the non-\exxm portions of \texttt{QE}.

\begin{figure}[t!]
  \includegraphics[width=\linewidth]{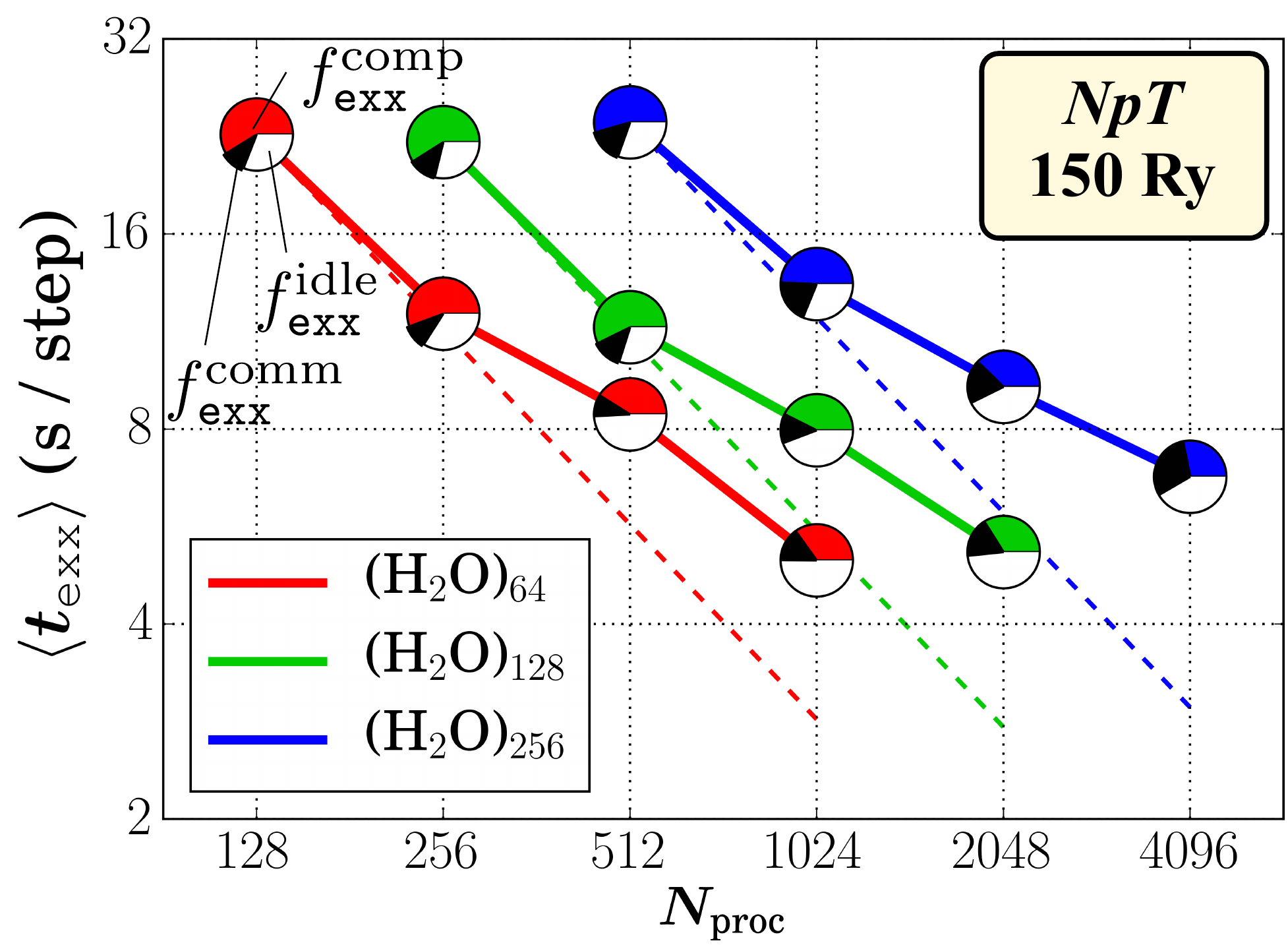}
  \caption{
  Strong-scaling analysis of the extended \exxm module in \texttt{QE} during $NpT$ CPMD simulations of liquid water at the hybrid PBE0 level on \textit{Mira} IBM Blue Gene/Q.
  For a fixed system size ($N_{\rm water} = 64$ (red line), $128$ (green line), $256$ (blue line)), the mean wall times (averaged over $50$ CPMD steps) spent in the \exxm module ($\braket{t_{\exxm}}$ in s/step) are plotted against the number of \mpi{} processes ($N_{\rm proc}$, varied \via $\zeta \equiv  N_{\rm proc}/N_{o} \in \{\sfrac{1}{2},1,2,4\}$).
  For comparison, ideal strong-scaling wall times (dashed lines) were computed with respect to the $\zeta_{\rm ref}=\sfrac{1}{2}$ case (see Eq.~\eqref{eq:eff_mpi_st}).
  Inset pie charts also depict the fraction/percent of $\braket{t_{\exxm}}$ dedicated to computation ($f_{\exxm}^{\rm comp}$, colored), communication ($f_{\exxm}^{\rm comm}$, black), and processor idling ($f_{\exxm}^{\rm idle}$, white).
  }
  \label{fig:mpi_scaling_st}
\end{figure}
When compared to the previous strong-scaling tests of \exxm on liquid water in the $NVT$ ensemble (with a \textit{fixed} simulation cell and real-space grid compatible with an $85$~Ry planewave cutoff, see Fig.~8 in \pI~\cite{paper1}), we again observe similar \mpi{} performance for the extended \exxm module in the $NpT$ ensemble (with a \textit{fluctuating} simulation cell and real-space grid compatible with the significantly larger $150$~Ry planewave cutoff, see Fig.~\ref{fig:mpi_scaling_st}).
For a given (and fixed) system size, we follow \pI~\cite{paper1} and define the strong-scaling efficiency of \exxm with respect to a reference $\zeta$ value (\ie $\zeta_{\rm ref}=\sfrac{1}{2}$, a commonly used setting for AIMD simulations of liquid water) as:
\begin{align}
  \eta^{\rm strong}_{\mpi{}}(\zeta) &\equiv 
  \frac{\zeta_{\rm ref} \cdot \braket{t_{\exxm}}_{\zeta_{\rm ref}}}{\zeta \cdot \braket{t_{\exxm}}_{\zeta}} = \frac{ \frac{1}{2} \cdot \braket{t_{\exxm}}_{\zeta = 1/2}}{\zeta \cdot \braket{t_{\exxm}}_{\zeta}} ,
  \label{eq:eff_mpi_st}
\end{align}
in which $\braket{t_{\exxm}}_{\zeta}$ is the wall time spent in \exxm when using a specific $\zeta$ value.
For $\zeta > \sfrac{1}{2}$, we find that $\eta^{\rm strong}_{\mpi{}}$ (when averaged over \ce{(H2O)64}, \ce{(H2O)128}, and \ce{(H2O)256}) decreases to $\approx 93\%$ ($\zeta=1$), $\approx 67\%$ ($\zeta=2$), and $\approx 52\%$ ($\zeta=4$).
Quite interestingly, the strong-scaling performance of \exxm in the more demanding $NpT$ ensemble is nearly identical to that observed for the same systems in the $NVT$ ensemble (see Fig.~8 and the surrounding discussion in \pI~\cite{paper1}), where we reported $\eta^{\rm strong}_{\mpi{}}$ values of $\approx 93\%$ ($\zeta=1$), $\approx 66\%$ ($\zeta=2$), and $\approx 50\%$ ($\zeta=4$).
In general, the \exxm module is more efficient for smaller $\zeta$ values (\ie $\zeta \leq 1$) since the use of massively parallel HPC resources ($\zeta \gg 1$) is intrinsically more susceptible to processor idling (due to the larger computational workload imbalance associated with more \mpi{} processes) and also requires additional/duplicate MLWF communication across the larger pool of \mpi{} processes.
See below for a more detailed breakdown of $\braket{t_{\exxm}}$ into computation, communication, and processor idling, as well as a discussion on how these components influence the strong-scaling efficiency of \exxm.

\begin{figure}[t!]
  \includegraphics[width=\linewidth]{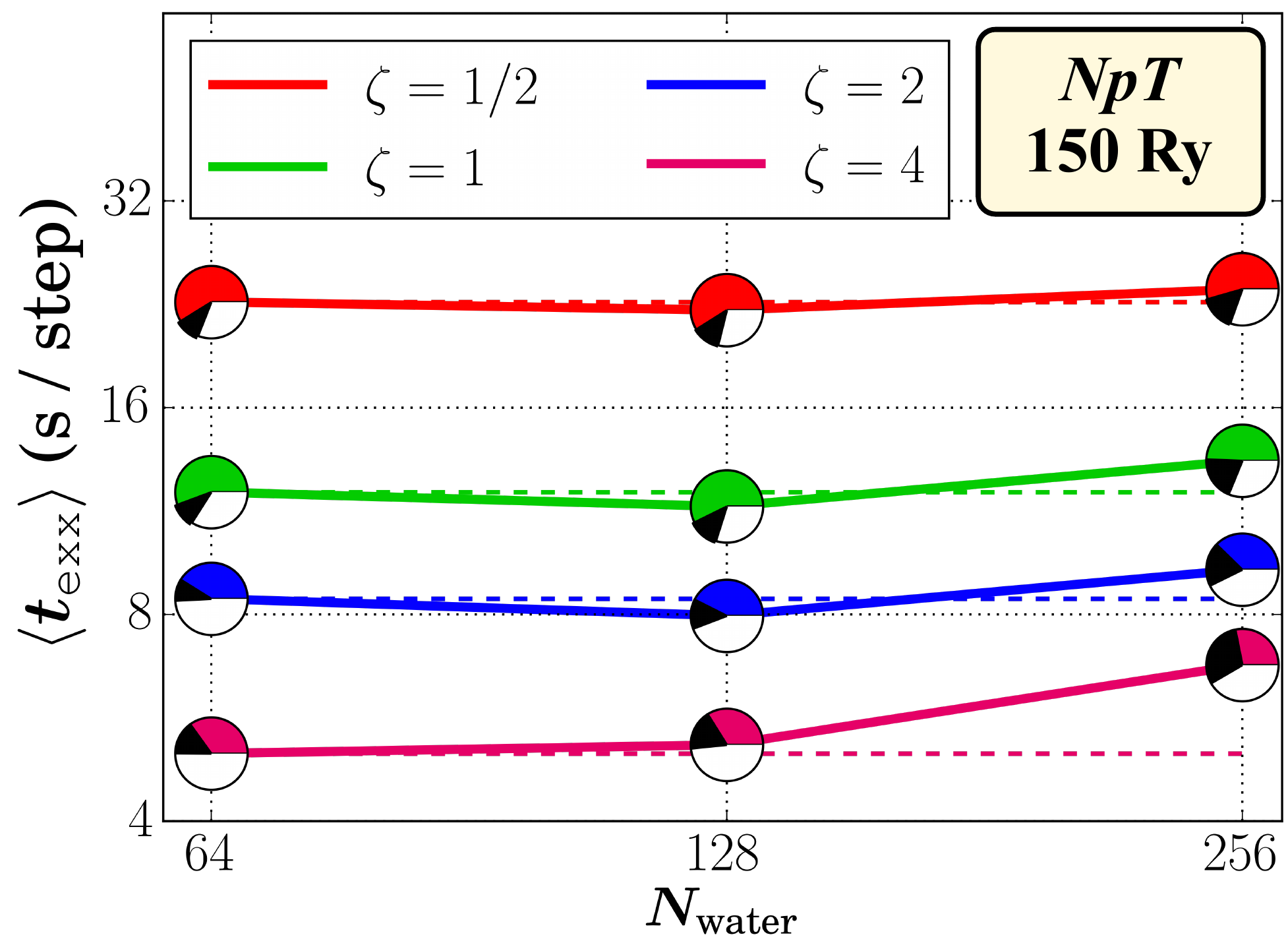}
  \caption{
  Weak-scaling analysis of the extended \exxm module in \texttt{QE} during $NpT$ CPMD simulations of liquid water at the hybrid PBE0 level on \textit{Mira} IBM Blue Gene/Q.
  For a fixed ratio of system size to number of processing elements ($\zeta = \sfrac{1}{2}$ (red line), $1$ (green line), $2$ (blue line), and $4$ (magenta line)), the mean wall times (averaged over $50$ CPMD steps) spent in the \exxm module ($\braket{t_{\exxm}}$ in s/step) are plotted against the system size ($N_{\rm water}$, varied to include \ce{(H2O)64}, \ce{(H2O)128}, and \ce{(H2O)256}).
  For comparison, ideal weak-scaling wall times (dashed lines) correspond to linear (or $\mathcal{O}(N)$) scaling and were computed with respect to the $N_{\rm water}=64$ case (see Eq.~\eqref{eq:eff_mpi_wk}).
  Inset pie charts again depict the fraction/percent of $\braket{t_{\exxm}}$ dedicated to computation ($f_{\exxm}^{\rm comp}$, colored), communication ($f_{\exxm}^{\rm comm}$, black), and processor idling ($f_{\exxm}^{\rm idle}$, white).
  }
  \label{fig:mpi_scaling_wk}
\end{figure}
When compared to the previous weak-scaling tests of \exxm (on liquid water in the $NVT$ ensemble, see Fig.~9 in \pI~\cite{paper1}), however, we observe a substantial improvement in the \mpi{} performance of the extended \exxm module during large-scale $NpT$ simulations (see Fig.~\ref{fig:mpi_scaling_wk}).
For a given (and fixed) $\zeta$ value, we again follow \pI~\cite{paper1} and define the weak-scaling efficiency of \exxm with respect to a reference system size (\ie $N_{\rm water}^{\rm ref}=64$, a commonly used system size for AIMD simulations of liquid water) as:
\begin{equation}
  \eta^{\rm weak}_{\mpi{}}(N_{\rm water}) \equiv \frac{\braket{t_{\exxm}}_{N_{\rm water}^{\rm ref}}}{\braket{t_{\exxm}}_{N_{\rm water}}} = \frac{\braket{t_{\exxm}}_{N_{\rm water}=64}}{\braket{t_{\exxm}}_{N_{\rm water}}} , 
  \label{eq:eff_mpi_wk}
\end{equation}
in which $\braket{t_{\exxm}}_{N_{\rm water}}$ is the the wall time spent in \exxm for a specific $N_{\rm water}$.
For $N_{\rm water} > 64$, we find that $\eta^{\rm weak}_{\mpi{}}$ (when averaged over $\zeta \in \{\sfrac{1}{2},1,2,4\}$) first slightly increases to $\approx 103\%$ ($N_{\rm water} = 128$) and then decreases to $\approx 89\%$ ($N_{\rm water} = 256$).
These weak-scaling efficiencies are marked improvements over the $NVT$ values of $\approx 89\%$ ($N_{\rm water}=128$) and $\approx 81\%$ ($N_{\rm water}=256$) reported in Fig.~9 (and the surrounding discussion) in \pI~\cite{paper1}, and demonstrate that the extended \exxm module is exhibiting close to linear (or $\mathcal{O}(N)$) scaling behavior in the \ce{(H2O)64}--\ce{(H2O)256} system size regime in the more demanding $NpT$ ensemble.
Here, we note in passing that the observed $\eta^{\rm weak}_{\mpi{}}$ value exceeding $100\%$ for \ce{(H2O)128} is merely an artifact of choosing \ce{(H2O)64} as the reference system size as well as averaging over all four $\zeta$ values; as such, we interpret this result as a simple indication that \exxm is scaling nearly ideally when the system is doubled from \ce{(H2O)64} to \ce{(H2O)128}.
When using a relatively low amount of computational resources (\eg $\zeta=1/2,1$), we find that the weak-scaling behavior of \exxm is quite close to linear scaling in the \ce{(H2O)64}--\ce{(H2O)256} system size regime (Fig.~\ref{fig:mpi_scaling_wk}).
However, the scalability starts to degrade for \ce{(H2O)256} at the $\zeta \ge 2$ level, which we largely attribute to: (\textit{i}) increased communication due to the underlying \texttt{ALL-TO-ALL} \mpi{} operations in the data redistribution steps (Steps I and VI in Fig.~\ref{fig:flowchart}), and (\textit{ii}) increased processor idling due to the inherent difficulty with balancing the workload across a larger number of \mpi{} processes (see below and Sec.~IV B in \pI~\cite{paper1}).
Furthermore, we also note that the weak-scaling efficiency of \exxm (in both the $NVT$ and $NpT$ ensembles) is significantly better than its strong-scaling efficiency; however, this result is not surprising as it is (in general) more efficient to distribute the additional workload associated with an increased system size over a larger number of processing units rather than use the increased processing resources to reduce the overall time to solution for a fixed system size.

Here, we remind the reader that the \exxm module only represents one portion of an overall hybrid DFT calculation: input into \exxm is the current set of MLWFs at a given CPMD step; output from \exxm is $\exx$, $\{ \dxx{i} \}$, and $\sxx$.
As such, several other modules in \texttt{QE} (some of which are not necessarily linear scaling) are required to perform the remaining non-\exxm tasks (\ie all other GGA-DFT operations as well as MLWF localization), and will ultimately dominate the overall scalability of a hybrid DFT calculation.
For instance, the cost associated with MLWF localization, which contains some cubic-scaling matrix operations, can become more substantial for larger system sizes (\eg $\approx 10\mathrm{-}20\%$ of the total wall time for \ce{(H2O)256}); see Table~1 in \pI~\cite{paper1} and the surrounding text for a more detailed discussion.
As such, incorporating the \exxm module into an overall linear-scaling GGA code---in conjunction with a more efficient \textit{on-the-fly} orbital localization procedure---could be a viable strategy for achieving a fully (overall) linear-scaling hybrid DFT approach.

For the largest systems considered in this work (\ie \ce{(H2O)128} and \ce{(H2O)256}), the extended \exxm module can evaluate all EXX-related quantities required to propagate the constant-pressure CPMD equations of motion in Eqs.~\eqref{eq:cpE}--\eqref{eq:cpC} in $\approx 5.2$~s/step for \ce{(H2O)128} and $\approx 6.8$~s/step for \ce{(H2O)256} using massively parallel HPC resources (\ie $\zeta = 4$) on the \textit{Mira} IBM Blue Gene/Q platform.
When compared to $NVT$ simulations of liquid water using \exxm and the same computational resources (\cf $\approx 2.0$~s/step for \ce{(H2O)128} and $2.4$~s/step for \ce{(H2O)256}, see Table~1 of \pI~\cite{paper1}), the increased wall times observed here mainly originate from the larger planewave cutoff (\cf $150$~Ry for $NpT$ {vs.} $85$~Ry for $NVT$) and hence the larger number of points in the real-space grid (\textit{vide infra}).
In practice, $50$~ps $NpT$ simulations of large systems like \ce{(H2O)128} and \ce{(H2O)256} would therefore require $\approx 1.0\mathrm{-}1.3$ months using similar HPC resources and a more conventional CPMD time step of $0.10$~fs. 
As such, the extended \exxm module enables very challenging large-scale $NpT$ simulations for extended length scales at the hybrid DFT level of theory.
\begin{table*}[ht!]
  \centering
  \caption{
  Computational timings profile for $NpT$ CPMD simulations of liquid water at the hybrid PBE0 level on \textit{Mira} IBM Blue Gene/Q, \textit{Cori} Haswell, and \textit{Cori} KNL using the extended \exxm module in \texttt{QE}.
  All timings (in s/step) were averaged over $50$ CPMD steps, and correspond to the mean wall times associated with completing all GGA (non-\exxm) contributions to the simulation ($\braket{t_{\rm GGA}}$), optimizing the Marzari-Vanderbilt functional ($\braket{t_{\rm MLWF}}$, needed to re-localize the MLWFs between each CPMD step), running through the entire \exxm module ($\braket{t_{\exxm}}$, \ie Steps I--VI in Fig.~\ref{fig:flowchart}), as well as performing a given CPMD step ($\braket{t_{\rm total}}$)~\cite{note_no_invfft_time}.
  Also included are the $\braket{t_{\exxm}}/\braket{t_{\rm GGA}}$ ratios, as well as a breakdown of $\braket{t_{\exxm}}$ into the following components: computation ($\braket{t_{\exxm}^{\rm comp}}$), communication ($\braket{t_{\exxm}^{\rm comm}}$), and processor idling ($\braket{t_{\exxm}^{\rm idle}}$); for convenience, the fraction/percent of $\braket{t_{\exxm}}$ dedicated to each of these components (\ie $f_{\exxm}^{\rm comp}$, $f_{\exxm}^{\rm comm}$, and $f_{\exxm}^{\rm idle}$) are reported as percentages of $\braket{t_{\exxm}}$.
  All timings were obtained during $NpT$ simulations of \ce{(H2O)128} with $\zeta = 1$, $N_{\rm tg} = 2$, and $512$ nodes on each architecture (using one \mpi{} process and all available physical cores per node).
  Hyperthreading was fully activated on each physical core, except for \textit{Cori} KNL, where hyperthreading was disabled to prevent performance degradation (\cf Table~2 in \pI~\cite{paper1}).
  See text for more details. 
  }
  \begin{tabular}{c|ccccc|ccc}
    \hline\hline
    \multirow{2}{*}{Architecture} & \multicolumn{5}{c|}{\texttt{QE} Module Timings} & \multicolumn{3}{c}{Breakdown of $\braket{t_\exxm}$} \\
    \cline{2-9}
    & $\braket{t_{\rm GGA}}$ & $\braket{t_{\rm MLWF}}$ & $\braket{t_{\exxm}}$ & $\braket{t_{\rm total}}$ & $\braket{t_{\exxm}}\!/\!\braket{t_{\rm GGA}}$ & $\braket{t_{\exxm}^{\rm comp}}$ ($f_{\exxm}^{\rm comp}$) & $\braket{t_{\exxm}^{\rm comm}}$ ($f_{\exxm}^{\rm comm}$) & $\braket{t_{\exxm}^{\rm idle}}$ ($f_{\exxm}^{\rm idle}$) \\
    \hline
    \textit{Mira} IBM Blue Gene/Q          & $\phantom{0}2.79$ & $0.59$ & $11.49$ & $14.87$ & $4.1$ & $6.58$\quad($57.3$) & $1.47$\quad($12.8$) & $3.43$\quad($29.9$) \\
    \textit{Cori} Haswell                 & $\phantom{0}1.16$ & $1.26$ & $\phantom{0}5.37$ & $\phantom{0}7.79$ & $4.6$ & $2.96$\quad($55.1$) & $0.77$\quad($14.3$) & $1.65$\quad($30.6$) \\
    \textit{Cori} KNL (no hyperthreading$^a$) & $12.25$ & $3.71$ & $\phantom{0}9.16$           & $25.12$ & $0.7$ & $4.85$\quad($52.9$) & $2.18$\quad($23.8$) & $2.13$\quad($23.3$) \\
    \textit{Cori} KNL (no hyperthreading$^b$) & $\phantom{0}5.10$  & $1.98$ & $\phantom{0}8.20$ & $15.28$ & $1.6$ & $4.61$\quad($56.2$) & $1.60$\quad($19.5$) & $1.99$\quad($24.3$) \\
    \hline\hline
    \multicolumn{9}{l}{\scriptsize{$^{a}$Using default \omp{} settings (\ie the same settings used in \pI~\cite{paper1}). $^{b}$Using \texttt{OMP\_PROC\_BIND = true} and \texttt{OMP\_PLACES = cores}.}}
  \end{tabular}
  \label{tab:arch_bench}
\end{table*}

Similar to \pI~\cite{paper1}, we further investigate the \exxm wall times by breaking $\braket{t_{\exxm}}$ into the following contributions: computation events ($\braket{t_{\exxm}^{\rm comp}}$), communication overhead ($\braket{t_{\exxm}^{\rm comm}}$), and processor idling  due to workload imbalance ($\braket{t_{\exxm}^{\rm idle}}$).
For convenience, the fraction/percent of $\braket{t_{\exxm}}$ dedicated to each of these components (\ie $\braket{f_{\exxm}^{\rm comp}}$, $\braket{f_{\exxm}^{\rm comm}}$, and $\braket{f_{\exxm}^{\rm idle}}$) are depicted as pie charts in Figs.~\ref{fig:mpi_scaling_st} and \ref{fig:mpi_scaling_wk}. 
For the $12$ $NpT$ simulations performed in this work, we find that all three of these components are larger in magnitude than in the $NVT$ case, and still represent sizable contributions to $\braket{t_{\exxm}}$.
As mentioned above, the increased wall times reported herein are a direct consequence of the larger planewave cutoffs employed during constant-pressure $NpT$ simulations; by increasing the cutoff from $85$~Ry ($NVT$) to $150$~Ry ($NpT$), the density of real-space grid points in $\Omega$ (as well as $\pe{\gC{ij}}$ and $\me{\gC{ij}}$) increases by a factor of $2.5\mathrm{-}2.7\times$. 
For the computational cost, the larger $\npe$ increases the number of steps (as well as the computational complexity per step) during the iterative CG solution to the PE (see Sec.~\ref{performance:latt_sym_ices}), while the larger $\nme$ increases the cost of the ME.
For the communication overhead, the larger grid density requires sending/receiving larger chunks of data during the forward/backward redistribution (\eg Steps I and VI in Fig.~\ref{fig:flowchart}, to maintain compatibility with \texttt{QE}) as well as the internal communication needed to compute each $\braket{ij}$ contribution to the energy, wavefunction forces, and stress tensor (\eg Steps III--V).
With an increased computational cost per overlapping MLWF pair, the larger $\npe$ and $\nme$ also lead to more extended processor idling times due to the intrinsic imperfect distribution of $\braket{ij}$ pairs across \mpi{} processes (see Secs.~III C 2 and IV B 1 in \pI~\cite{paper1}).
Cell fluctuations during $NpT$ simulations further impact the processor idling in \exxm by introducing larger variability in the time to solution for each PE (due primarily to variable-quality guesses based on previous CPMD steps) as well as additional imbalance in the computational workload (due to the more diverse local environments sampled by each MLWF).

For small $\zeta$ values ($\zeta \le 1$), we find that \exxm is technically computation-bound, with $\braket{f_{\exxm}^{\rm comp}} \approx 56\%$, $\braket{f_{\exxm}^{\rm comm}} \approx 13\%$, and $\braket{f_{\exxm}^{\rm idle}} \approx 31\%$ (when averaged over $\zeta = \sfrac{1}{2}$ and $\zeta = 1$ for \ce{(H2O)64}, \ce{(H2O)128}, and \ce{(H2O)256}), although the wall time associated with communication overhead and processor idling ($\approx 44\%$) still remains substantial. 
With HPC resources ($\zeta \gg 1$), the balance among computation and processor idling is now switched, with $\braket{f_{\exxm}^{\rm comp}} \approx 36\%$, $\braket{f_{\exxm}^{\rm comm}} \approx 18\%$, and $\braket{f_{\exxm}^{\rm idle}} \approx 46\%$ (when averaged over $\zeta = 2$ and $\zeta = 4$ for \ce{(H2O)64}, \ce{(H2O)128}, and \ce{(H2O)256}), but the combined computation and communication cost ($\approx 54\%$) is technically dominant.
In this limit, we have previously observed a roughly equal distribution of $f_{\exxm}^{\rm comm} \approx f_{\exxm}^{\rm comp} \approx f_{\exxm}^{\rm idle} \approx 33\%$ during large-scale $NVT$ simulations of liquid water (\ie \ce{(H2O)256} with $\zeta = 4$, see Table~1 and Figs.~8--9 in Sec.~IV B 1 of \pI~\cite{paper1}); in the more challenging $NpT$ case investigated here, the role of processor idling has become even more prominent in determining the overall time to solution, while the (albeit reduced) relative contributions from computation and communication are still considerable.
As such, we are in the process of developing a comprehensive three-pronged theoretical and algorithmic approach (\ie the $\beta$ version of \exxm) that specifically addresses each of these sizable contributions to $\braket{t_{\exxm}}$ and will enable hybrid-DFT based $NpT$ simulations of even larger systems and longer durations.

We complete this section with a brief discussion on intranode \omp{} parallelization efficiency as well as the overall performance of \exxm when performing large-scale $NpT$ simulations on different HPC architectures.
Regarding the \omp{} strong-scaling efficiency, we point the reader to Fig.~10 (as well as the surrounding text in Sec.~IV B 2) in \pI~\cite{paper1}, where we specifically investigated the performance of \exxm during Step~IV (the computational bottleneck of \exxm) using two different planewave cutoffs: $85$~Ry and $150$~Ry (to mimic the typical settings employed during $NVT$ and $NpT$ simulations)~\cite{note_openmp}.
When performing these simulations, we found that \exxm maintains high strong-scaling efficiencies with $\eta^{\rm strong}_{\omp{}}$ values (see Eq.~(40) in \pI~\cite{paper1}) of $\approx 84\%$ ($85$~Ry) and $\approx 92\%$ ($150$~Ry) as the number of \omp{} threads was increased from one (single-thread limit) to $16$ (complete activation of all physical cores) per \textit{Mira} IBM Blue Gene/Q node (with a further $30\mathrm{-}40\%$ boost when all $64$ hyperthreads were activated).
Since the computational workload assigned to each thread increases with the planewave cutoff, the \omp{} efficiency of \exxm generally increases during large-cutoff ($NVT$ or $NpT$) simulations; as such, we expect that \exxm will also benefit from the use of advanced vectorization techniques as well as offloading to graphics processing units (GPUs).

As a final assessment of the extended \exxm module, we repeated the $\zeta=1$ $NpT$ CPMD simulations of \ce{(H2O)128} on the \textit{Cori} Haswell and \textit{Cori} KNL supercomputer architectures located at the National Energy Research Scientific Computing Center (NERSC).
In analogy to the $NVT$ timing profiles provided in Table~2 of \pI~\cite{paper1}, Table~\ref{tab:arch_bench} shows that there exists some variability in the individual \texttt{QE} module timings across all three architectures, with $\braket{t_{\rm GGA}}$ ranging from $\approx 1.2$~s/step (Haswell) to $\approx 12.3$~s/step (KNL), and $\braket{t_{\exxm}}$ ranging from $\approx 5.4$~s/step (Haswell) to $\approx 11.5$~s/step (IBM Blue Gene/Q).
With $\braket{t_{\exxm}}/\braket{t_{\rm GGA}} = 0.7\mathrm{-}4.6$, we again observe that the extended \exxm module requires a wall time cost that is comparable to semi-local DFT, and therefore enables large-scale constant-pressure AIMD simulations at the hybrid DFT level.
Here, we note in passing that the performance of \exxm (as well as the non-\exxm portions of \texttt{QE}) on \textit{Cori} KNL is quite sensitive to the \omp{} settings as well as the use of hyperthreading.
For instance, refining the default \omp{} settings on \textit{Cori} KNL (by specifying \texttt{OMP\_PROC\_BIND = true} and \texttt{OMP\_PLACES = cores}) leads to an $\approx 60\%$ reduction in $\braket{t_{\rm GGA}}$ from $12.3$~s/step to $5.1$~s/step, accompanied by a more modest (but still noticeable) $\approx 10\%$ reduction in $\braket{t_{\exxm}}$ from $9.2$~s/step to $8.2$~s/step.
In fact, a $150$~Ry $NpT$ simulation of \ce{(H2O)128} on \textit{Cori} KNL using these refined settings ($\braket{t_{\rm GGA}} = 5.1$~s/step) can actually be performed faster than an $85$~Ry $NVT$ simulation of the same system using the default \omp{} settings ($\braket{t_{\rm GGA}} = 5.4$~s/step, see Table~2 of \pI~\cite{paper1}).
Within the \exxm module, we find that the breakdown of $\braket{t_{\exxm}}$ into computation, communication, and processor idling is very similar across these three HPC architectures and quite consistent with that reported in Table~2 of \pI~\cite{paper1} for the analogous $NVT$ case.
In this system size and $\zeta$ regime, \exxm is technically computation-bound ($f_{\exxm}^{\rm comp} = 55.4 \pm 1.9\%$), with communication ($f_{\exxm}^{\rm comm} = 17.6 \pm 5.0\%$) and processor idling ($f_{\exxm}^{\rm idle} = 27.0 \pm 3.8\%$) accounting for the remainder of the time spent in the \exxm module.
With sizable contributions from all three components, this observation once again reiterates the need for a comprehensive three-pronged strategy in the next-generation \exxm codebase.
Hence, the combination of the current (and next-generation) \exxm codebase---along with an overall linear-scaling GGA implementation and a more efficient \textit{on-the-fly} orbital localization scheme---could be a viable route towards a fully linear-scaling hybrid DFT approach.

\section{Conclusions and Future Outlook \label{sec:conclusion}}

In this work, we present several theoretical and algorithmic developments to our linear-scaling and real-space MLWF-based EXX approach~\cite{paper1} that enable constant-pressure CPMD simulations (in the $NpH$ and/or $NpT$ ensembles) of large-scale finite-gap condensed-phase systems in general/non-orthogonal cells at the hybrid DFT level.
For the theoretical extension to this approach, we derived an analytical expression for the EXX contribution to the stress tensor for systems with general and fluctuating simulation cells with a computational complexity that scales linearly with system size.
When used in conjunction with the previously developed theoretical approaches for obtaining the EXX contribution to the energy and wavefunction forces,~\cite{paper1} this work provides the remaining ingredient needed for propagating the CPMD equations of motion under constant-pressure conditions, and hence an overall order-$N$ method for performing large-scale hybrid DFT based CPMD simulations in the $NVE$/$NVT$ as well as $NpH$/$NpT$ ensembles.
For the algorithmic extension to this approach, we have incorporated a number of new routines into the \exxm module in \qe (\texttt{QE}) that have been optimized to: (\textit{i}) provide generalized subdomains that handle both static and fluctuating simulation cells with non-orthogonal lattice symmetries, (\textit{ii}) solve Poisson's equation (PE) in general/non-orthogonal cells \via an automated selection of the auxiliary grid directions in the Natan-Kronik (NK) representation of the discrete Laplacian operator, and (\textit{iii}) evaluate the EXX contribution to the stress tensor using the analytical expression derived in this work.

This was followed by a case study demonstrating that one can use \exxm---with an appropriate choice of parameters---to tightly and simultaneously converge the EXX contributions to the energy and stress tensor for a wide variety of condensed-phase systems (including liquid \ce{(H2O)128}, the monoclinic benzene-II polymorph, and semi-conducting \ce{Si216} crystal).
We also provided a critical assessment of the computational performance of the extended massively parallel hybrid \mpi{}/\omp{} based \exxm module across several different HPC architectures (\eg \textit{Mira} IBM Blue Gene/Q, \textit{Cori} Haswell, and \textit{Cori} KNL) \via detailed case studies on: (\textit{i}) the computational complexity due to lattice symmetry during short $NpT$ simulations of the ice I$h$, II, and III polymorphs at their corresponding triple point, and (\textit{ii}) the strong- and weak-scaling of \exxm during large-scale $NpT$ simulations of ambient liquid water ranging from \ce{(H2O)64} to \ce{(H2O)256}.
In doing so, we found that evaluation of the EXX contribution to the stress tensor required negligible ($<1\%$) computational overhead for all systems tested, thereby providing a simultaneously more accurate and more computationally efficient approach than direct numerical differentiation of $\exx$ with respect to $\bm{h}$.
We also demonstrate that the extended \exxm module remains quite robust and highly scalable when performing challenging $NpT$ simulations of liquid water (with a very tight $150$-Ry planewave cutoff); here, we found that the \mpi{} strong scaling behavior remains essentially the same as that observed during $85$~Ry $NVT$ simulations in \pI~\cite{paper1}, while the \mpi{} weak scaling efficiency of \exxm becomes noticeably improved.
With these theoretical and algorithmic advances, the extended \exxm module brings us another step closer to routinely performing high-fidelity hybrid DFT based AIMD simulations of sufficient duration for complex and large-scale condensed-phase systems across a wide range of thermodynamic conditions.

Moving forward, our group is in the process of further improving the strong and weak scaling efficiencies of \exxm by implementing a comprehensive three-pronged strategy that simultaneously attacks the remaining contributions from computation, communication, and processor idling to the wall time cost.
Our group is also actively working on a variable subdomain generalization of the \exxm module for an accurate and computationally efficient treatment of EXX in heterogeneous systems with multiple phases and/or components, which is needed for the study of physical processes and chemical reactions in diverse environments and complex interfaces.
Other future research directions include optimizing \exxm for performing high-throughput calculations needed for machine-learning intra-/inter-molecular potentials of condensed-phase systems, as well as extending \exxm to sample other statistical ensembles (\ie $\mu VT$) needed for simulating even larger swaths of experimental conditions at the hybrid DFT level.

\begin{acknowledgements}
All authors thank Roberto Car, Amir Natan, Tatsuhiro Onodera, and Leeor Kronik for helpful scientific discussions.
This material is based upon work supported by the National Science Foundation under Grant No. CHE-1945676.
RAD also gratefully acknowledges financial support from an Alfred P. Sloan Research Fellowship.
This research used resources of the National Energy Research Scientific Computing Center, which is supported by the Office of Science of the U.S. Department of Energy under Contract No. DE-AC02-05CH11231.
This research used resources of the Argonne Leadership Computing Facility at Argonne National Laboratory, which is supported by the Office of Science of the U.S. Department of Energy under Contract No. DE-AC02-06CH11357.
\end{acknowledgements}

\providecommand{\latin}[1]{#1}
\makeatletter
\providecommand{\doi}
  {\begingroup\let\do\@makeother\dospecials
  \catcode`\{=1 \catcode`\}=2 \doi@aux}
\providecommand{\doi@aux}[1]{\endgroup\texttt{#1}}
\makeatother
\providecommand*\mcitethebibliography{\thebibliography}
\csname @ifundefined\endcsname{endmcitethebibliography}
  {\let\endmcitethebibliography\endthebibliography}{}


\begin{mcitethebibliography}{144}
\bibitem[Upp()]{UpperGnote}
For increased computational efficiency, sweeps over grid-point shells account
  for inversion symmetry in the real-space grid, and only include non-axial
  grid points with $\xi_{3} - (\xi_{0})_{3} > 0$ and $\xi_{2} - (\xi_{0})_{2} >
  0$ when $\xi_{3} - (\xi_{0})_{3} = 0$. As such, there are only $10$ grid
  points in $\mathbb{G}_{1}$.\relax
\mciteBstWouldAddEndPunctfalse
\mciteSetBstMidEndSepPunct{\mcitedefaultmidpunct}
{}{\mcitedefaultseppunct}\relax
\EndOfBibitem
\bibitem[not()]{note_our_code_repo}
A development version of the code used in this work is available for download upon request.\relax
\mciteBstWouldAddEndPuncttrue
\mciteSetBstMidEndSepPunct{\mcitedefaultmidpunct}
{\mcitedefaultendpunct}{\mcitedefaultseppunct}\relax
\EndOfBibitem
\bibitem[not()]{note_no_invfft_time}
As discussed in \pI~\cite{paper1}, we found a deprecated \texttt{invFFT} call
  in the current version of \texttt{QE} that unnecessarily recomputes the MLWFs
  in real space. As such, the $\braket{t_{\rm Total}}$ values reported in
  Table~\ref{tab:arch_bench} do not include the computational cost associated
  with this routine.\relax
\mciteBstWouldAddEndPunctfalse
\mciteSetBstMidEndSepPunct{\mcitedefaultmidpunct}
{}{\mcitedefaultseppunct}\relax
\EndOfBibitem
\bibitem[not()]{note_openmp}
Although these tests were based on short ($50$-step) $NVT$ simulations, the
  average number of floating point operations executed during each CPMD step
  (for the simulation with the $150$~Ry cutoff) is essentially the same as that
  during a (similarly short) $NpT$ simulation using the same cutoff. Taken
  together with the fact that the evaluation of $\sxx$ introduces minor ($<
  1\%$) computational overhead (see Sec.~\ref{performance:latt_sym_ices}), we
  believe that this test is an accurate and reliable assessment of the \omp{}
  strong-scaling efficiency of the extended \exxm module introduced in this
  work.\relax
\mciteBstWouldAddEndPunctfalse
\mciteSetBstMidEndSepPunct{\mcitedefaultmidpunct}
{}{\mcitedefaultseppunct}\relax
\EndOfBibitem
\end{mcitethebibliography}
\end{document}